\documentclass[twocolumn]{aastex631}

\usepackage{bm}
\usepackage{amsmath}
\usepackage{graphicx}       
\usepackage{multirow}
\usepackage{soul}
\usepackage{enumerate}

\begin{document}

\title{Low-Acceleration Gravitational Anomaly from Bayesian 3D Modeling of Wide Binary Orbits: Methodology and Results with Gaia Data Release 3}

\correspondingauthor{Kyu-Hyun Chae}
\email{chae@sejong.ac.kr, kyuhyunchae@gmail.com}

\author[0000-0002-6016-2736]{Kyu-Hyun Chae}
\affiliation{Department of Physics and Astronomy, Sejong University, 209 Neungdong-ro Gwangjin-gu, Seoul 05006, Republic of Korea}



\begin{abstract}
Isolated wide binary stars provide natural laboratories to directly probe gravity for Newtonian acceleration of $g_{\rm{N}}\lesssim 10^{-9}$~m\,s$^{-2}$. Recent statistical analyses of wide binaries have been performed only with sky-projected relative velocities $v_p$ in the pairs. A new method of Bayesian orbit modeling exploiting three relative velocity components including the radial (line-of-sight) component $v_r$ is developed to measure the gravitational anomaly parameter $\Gamma\equiv\log_{10}\sqrt{\gamma_g}\equiv\log_{10}\sqrt{G_{\rm{eff}}/G_{\rm{N}}}$ where $G_{\rm{eff}}$ is the effective gravitational constant for pseudo-Newtonian elliptical orbits, while $G_{\rm{N}}$ is Newton's constant. The method infers individual probability distributions of $\Gamma$ and then combines the independent distributions to obtain a consolidated distribution in a specific range of $g_{\rm{N}}$. Here the method is described and applied to a sample of 312 wide binaries in a broad dynamic range of $10^{-11.0}\la g_{\rm{N}}\la 10^{-6.7}$~m\,s$^{-2}$ with $v_r$ uncertainties in the range $168<\sigma_{v_r}<380$~m\,s$^{-1}$ selected from the Gaia Data Release 3 database. The following results are obtained: $\Gamma = 0.000\pm 0.011$ or $\gamma_g=1.00\pm 0.05$ ($N_{\rm{binary}}=125$) for a high acceleration regime ($10^{-7.9} \lesssim g_{\rm{N}} \lesssim 10^{-6.7}$~m\,s$^{-2}$) agreeing well with Newton, but $\Gamma = 0.085\pm 0.040$ or $\gamma_g=1.48_{-0.23}^{+0.33}$ (35) for the modified Newtonian dynamics (MOND) regime ($10^{-11.0}\lesssim g_{\rm{N}}\lesssim 10^{-9.5}$~m\,s$^{-2}$) and $\Gamma = 0.063\pm 0.015$ or $\gamma_g=1.34_{-0.08}^{+0.10}$ (111) for the MOND+transition regime ($10^{-11.0}\lesssim g_{\rm{N}}\lesssim 10^{-8.5}$~m\,s$^{-2}$). These results show that gravitational anomaly is evident for $g_{\rm{N}}\lesssim 10^{-9}$~m\,s$^{-2}$ and $\Gamma$ in the MOND regime ($\lesssim 10^{-9.5}$~m\,s$^{-2}$) agrees with the first-tier prediction ($\Gamma \approx 0.07$ or $\gamma_g \approx 1.4$) of MOND gravity theories.
\end{abstract}



\section{Introduction} \label{sec:intro}

The nature of gravity in the low acceleration limit of the nonrelativistic regime is of great importance because any deviation from standard gravity of Newton and Einstein represented by Poisson's equation will have profound implications for the concept of dark matter, dynamics of astrophysical systems, cosmological theories, and fundamental theories of physics. For the past decade or so, wide binary stars have been recognized \citep{hernandez2012,pittordis2018,banik2018,hernandez2022} as potentially decisive testbeds of gravity at low acceleration because a significant number of wide binaries can be found in the solar neighborhoods, permitting relatively direct observational studies and their internal dynamics is free from the effects of hypothetical dark matter.

Several statistical analyses of wide binary internal motions were carried out during 2023 - 2024 based on Gaia Data Release 3 \citep[DR3;][]{dr3}, by Chae \citep{chae2023a,chae2024a,chae2024c} and independently by Hernandez et al. \citep{hernandez2023,hernandez2024a,hernandez2025}, wich indicate that standard gravity is not only broken at low internal Newtonian acceleration ($g_{\rm{N}}\la 10^{-9}$~m~s$^{-2}$) but gravity is boosted by approximately 40\% at $g_{\rm{N}}\la 10^{-10}$~m~s$^{-2}$ in agreement with the generic prediction of modified gravity theories \citep{bekenstein1984,milgrom2010} in the paradigm of modified Newtonian dynamics \citep[MOND;][]{milgrom1983}. Although divergent results were published during the same period, the recent critical review by \cite{hernandez2024b} (see also \citealt{chae2024c} for more details) addresses the origin of the divergent results, in particular the lack of the self-calibration procedure for the fraction of hierarchical systems in an impure binary sample (i.e., including hierarchical systems).

All recently published results are based on the relative sky-projected (i.e., transverse) velocities $v_p$ in the pairs rather than the relative physical velocities $\mathbf{v}$ (or its magnitude $v$). This is because the Gaia DR3 database provides precise proper motions and parallaxes (so precise $v_p$) but far less precise or no relative radial (i.e., line-of-sight) velocities $v_r$. Existing tests with $v_p$ alone are less direct than using $\mathbf{v}$ itself, and thus involve more complications. Furthermore, gravity cannot be individually constrained for each binary, but only a statistical population can be compared with the corresponding population predicted by a gravity model in conjunction with statistical correlations/assumptions of the binary population (see \citealt{chae2023a}).

For an ideal individual measurement of gravity (with known mass or vice versa) with a binary system, its orbit would have to be traced for a significant fraction of the period with well-measured $\mathbf{v}$ values. In reality, due to the long periods $\sim 10^{5-6}$~yr for wide binaries of interest (see, e.g., Figure~2 of \citealt{chae2024a}), we will have to work essentially with one value of $\mathbf{v}$ at a random phase of the orbit. It is then unclear how well gravity can be individually constrained by such a snapshot observation of the orbital motion even if $\mathbf{v}$ is well measured. One may imagine that any individual inference of gravity will cover a broad range of possibilities, if any meaningful inference is feasible at all. To address this question in search of a more direct test of weak gravity than existing analyses, here I develop a Bayesian orbit modeling method with snapshot measurements of wide binary internal motions for hundreds of exceptional systems from the Gaia DR3 as well as mock wide binaries with controlled precision of $v_r$ inspired by ongoing and future observations. Note that, as details will be given below, even the exceptional systems from the Gaia DR3 have much less precise $v_r$ than $v_p$. 

Not surprisingly, as will be shown in this work, Bayesian-inferred individual probability density functions (PDFs) of gravity with the Gaia DR3 database are too broad to distinguish gravity theories of current interest (in particular, Newton and MOND). This will, in general, be true even for future data with much more precise $v_r$ due to the snapshot nature of the observations. However, when individual PDFs represent independent inferences of the same quantity (a gravity parameter in the present case) with different data, the PDFs may be combined to maximize the information from all data \citep{hill2011a,hill2011b}. With such a statistical combination/consolidation method called ``conflation'', we will see that even the current data clearly indicate a gravitational anomaly in the low-acceleration regime, in line with the results of 2023-2024 based on $v_p$. This holds great promise for future studies with a significant number of precise values of $v_r$.

This paper is structured as follows. In Section~\ref{sec:method}, I describe the methodology and data. In detail, I first develop a Bayesian methodology to infer PDFs of gravity from individual binaries (Section~\ref{sec:bayes}), and then describe the method of statistically combining individual PDFs to obtain a consolidated probability distribution by maximizing the information (Section~\ref{sec:demo}). Section~\ref{sec:sample} describes the selection of a pure binary sample from the Gaia DR3 database that will be used for individual Bayesian modeling. The selection of the sample will be more stringent than previous samples that were used mainly for median statistics in bins of accelerations (or equivalent variables) by \cite{chae2024a,chae2024c}.  In Section~\ref{sec:result}, I describe the results of Bayesian modeling based on the Gaia sample. In Section~\ref{sec:result_intro}, I first examine the individual Bayesian results and the overall properties of the results. In Section~\ref{sec:result_newt}, I present the consolidated probability distributions of gravity with wide binaries in a Newtonian (relatively strong internal acceleration) regime. In Section~\ref{sec:result_mond}, I present the main results of this work, i.e., the consolidated probability distributions of gravity in low-acceleration regimes. In Section~\ref{sec:disc}, I discuss the meaning of the results and future wide binary gravity research. In Section~\ref{sec:conc}, I present the current conclusion and prospects with future data on radial velocities (RVs). In Appendix~\ref{sec:Euler}, a general transformation with the Euler angles between the orbital plane and the observer's frame is described as it is needed to produce mock samples. {In Appendix~\ref{sec:pers}, perspective effects in PMs and RVs of wide binary stars are investigated to explore any possible biases in the low-acceleration regime data.}

For a quick grasp of this rather lengthy paper, the reader may read Sections~\ref{sec:bayes}, \ref{sec:demo}, \ref{sec:result_newt}, \ref{sec:result_mond}, \ref{sec:meaning}, and \ref{sec:conc}.  {The wide binary samples and the main Bayesian modeling results are available on Zenodo under an open-source Creative Commons Attribution license: 
\dataset[10.5281/zenodo.14900456]{https://doi.org/10.5281/zenodo.14900456}. The key Python codes are also available \citep{chae2025}.}

\section{Methodology and Data} \label{sec:method}

\subsection{Bayesian 3D modeling} \label{sec:bayes}

\begin{figure*}
  \centering
  \includegraphics[width=0.8\linewidth]{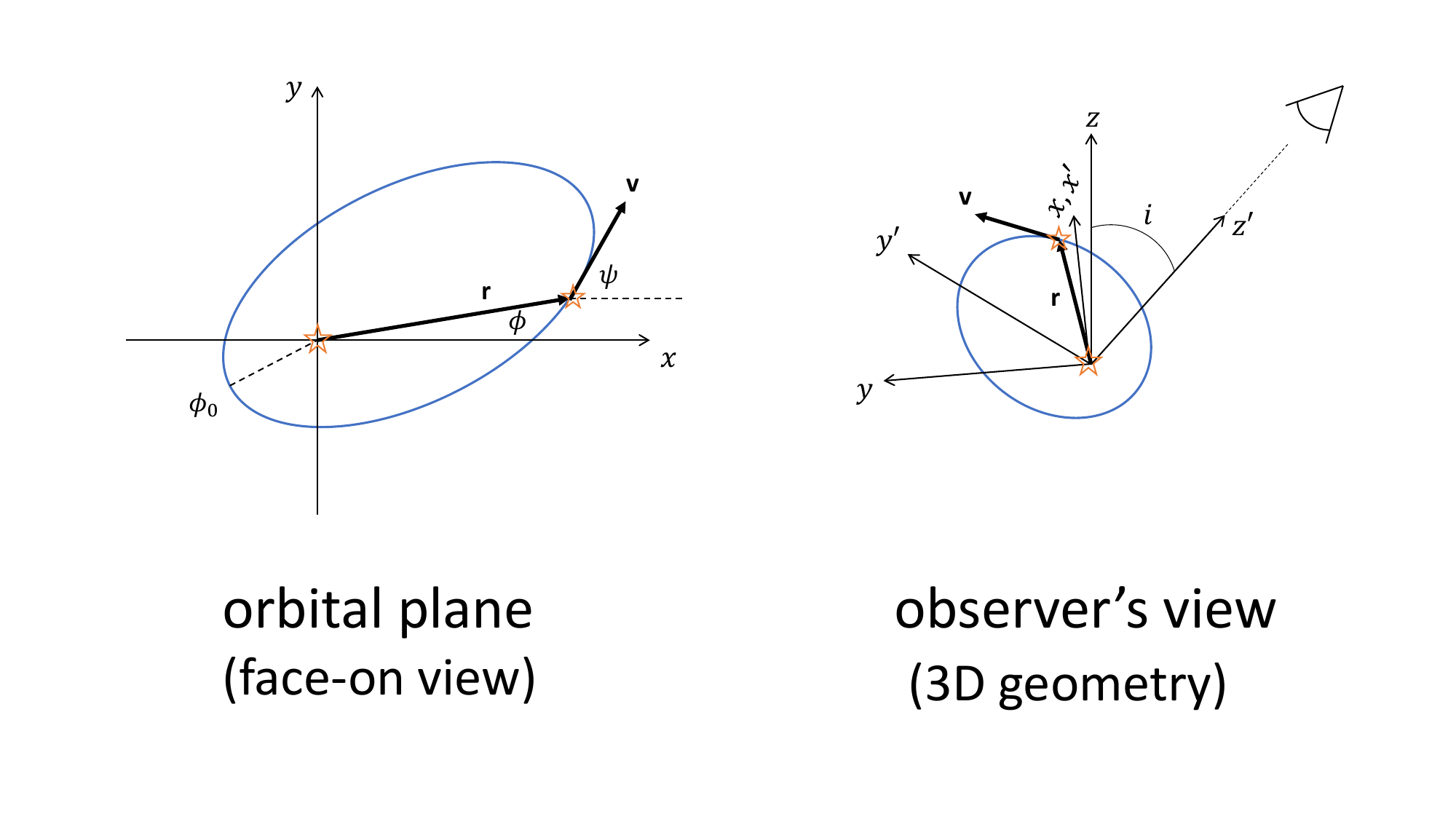}
    \vspace{-0.2truecm}
    \caption{\small 
    Adapted from \cite{chae2023a}. The left panel shows the motion of one star relative to the other, which can also be described by a one-particle equivalent motion of a reduced mass of the two stars in Newtonian and pseudo-Newtonian dynamics. The right panel defines the observer's plane of the sky and its rotation about the $x$-axis (an ``inclination'') $i$ with respect to the orbital plane.
    } 
   \label{fig:orbit}
\end{figure*} 

A gravitational anomaly with respect to the prediction of Newtonian gravity can be defined and tested/measured in various ways. Because of the extremely long periods of wide binaries compared with the observational time span, only one relative velocity is observed in effect. In Newtonian gravitational dynamics, the relative motion between the pair in an isolated binary system follows an ellipse. In nonstandard theories of gravity such as MOND gravity (e.g., \citealt{bekenstein1984,milgrom2010}), the orbits of wide binaries under the external field of the Milky Way oriented at an arbitrary angle with respect to the orbital plane may not be exact ellipses. However, whether the actual orbits are exact ellipses or not, I will work with elliptical orbits. The assumption of elliptical orbits will have no impact on testing Newtonian gravity, i.e., whether a gravitational anomaly is present or not. However, when a gravitational anomaly is present, the quantified gravitational anomaly must be understood as an effective quantity for elliptical orbits. 

I consider an effective gravitational constant $G_{\rm{eff}}$ (to be also referred to as $G$ for simplicity) for pseudo-Newtonian gravity and use a gravitational anomaly parameter taken from \cite{chae2024c} as
\begin{equation}
    \Gamma \equiv \log_{10}\sqrt{\gamma_g} \equiv \log_{10}\sqrt{G_{\rm{eff}}/G_{\rm{N}}},
    \label{eq:anomaly}   
\end{equation}
where $\gamma_g$ is the gravitational boost factor. 

Consider Figure~\ref{fig:orbit}, which connects the observer's primed frame with the unknown orbital plane. Note that the $x$ and $x^\prime$ axes coincide by the definition of the two coordinate systems in this figure. (See Appendix~\ref{sec:Euler} for fully general transformations.) As in Gaia observations, suppose that observed quantities for the two stars of each binary are the R.A. ($\alpha_k$ with $k=A,B$ referring to the brighter and fainter components hereafter) and decl. ($\delta_k$) given in units of degree, PMs ($\mu^\star_{\alpha,k}$ and $\mu_{\delta,k}$) given in units of mas\,yr$^{-1}$, and RVs ($\text{RV}_{k}$) given in units of km\,s$^{-1}$. What matter for internal dynamics are the relative positions and relative velocities between the pair. 

The relative positions on the plane of the sky are given by
\begin{equation}
\left. \begin{array}{cl}
   \Delta x^\prime = & -3600 d_M \cos(0.5(\delta_A+\delta_B)\pi/180)  \Delta\alpha \text{ au} \\ 
   \Delta y^\prime = & 3600 d_M \Delta\delta \text{ au} 
\end{array} \right\},
\label{eq:relpos}
\end{equation}
where $\Delta\alpha = \alpha_B-\alpha_A$, $\Delta\delta = \delta_B-\delta_A$, and hereafter $d_M$ is the error-weighted mean of the two distances given in units of pc. From Equation~(\ref{eq:relpos}) the projected separation between the pair is given by 
\begin{equation}
    s = \sqrt{(\Delta x^\prime)^2 + (\Delta y^\prime)^2}.
\label{eq:sep}    
\end{equation}
The relative velocity components on the plane of the sky are given by
\begin{equation}
\left. \begin{array}{cl}
  v_{x^\prime} = & -4.7404 d_M (\mu^\star_{\alpha,B}-\mu^\star_{\alpha,A})\text{ m s}^{-1} \\ 
  v_{y^\prime} = & 4.7404 d_M (\mu_{\delta,B}-\mu_{\delta,A})\text{ m s}^{-1} 
\end{array} \right\},
\label{eq:relvel}
\end{equation}
from which the relative projected velocity follows as
\begin{equation}
    v_p = \sqrt{ v_{x^\prime}^2 + v_{y^\prime}^2}.
\label{eq:vpobs}    
\end{equation}
The ratio of the two projected velocity components will be denoted by
\begin{equation}
    \beta_{p,\rm{obs}} \equiv \frac{v_{y^\prime}}{v_{x^\prime}},
\label{eq:betpobs}    
\end{equation}
which will be used as an observational constraint on the orbit. The relative RV is given by
\begin{equation}
   v_r \equiv v_{z^\prime} = -1000(\text{RV}_B-\text{RV}_A)\text{ m s}^{-1}.
\label{eq:vrobs}
\end{equation}

 {The above equations (Equations~(\ref{eq:vpobs}) and (\ref{eq:vrobs})) for the relative velocity components based on the observed PMs and RVs do not include corrections for the ``perspective effects'' \citep{shaya2011}, which are apparent relative velocities between the pair in the observer's spherical coordinates due to the motion of the barycenter of the system in the 3D space. For wide binaries in the solar neighborhood, perspective effects can be relevant for $s > 0.1$~pc (i.e., 20~kau; \citealt{elbadry2019}).}

 {Almost all binaries considered in this work have $s<20$~kau (99\% satisfy $s<15.5$~kau) meaning that perspective effects are minimal, and thus their corrections are not warranted when RV uncertainties are large. Moreover, for an accurate estimate of the perspective effects, the difference in the distances of the two stars is needed. Unfortunately, observed distance uncertainties are (usually much) larger than $s$ and one has to assume zero (or reasonably small) difference in the distances to obtain a first-order estimate of the perspective effects. The default choice of the present work will be to ignore any perspective effects. However, I will also consider the case of correcting for first-order perspective effects estimated by a numerical procedure described in Appendix~\ref{sec:pers}.   }

I define the ratio of the RV to a projected velocity component as 
\begin{equation}
    \tau_{\rm{obs}} \equiv - \frac{v_r}{v_{y^\prime}},
\label{eq:tauobs}    
\end{equation}
which is related to the ``inclination'' angle $i$ by $\tau_{\rm{obs}} = \tan i$.\footnote{Here the parameter $i$ ﻿is﻿ the same as the angle between the orbital plane and the plane of the sky only﻿ in the special case of Figure 1. In general cases where the sky plane is rotated by an arbitary angle about the $z^\prime$-axis as shown Appendix~\ref{sec:Euler}, $i$ is not the true inclination angle but a mere parameter.} Finally, the magnitude of the relative physical velocity in the 3D space is then given by 
\begin{equation}
    v_{\rm{obs}} = \sqrt{ v_p^2 + v_r^2}.
\label{eq:vobs}    
\end{equation}
The above three quantities given by Equations~(\ref{eq:betpobs}), (\ref{eq:tauobs}), and (\ref{eq:vobs}) provide observational constraints.

For elliptical orbits obeying pseudo-Newtonian dynamics the magnitude of the relative physical velocity between the pair in the 3D space is predicted as 
\begin{equation}
    v_{\rm{mod}} = \sqrt{\frac{\gamma_g G_{\rm{N}} f_M M_{\rm{tot,obs}}}{s/\sqrt{\cos^2\phi+\cos^2 i \sin^2\phi}} \left( 2 - \frac{1-e^2}{1+e\cos(\phi-\phi_0)} \right) },
  \label{eq:vmod}
\end{equation}
where $s$ is the projected separation between the pair, $M_{\rm{tot,obs}}$ is an observational value for the total mass of the binary (see below), and 
\begin{equation}
    f_M \equiv \frac{M_{\rm{tot}}}{M_{\rm{tot,obs}}}
    \label{eq:fM} 
\end{equation}
is a parameter introduced to allow for the Bayesian inferred value $M_{\rm{tot}}$ to vary from $M_{\rm{tot,obs}}$. 

The ratio of the two projected velocities is predicted as 
\begin{equation}
    \beta_{p,\rm{mod}} = -\cos i \frac{\cos\phi + e\cos\phi_0}{\sin\phi+e\sin\phi_0},
\label{eq:betpmod}    
\end{equation}
and the model prediction of $v_r/v_{y^\prime}$ is given as
\begin{equation}
    \tau_{\rm{mod}} = \tan i.
\label{eq:taumod}    
\end{equation}

In Equations~(\ref{eq:vmod}), (\ref{eq:betpmod}), and (\ref{eq:taumod}), $e$ is the eccentricity, $i$ is the inclination parameter, $\phi_0$ is the longitude of the periastron, and $\phi$ is the phase angle given by
\begin{equation}
    \phi=\tan^{-1}((\Delta y^\prime/\Delta x^\prime)/\cos i),
    \label{eq:phi}
\end{equation}
 {which is essentially a function of only $i$ in the observer's frame because $\Delta x^\prime$ and $\Delta y^\prime$ are precisely measured.} In Equation~(\ref{eq:vmod}), $\gamma_g$ is the gravity boost factor defined in Equation~(\ref{eq:anomaly}). In Equation~(\ref{eq:fM}), $M_{\rm{tot}}$ is the total mass of the binary system to be constrained,  {while $M_{\rm{tot,obs}}$ is the observational estimate (see Section~\ref{sec:sample}).} The set of free parameters for each binary is thus given by $\mathbf{\Theta} = \{e, \phi_0, i, \log_{10}(f_M), \Gamma\}$.

In the Bayesian approach, the likelihood $\mathcal{L}$ is defined by
\begin{equation}
    \ln\mathcal{L} = -\frac{1}{2}\sum_j \left[ \left(\frac{X_{j,\rm{mod}}(\mathbf{\Theta})-X_{j,\rm{obs}}}{\sigma_j} \right)^2 +\ln(2\pi\sigma_j^2) \right],
\label{eq:likelihood}    
\end{equation}
where $\{X_{j,\rm{mod}}(\mathbf{\Theta})\}$ ($j=1,2,3$) are the model predictions given by Equations~(\ref{eq:vmod}), (\ref{eq:betpmod}), and (\ref{eq:taumod}) and $\{X_{j,\rm{obs}}\}$ are the three observational constraints given by Equations~(\ref{eq:vobs}), (\ref{eq:betpobs}), and (\ref{eq:tauobs}).\footnote{In principle, $\Delta d (=d_B-d_A)$ can provide another observational constraint, but its observational uncertainty is usually much larger than the line-of-sight separation between the stars, so that it is not useful. Nevertheless, the Bayesian modeling code \citep{chae2025} has the option to include the constraint.} In Equation~(\ref{eq:likelihood}), $\{\sigma_{j}\}$ are the observational uncertainties estimated as follows.

The uncertainties for $\beta_{p,\rm{obs}}$ and $\tau_{\rm{obs}}$ are estimated by
\begin{equation}
    \sigma_{\beta_p} = \beta_{p,\rm{obs}} \sqrt{\sigma_{v_{x^\prime}}^2/v_{x^\prime}^2+\sigma_{v_{y^\prime}}^2/v_{y^\prime}^2},
\label{eq:betperr}    
\end{equation}
and 
\begin{equation}
    \sigma_{\tau} = \left|\frac{v_r}{v_{y^\prime}}\right| \sqrt{\sigma_{v_{r}}^2/v_{r}^2+\sigma_{v_{y^\prime}}^2/v_{y^\prime}^2}.
\label{eq:tauerr}    
\end{equation}
where $v_{x^\prime}$ and $v_{y^\prime}$ are the relative projected velocity components given by Equation~(\ref{eq:relvel}) while $\sigma_{v_{x^\prime}}$ and $\sigma_{v_{y^\prime}}$ are their respective uncertainties arising from PM uncertainties, and $v_r$ is the relative RV given by Equation~(\ref{eq:vrobs}) while $\sigma_{v_{r}}$ is its uncertainty arising from the uncertainties of RV$_A$ and RV$_B$. The above uncertainties are based on the standard error propagation procedure.

The uncertainty of the physical velocity $v_{\rm{obs}}$ (Equation~(\ref{eq:vobs})) is not estimated by the standard error propagation procedure because the uncertainty of $v_r$ can be much larger than that of $v_p$ based on the Gaia database. Instead, I use a Monte Carlo simulation with the uncertainties of $v_r$ and $v_p$. The uncertainty of $v_p$ is estimated using the standard error propagation procedure.

With the likelihood specified above the posterior probability $p(\mathbf{\Theta})$ for the parameters $\mathbf{\Theta}$ is given by
\begin{equation}
    \ln p(\mathbf{\Theta}) = \ln\mathcal{L} + \sum_l \ln\text{Pr}(\Theta_l),
\label{eq:postprob}    
\end{equation}
where Pr($\Theta_l$) ($l=1,\cdots,5$) is the prior probability for parameter $\Theta_l$ which is specified as follows. 

For eccentricity $e$ I consider the range $0.001<e<0.999$ with  {a prior probability distribution. For the prior, I consider three choices: a flat prior, a ``thermal'' probability density $p(e)=2e$, or a separation($s$)-dependent probability density $p(e)=(1+\alpha)e^\alpha$ with $\alpha=\alpha(s)$ from \cite{hwang2022} (hereafter referred to as the ``H2022'' prior) as described by Equation~(18) of \cite{chae2024a}.  } If the data can determine the orbit well, the flat prior may be sufficient. However, with snapshot data, $e$ is ill constrained in particular for relatively large uncertainties of $v_r$ as in many cases based on the current Gaia DR3 database. When $e$ is ill constrained (or essentially unconstrained in some cases), the prior matters because the flat prior would be equivalent to assuming that the median eccentricity is $0.5$ while the thermal prior has a median of $\sqrt{0.5}(\approx 0.71)$ more consistent with the observed statistical properties.  {The H2022 prior has $\alpha\approx 1.3$ for $s >1$~kau(kilo astronomical unit) implying a median of $\approx 0.74$ that differs from the thermal case by just $0.03$, indicating that two priors would make little difference. Moreover, because individual data themselves have some constraining power in this work, an eccentricity probability distribution plays a weaker role as a prior than a fixed input as in previous studies.} For this reason, the thermal prior will be the default choice.

For $\phi_0$ I assume a flat prior on the orbital phase in terms of time, thus a prior given by\footnote{See, e.g., Section~3.2 of \cite{poisson2014}.}
\begin{equation}
    \text{Pr}(\phi_0) = \frac{(1-e^2)^{3/2}}{2\pi} \frac{1}{[ 1+ e \cos(\phi-\phi_0) ]^2}, 
\label{eq:prphi0}   
\end{equation}
where  {$\phi$ is given by Equation~(\ref{eq:phi})}. The parameter $\phi_0$ is allowed to vary freely within the range $(-360^\circ,360^\circ)$. 

For $f_M$, I assume a Gaussian probability density for $\log_{10}(f_M)$ with $(\mu,\sigma)=(0,0.021)$ allowing a 5\% scatter of $M_{\rm{tot}}$, which covers uncertainties of the mass estimates (see Section~\ref{sec:sample}). For $i$ and $\Gamma$, I assume flat priors in the ranges: $-90^\circ <i< 90^\circ$ and $-1<\Gamma<1$.

\subsection{Demonstration of the Bayesian inference and consolidation of probability distributions with illustrative examples based on mock Newtonian binaries}  \label{sec:demo}

In this subsection, the Bayesian method described in Section~\ref{sec:bayes} is applied to mock Newtonian binary systems to investigate the properties of individual PDFs of $\Gamma$ (Equation~(\ref{eq:anomaly})). Also, I investigate how individual PDFs of $\Gamma$ based on independent data (i.e., snapshot motions) can be combined to give a consolidated probability distribution.

For the orbital motion shown in Figure~\ref{fig:orbit}, consider binary systems with total mass $M_{\rm{tot}}=2.0 M_\odot$, eccentricity $e=0.75$, and $\phi_0=-135^\circ$. For the semi-major axis of the elliptical orbit, I consider two cases: $a=0.8$~kau and $8$~kau to represent binaries in the Newtonian ($g_{\rm{N}}\gtrsim 10^{-8}$~m\,s$^{-2}$) and MOND ($g_{\rm{N}}\lesssim 10^{-9.5}$~m\,s$^{-2}$) regimes. Consider a viewing angle $i=60^\circ$ in the geometry of Figure~\ref{fig:orbit}. For the sake of illustration, assume that both binaries obey Newtonian gravity regardless of their separation (i.e., $\gamma_g =1$ or $\Gamma=0$). 

Suppose that 20 independent snapshot observations are made for instant motions uniformly spaced in normalized time at $t/P=\{0.05,0.10,\cdots,0.95,1.00\}$ of each binary, where $t$ is the time from the passage of the periastron and $P$ is the orbital period. For each value of $t/P$ the value of phase $\phi$ is found by numerically solving the integral equation
\begin{equation}
    t/P = \int_{\phi_0}^\phi d\phi^\prime f(\phi^\prime;\phi_0), 
    \label{eq:integralphi} 
\end{equation}
where $f(\phi;\phi_0)$ is the right-hand side of Equation~(\ref{eq:prphi0}). 

The Newton-predicted relative positions and velocities are calculated as follows. The physical separation $r$ between the pair is given by
\begin{equation}
    r(\phi) = \frac{a(1-e^2)}{1+e\cos(\phi-\phi_0)},
    \label{eq:r}
\end{equation}
from which $x=r\cos\phi$ and $y=r\sin\phi$, and the relative position components 
in the observer's frame are given by
\begin{equation}
  \left.  \begin{array}{ccl}
    x^\prime & = & r \cos\phi \\
    y^\prime & = & r \cos i\sin\phi \\
    z^\prime & = & -r \sin i\sin\phi
    \end{array} \right\}.
\label{eq:positioncomponents}  
\end{equation}
The physical relative velocity between the pair is given by
\begin{eqnarray}
   v(r) & = & \sqrt{\frac{G_{\rm N} M_{\rm{tot}}}{r}\left(2-\frac{r}{a}\right)} \nonumber \\
     & = & 941.9\sqrt{\frac{M_{\rm{tot}}/M_\odot}{r/\text{kau}}\left(2-\frac{r}{a}\right)}\text{ m s}^{-1},
\label{eq:v3D}  
\end{eqnarray}
and the three velocity components in the observer's frame are given by
\begin{equation}
   \left.   \begin{array}{ccl}
    v_{x^\prime} & = & v(r) |\cos\psi| {\rm{sgn}}(dx/d\phi) \\
    v_{y^\prime} & = & v(r) \cos i \tan\psi |\cos\psi| {\rm{sgn}}(dx/d\phi) \\
    v_{z^\prime} & = & -v(r) \sin i \tan\psi |\cos\psi| {\rm{sgn}}(dx/d\phi)
    \end{array} \right\},
    \label{eq:vcomponents}  
\end{equation}
where\footnote{The presence/absence of the factor $\pi$ varies, but for Equation~(\ref{eq:vcomponents}) it is irrelevant.}
\begin{equation}
    \psi = \tan^{-1}\left(-\frac{\cos\phi + e\cos\phi_0}{\sin\phi+e\sin\phi_0} \right) (+\pi),
    \label{eq:psi}
\end{equation}
and ${\rm{sgn}}(dx/d\phi)$ is the sign function of
\begin{equation}
    \frac{dx}{d\phi} = \frac{a(1-e^2)e\sin(\phi-\phi_0)\cos\phi}{\left[1+e\cos(\phi-\phi_0)\right]^2} - r \sin\phi.
\end{equation}
See Appendix~\ref{sec:Euler} for the transformation of the position and velocity vectors from the orbital plane to the observer's frame. 

In passing, I note that the first two equations of Equation~(\ref{eq:vcomponents}) are different from Equation~(A1) of \cite{chae2024a} only in signs for some range of $\phi$. In previous studies carried out so far, only the magnitude of the sky-projected velocity $v_p = \sqrt{v_{x^\prime}^2 + v_{y^\prime}^2}$ was needed, meaning that the signs of $v_{x^\prime}$ and $v_{y^\prime}$ did not matter. In the present study, the correct signs are needed because all three velocity components are used. 

\begin{figure*}
      \centering
      \includegraphics[width=1.\linewidth]{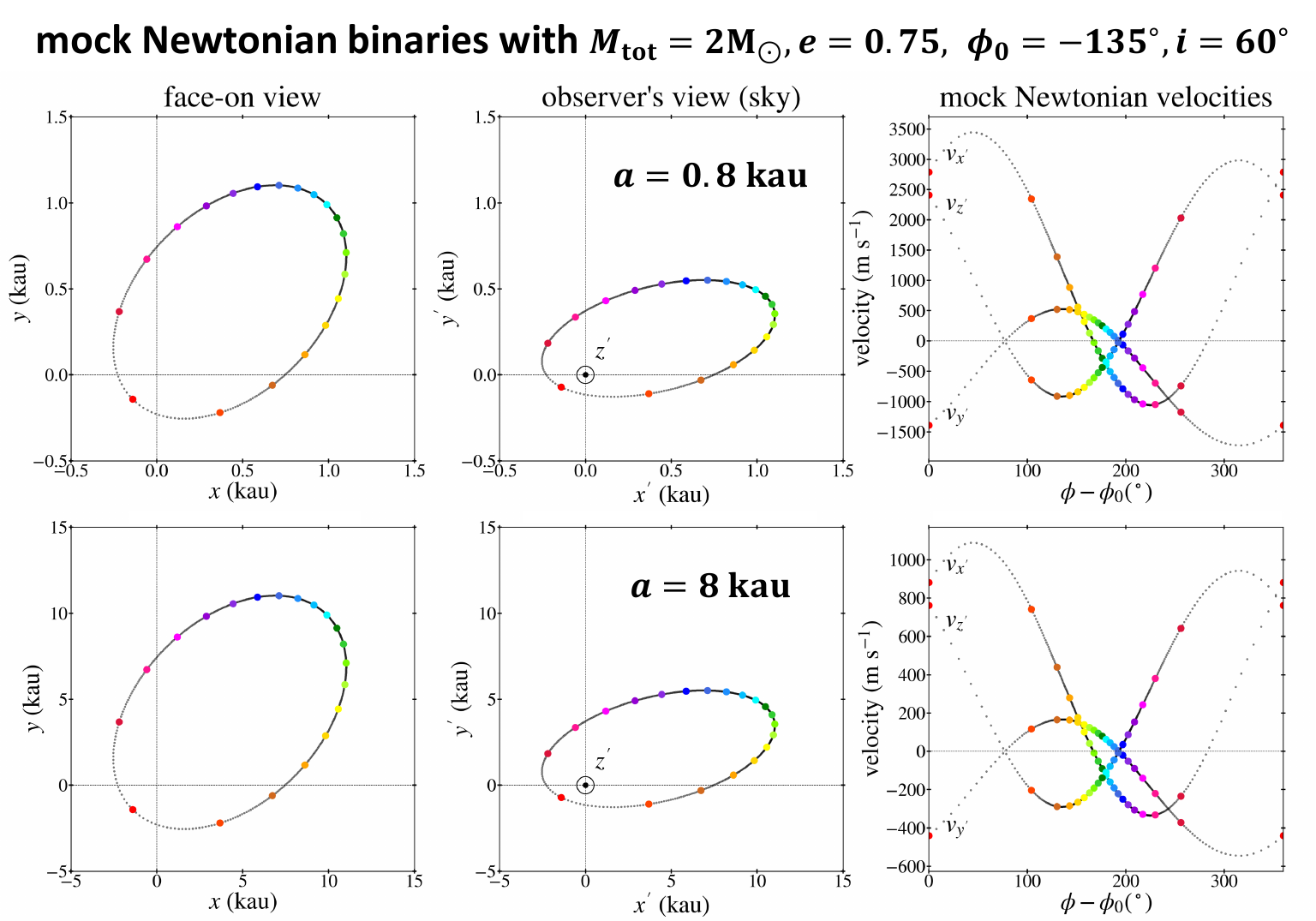}
      \caption{Each row shows a simulated Newtonian orbit with eccentricity $e=0.75$ and semimajor axis $a=0.8$ or 8~kau. All positions and velocities represent relative displacements and relative velocities between the pair of stars with a total mass of $M_{\rm{tot}}=2M_\odot$. See Figure~\ref{fig:orbit} for the geometry and the meanings of the other parameters. For each orbit the colored big dots represent 20 mock snapshot observations with uniformly spaced $t/P$ where $t$ is the time from the passage of the periastron (at $\phi=\phi_0$) and $P$ is the period. Because of long periods only snapshot observations are possible. The sky-projected velocities $v_{x^\prime}$ and $v_{y^\prime}$ would be observed as PMs in real observations while the line-of-sight (radial) velocity $v_{z^\prime}$ would be observed by spectroscopic observations.}
      \label{fig:orbit_simulation2}
  \end{figure*}

Figure~\ref{fig:orbit_simulation2} shows the Newton-predicted relative positions and velocities in the binaries. This figure shows that in random observations of binaries, the values of $|\phi-\phi_0|$ are centered around $180^\circ$ consistent with Equation~(\ref{eq:prphi0}). Each point shown in either row of Figure~\ref{fig:orbit_simulation2} represents an independent observation of the two components of the relative displacement (the middle column) and the three components of the relative velocity (the right column).

 Each dataset is modeled with the Bayesian method of Section~\ref{sec:bayes}. The three kinematic parameters in the likelihood (given by Equations~(\ref{eq:betpobs}), (\ref{eq:tauobs}), and (\ref{eq:vobs})) are assigned uncertainties based on the uncertainties of the mock velocities given as follows, motivated by typical Gaia DR3 values in a nominal sample that will be described in the following subsection. The uncertainties of $v_{x^\prime}$ and $v_{y^\prime}$ are randomly given small values of $18.5$ and $16.6$~m\,s$^{-1}$ as Gaia DR3 uncertainties of $v_p$ satisfy $\lesssim 35$~m\,s$^{-1}$. The uncertainties of $v_r(\equiv v_{z^\prime})$ are $\approx \sqrt{2}\times 200$~m\,s$^{-1}$ from the uncertainties of $200$~m\,s$^{-1}$ assigned to both RV$_A$ and RV$_B$. This is consistent with the median uncertainty of $\sigma_{v_r}\approx 280$~m\,s$^{-1}$ in the nominal sample. Thus, based on the current Gaia DR3 database, the uncertainty of $v_r$ will be the dominant source of uncertainty in the individual Bayesian inferences of $\Gamma$.

\begin{figure*}
      \centering
      \includegraphics[width=1.\linewidth]{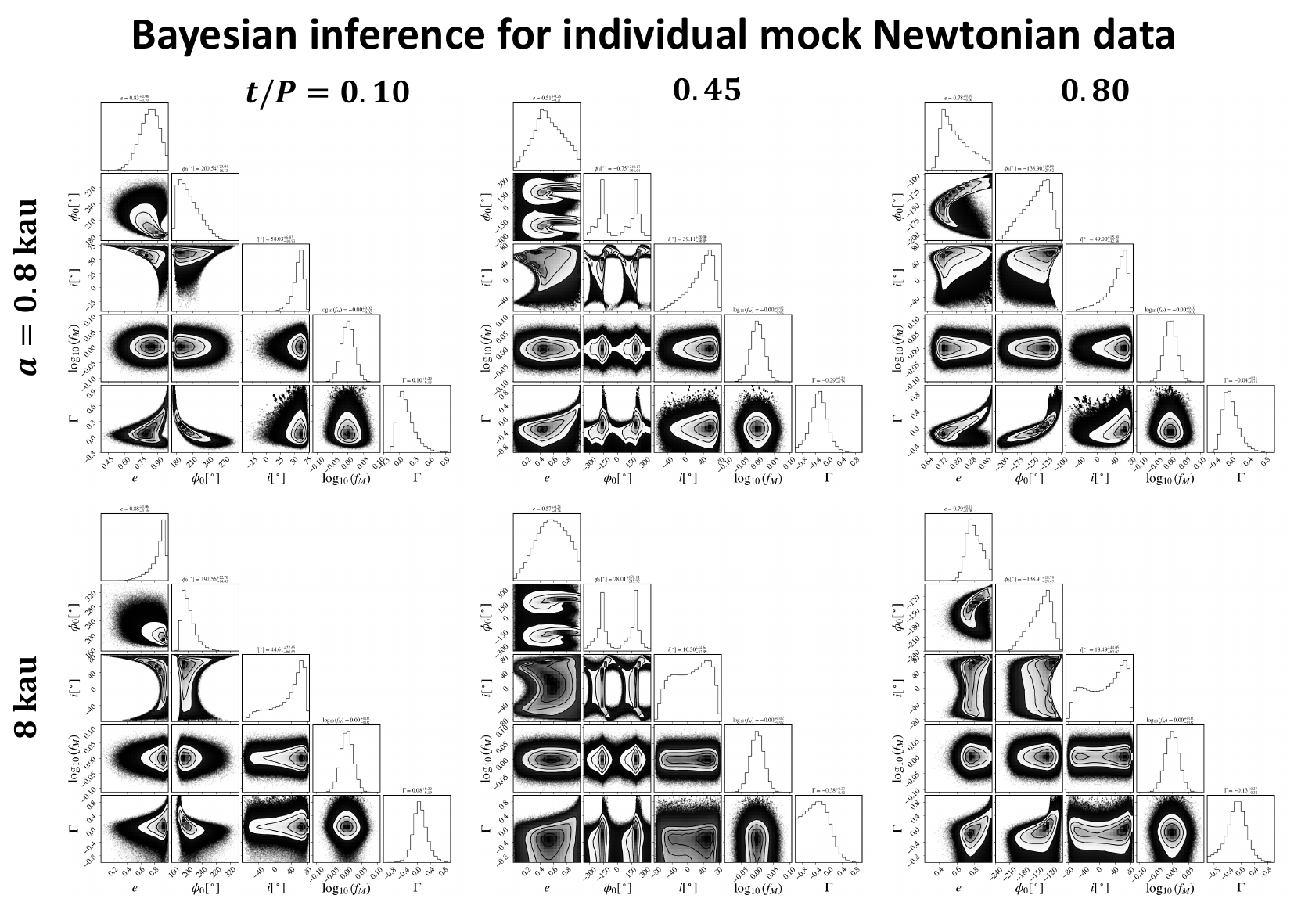}
      \caption{Each row shows three examples of the PDFs of the pseudo-Newtonian model parameters including $\Gamma$ (Equation~(\ref{eq:anomaly})) derived for the mock Newtonian (i.e., with $\Gamma=0$) data shown in Figure~\ref{fig:orbit_simulation2}. For the Bayesian modeling, an uncertainty of 200~m\,s$^{-1}$ is assumed for each radial velocity of a star, thus $\approx 283$~m\,s$^{-1}$ for $v_r$.}
      \label{fig:probdist_examples}
  \end{figure*}

Bayesian inferences are carried out through the Markov Chain Monte Carlo (MCMC) procedure with the public Python package {\tt emcee} \citep{emcee}. The numerical setup is as follows. With five parameters ({\tt ndim}$=5$), the number of random walkers is {\tt nwalkers}$=200$. I consider a very long chain of iterations with {\tt niter}$=250000$. The initial 50000 iterations are discarded ({\tt discard}$=50000$) to exclude (more than) the burn-in phase, and the rest of the 200000 iterations is thinned by 20 ({\tt thin}$=20$), so that a total of {\tt nwalkers} $\times 200000 / 20 = 2\times 10^6$ models are obtained to derive posterior PDFs of parameters. This setup seems more than necessary, and I find that much relaxed setups with less numerical burden produce qualitatively consistent PDFs.

Figure~\ref{fig:probdist_examples} shows examples of Bayesian inference with the above procedure for the mock Newtonian data of the two binaries with $a=0.8$~kau or $8$~kau shown in Figure~\ref{fig:orbit_simulation2}. These results represent some expectations for the current nominal Gaia statistical sample if gravity is Newtonian (or not too different from it). In general, the PDFs of the parameters except for $f_M$ are quite broad as expected considering the input sizable uncertainties (i.e., $\sigma_{v_r}\approx 283$~m\,s$^{-1}$) of the relative RV ($v_r$). This is particularly true for the data of the binary with $a=8$~kau because its $v_r$ values have absolute values even smaller than $\sigma_{v_r}$ in some mock observation phases. The PDF of $\log_{10}(f_M)$ is always equivalent to the prior Gaussian distribution because of its degeneracy with $\Gamma$, which has a flat prior in a broad range.

\begin{figure*}
      \centering
      \includegraphics[width=1.\linewidth]{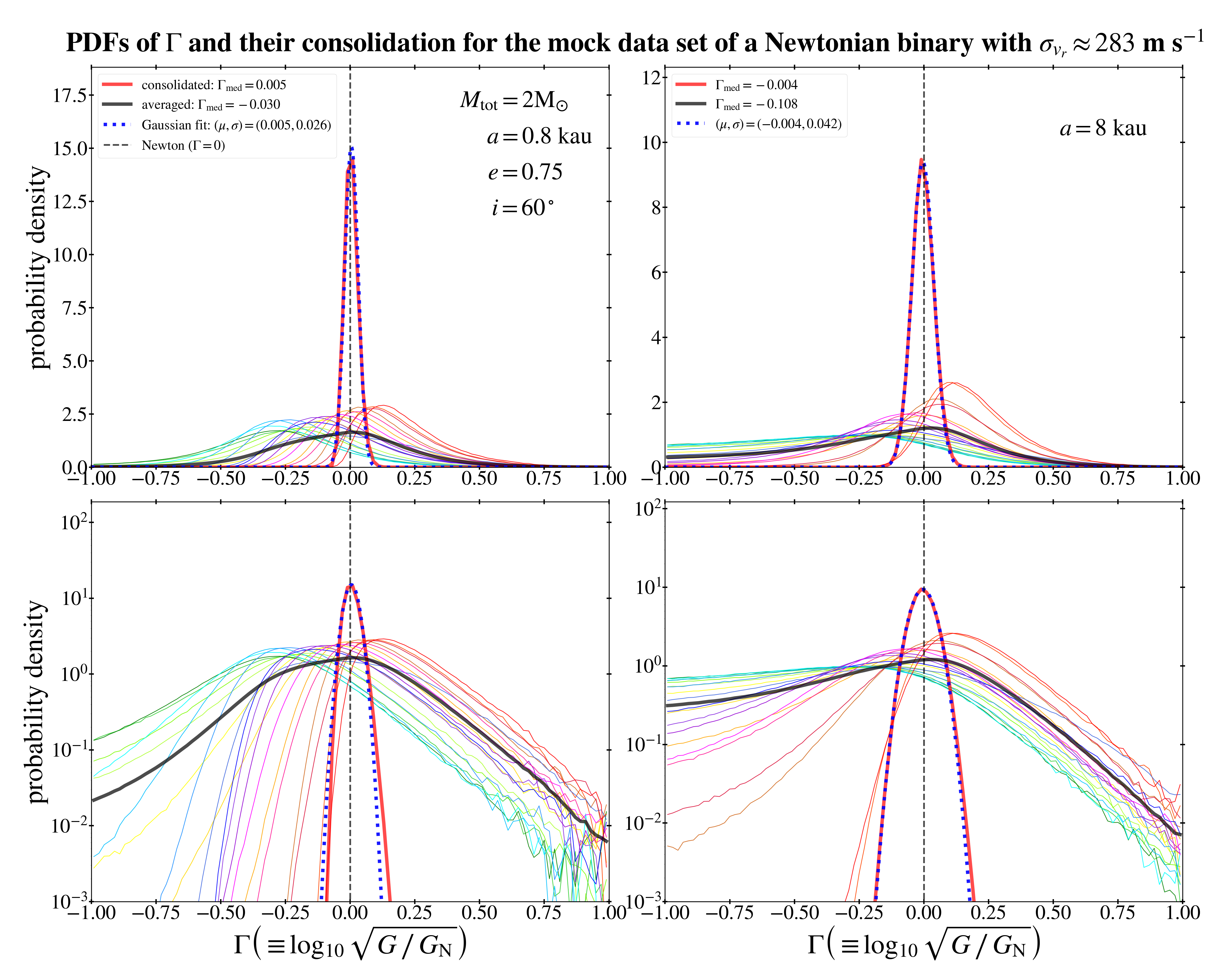}
      \caption{Each column displays all 20 PDFs of $\Gamma$ inferred for the mock data indicated by the colored dots in each row of Figure~\ref{fig:orbit_simulation2} using the same colors. The lower row is a reproduction of the upper row using a log scale for the ordinate for the purpose of increasing visibility. In the following, whenever multiple PDFs are displayed together, the log scale is used. Thick black curve represents the average of the individual PDFs while the thick red curve represents the PDF consolidated by ``conflation'' as detailed in the text. The consolidated PDFs are well described by the Gaussian function.}
      \label{fig:probdist_Gam_mockNewton}
  \end{figure*}

Although the individual PDFs of $\Gamma$ and the orbital and geometric parameters $e$, $\phi_0$, and $i$ are very broad, they are all consistent with the input values within reasonable confidence limits (e.g., $2\sigma$). Each column of Figure~\ref{fig:probdist_Gam_mockNewton} shows the individual PDFs of $\Gamma$ for the mock data set of the Newtonian binary with $a=0.8$~kau or $8$~kau. The lower rows use the logarithmic scale for probability density to enhance visibility. 

Figure~\ref{fig:probdist_Gam_mockNewton} reveals the key characteristics of the individual PDFs of $\Gamma$ for a set of evenly sampled data. First, in a majority of the 20 cases, the peaks and medians of individual PDFs tend to be biased toward values lower than the input value (i.e., $\Gamma=0$ in the present case) and the tails go to zero more slowly in the negative direction than in the positive direction, while the opposite is true in the rest of the minority cases. Second, each individual PDF tends to be broader in the majority of cases than in the minority of cases. Finally, due to the said properties of the individual PDFs the averaged PDF (the thick black curve) has a median biased toward a negative value. This bias is particularly pronounced for the data set of the $a=8$~kau binary.

The bias (i.e., the failure to reproduce the input value) of the averaged PDF indicates that it is a poor way of combining individual PDFs. The averaged PDF has another fatal property as a way of combining independent results. Its width is broader than individual PDFs meaning that many independent observations/experiments did not lead to any improvement in the measurement of $\Gamma$. This does not agree with common sense knowledge of the increased precision from $N$ independent measurements. 

Suppose that $N$ independent measurements of the same quantity $\mu$ were obtained: $x_i \pm \sigma_i$ ($i=1,\cdots,N$). Let $\mu^\prime$ and $\sigma_{\mu^\prime}$ denote an estimate of the quantity and its uncertainty. Let the likelihood be 
\begin{equation}
\mathcal{L}=\prod\limits_{i = 1}^N \left( \frac{1}{\sigma_i\sqrt{2\pi}} \right) \cdot\exp\left[ - \frac{1}{2}\sum\limits_{i = 1}^N \left(\frac{x_i-\mu^\prime}{\sigma_i}\right)^{2} \right]. 
\label{eq:likelihood_gauss}    
\end{equation}
Maximizing this likelihood gives the well-known mean of 
\begin{equation}
    \mu^\prime = \frac{\sum\limits_{i = 1}^N(x_i/\sigma_i^2)}{\sum\limits_{i = 1}^N(1/\sigma_i^2)},
\label{eq:mu_gauss}    
\end{equation}
whose uncertainty follows from the propagation of the uncertainties of $x_i$, i.e., $\sigma_i$, as
\begin{equation}
    \sigma_{\mu^\prime} = \frac{1}{\sqrt{\sum\limits_{i = 1}^N(1/\sigma_i^2)}}.
\label{eq:sigma_gauss}    
\end{equation}
For the ideal case of $x_i=\mu$ and $\sigma_i=\sigma$ (i.e., all measurements return the same value and the same uncertainty), we have $\mu^\prime=\mu$ and $\sigma_{\mu^\prime}=\sigma/\sqrt{N}$. Thus, as is well known, the rule of thumb is that with $N$ independent measurements of the same quantity the precision is improved by a factor of $\sqrt{N}$.

Given that the averaged PDF is not satisfactory, what would be the best way to extract the maximal information from the set of individual PDFs, or how to combine independent measurements of the same quantity? This is the question statisticians have asked, and have come up with reasonable ideas. One particular suggestion to follow here is the consolidation by \emph{conflation}\footnote{Throughout I will refer to this as ``consolidation by conflation'' or simply ``consolidation''.} due to T.\,P.\,Hill and J.\,Miller \citep{hill2011a,hill2011b}. In this approach the consolidated probability distribution of a variable $x$ is defined by
\begin{equation}
    p(x) \equiv \frac{\prod\limits_{i=1}^N p_i(x)}{\int_{-\infty}^{+\infty}dy\prod\limits_{i=1}^N p_i(y) }
\label{eq:conflation}   
\end{equation}
with the normalization property $\int_{-\infty}^{+\infty} p(x) dx = 1$. It is shown \citep{hill2011b} that the probability distribution given by Equation~(\ref{eq:conflation}) has a mean and width agreeing with the maximum likelihood estimates given by Equations~(\ref{eq:mu_gauss}) and (\ref{eq:sigma_gauss}) if the individual $p_i(x)$ are Gaussian.

The consolidation is shown to work well \citep{hill2011b} if the individual PDFs are independent and Gaussian. Under that condition, the consolidated distribution is guaranteed to be Gaussian and can be interpreted as a maximal extraction of information from all measurements. However, as Figure~\ref{fig:probdist_examples} shows, the individual PDFs of $\Gamma$ in the present problem are not Gaussian in general. In such non-Gaussian situations, there are no analytic results on the consolidated distribution, and it is unclear whether the consolidated distribution will be Gaussian or even approximately Gaussian. 

Lacking general analytic results in general non-Gaussian situations and considering the unknown nature of the present problem, here an empirical approach is followed, i.e., the consolidated distribution (Equation~(\ref{eq:conflation})) is tried for mock individual PDFs and its property is experimentally investigated. Figure~\ref{fig:probdist_Gam_mockNewton} shows examples of such a numerical experiment for the mock data sets of the two binaries shown in Figure~\ref{fig:orbit_simulation2}. Remarkably, the consolidated distributions are close to Gaussian in both cases despite individual non-Gaussianities. As will be shown later, the (near) Gaussianity of the consolidated distribution is always satisfied empirically for any sufficiently large set of individual PDFs of mock or real binaries. Figure~\ref{fig:probdist_Gam_mockNewton} indicates that the (near) Gaussianity is a consequence of the product (i.e., the conflation by Equation~(\ref{eq:conflation})) of the shapes of the individual PDFs (which were evenly sampled in $t/P$). As the right column of the figure shows this dramatic nature, although the majority of individual PDFs are biased toward the negative side (which is why the averaged PDF is also biased as discussed above), the minority rest is biased toward the positive side with narrower widths so as to make a ``correct'' balance with the majority.  

In addition to being (nearly) Gaussian, the consolidated distribution correctly reproduces the input value of $\Gamma=0$. As indicated in Figure~\ref{fig:probdist_Gam_mockNewton}, the consolidated distributions give $\Gamma=0.005\pm 0.026$ and $-0.004\pm 0.042$ for the two sets of data. Note that the former value is more precise because the relative uncertainties of $v_r$ are smaller for the data of the binary with $a=0.8$~kau for the fixed uncertainty of $\sigma_{v_r}\approx 283$~m\,s$^{-1}$. Although not shown here, precision can also be improved by increasing the sample size. These numerical experiments suggest that the consolidated distribution of $\Gamma$ from a sufficiently well-sampled individual PDFs have the potential to discriminate between Newtonian gravity and nonstandard gravities.

\begin{figure*}
      \centering
      \includegraphics[width=1.\linewidth]{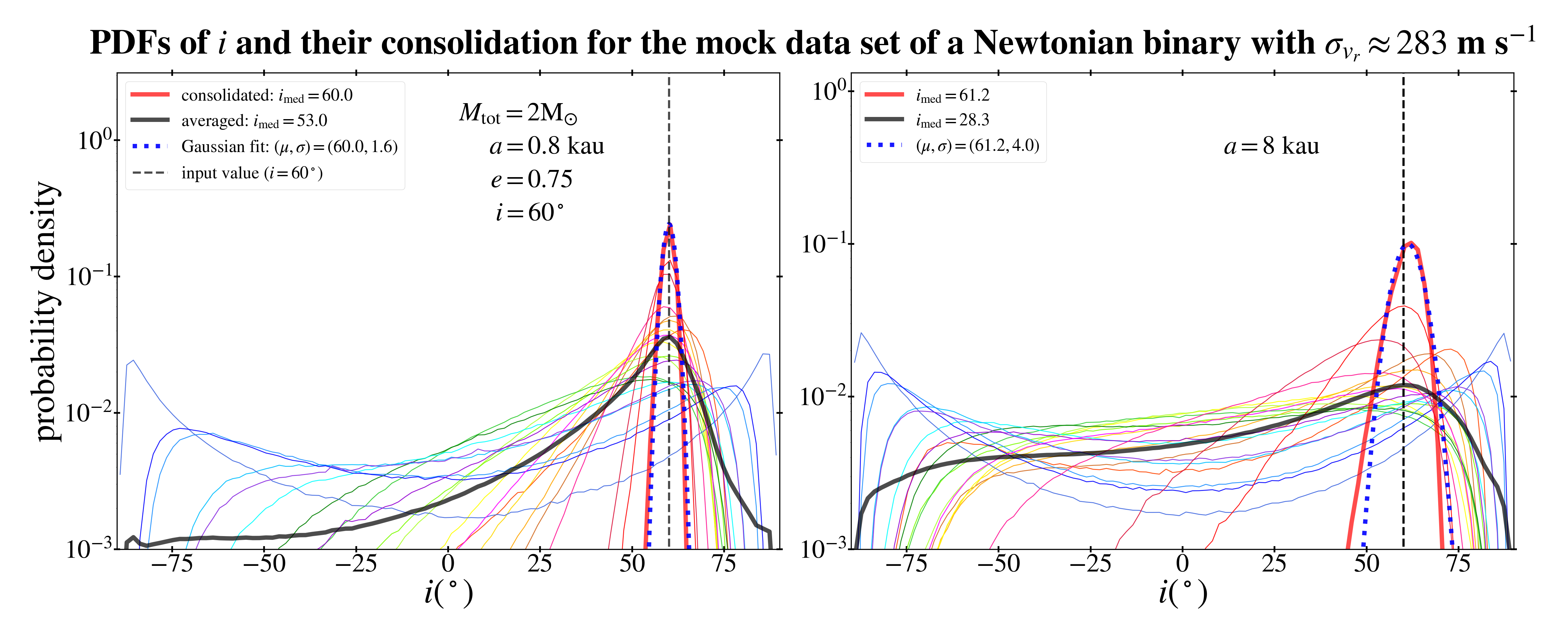}
      \caption{Same as the lower row of Figure~\ref{fig:probdist_Gam_mockNewton} but for the inclination parameter $i$.}
      \label{fig:probdist_inc}
\end{figure*}

In the present work, only the gravity anomaly parameter $\Gamma$ is of interest and can be consolidated for real data of wide binaries because the same gravity law is assumed in a certain acceleration range, but snapshot-observed wide binaries have different masses, orbital parameters, and inclinations, all of which are treated as nuisance parameters. However, the two specific sets of mock data shown in Figure~\ref{fig:orbit_simulation2} are special because each data set is composed of data for the same binary observed at sufficiently different times $t/P$, which of course cannot be realized in actual observations of wide binaries for obvious reasons.\footnote{Not to mention, such a data set can be obtained for very close binaries with $P$ sufficiently small, e.g.\ several years.} It is of some interest to consider the consolidation of the nuisance parameters for these mock data sets, not only as a matter of additionally testing the consolidation method, but also as a possible application of the method in very close binaries. Here I note that it is necessary to introduce and fit the rotation﻿ angle about the $z^\prime$-axis on the sky plane in general cases shown in Appendix A, if a PDF of the true inclination angle is desired. In the present case, it is not necessary because the rotation angle is zero by design in Figure~\ref{fig:orbit}. (In gravity tests with real binaries, the inclination angle is a nuisnace parameter and thus it is not necessary to introduce the rotation angle in the Bayesian modeling.)

\begin{figure*}
      \centering
      \includegraphics[width=1.\linewidth]{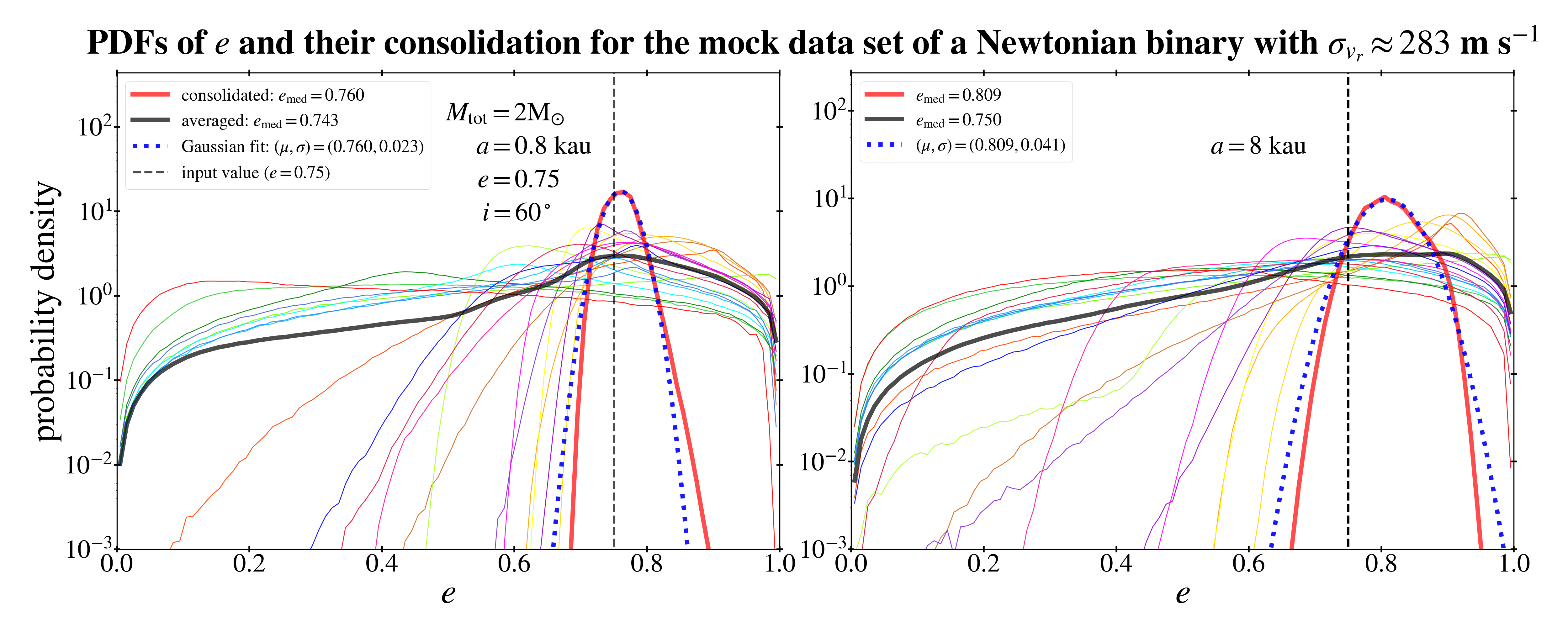}
      \caption{Same as the lower row of Figure~\ref{fig:probdist_Gam_mockNewton} but for the eccentricity parameter $e$.}
      \label{fig:probdist_eccen}
\end{figure*}

Figure~\ref{fig:probdist_inc} shows the individual PDFs and the consolidated distributions for the inclination parameter $i$ as defined in Figure~\ref{fig:orbit}. Individual PDFs of $i$ are generally not well localized, especially for the data of the $a=8$~kau binary. This is not surprising considering the assumed uncertainties of $v_r$. Nevertheless, the consolidated PDFs are well localized, nearly Gaussian, and reproduce the input value $i=60^\circ$ remarkably well. 

\begin{figure*}
      \centering
      \includegraphics[width=1.\linewidth]{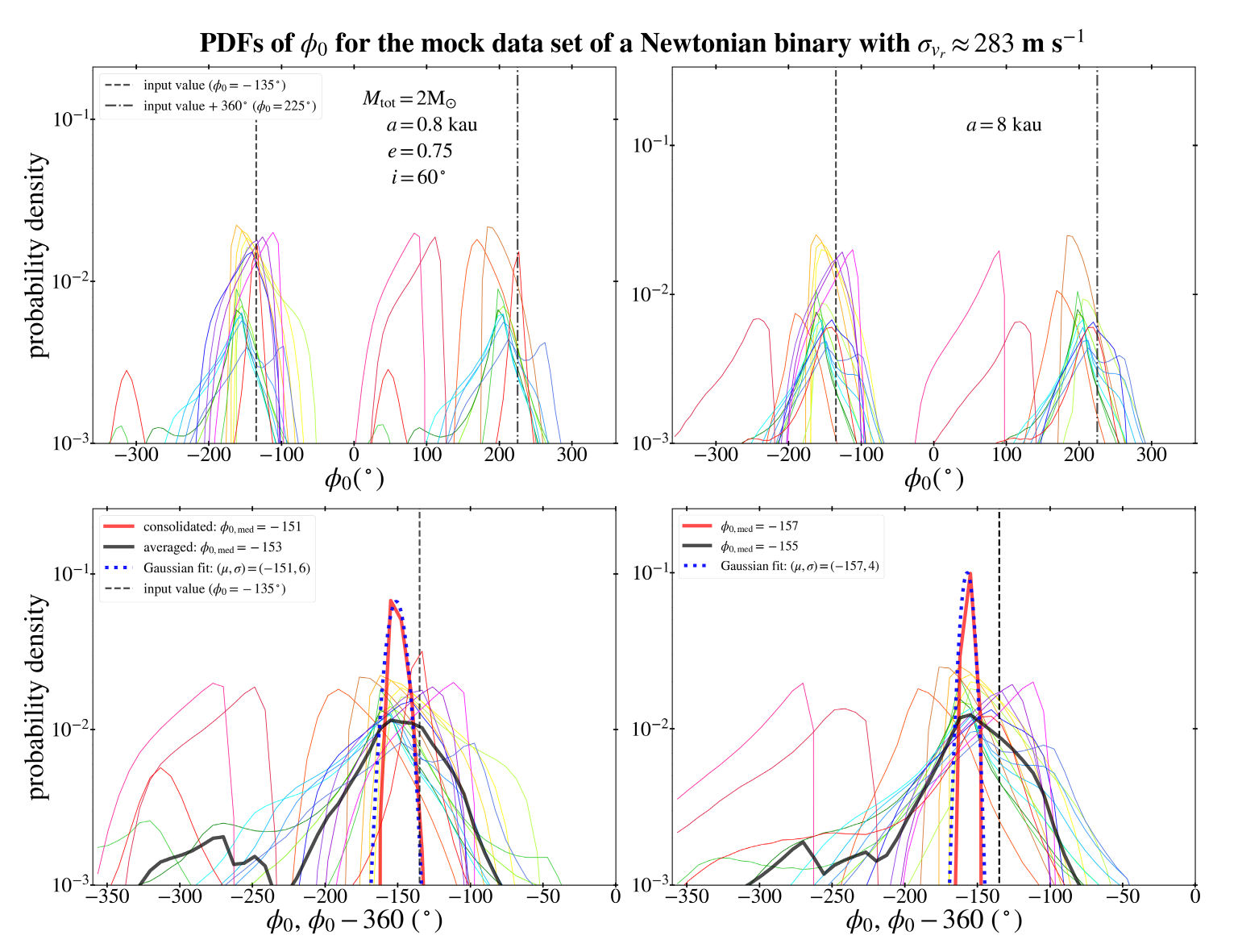}
      \caption{The upper row is same as the lower row of Figure~\ref{fig:probdist_Gam_mockNewton} but for the longitude of the periastron $\phi_0$. Because $\phi_0$ is allowed to vary within $(-360^\circ,360^\circ)$ as for real data to allow a smooth functional behavior when the distribution is broad, the PDFs are peaked at two values separated by $360^\circ$. In the lower row, those peaked at one value is shifted by $-360^\circ$.}
      \label{fig:probdist_phi0}
\end{figure*}

Similarly, Figure~\ref{fig:probdist_eccen} shows that the individual PDFs of the eccentricity $e$ are not well shaped in general, but the consolidated distributions are well localized, nearly Gaussian, and reproduce the input value $e=0.75$ remarkably well. 

Figure~\ref{fig:probdist_phi0} shows the results for $\phi_0$. Because $\phi_0$ was allowed to vary freely in the range $-360^\circ < \phi_0 < 360^\circ$ and the input value was $-135^\circ$, we expect that individual PDFs will be localized around $-135^\circ$ or $225^\circ$. Most PDFs are indeed so, but there are a few exceptions. The most pronounced are the PDFs for $t/P=0.90$ and $0.95$, which correspond to the two points on the second quadrant of the $x^\prime y^\prime$ plane in Figure~\ref{fig:orbit_simulation2}. Due to these outliers, the consolidated distributions (after transforming PDFs localized around $225^\circ$ by $-360^\circ$) are biased by $\approx -16^\circ$ and $\approx -22^\circ$, respectively for the two data sets of $a=0.8$ and $a=8$~kau, as can be seen in the lower panels. Compared with the net allowed range of $360^\circ$ the magnitude of these biases in $\phi_0$ is not large.  

The bias in the PDFs of $\phi_0$ can be understood as follows. Equations~(\ref{eq:vmod}) and (\ref{eq:betpmod}) have dependence on $\phi_0$, but when their observational constraints given by Equations~(\ref{eq:vobs}) and (\ref{eq:betpobs}) are not strong enough, the prior constraint given by Equation~(\ref{eq:prphi0}) will play a major role. This means that some data near the periastron (that is, when $\phi \sim \phi_{0,\rm{true}}$) can lead to PDFs localized near $\phi_0 \sim \phi\pm 180^\circ \sim \phi_{0,\rm{true}} \pm 180^\circ$ leading to bias. This appears to be the case for the two abovementioned outliers in each data set. 

\begin{figure*}
      \centering
      \includegraphics[width=1.\linewidth]{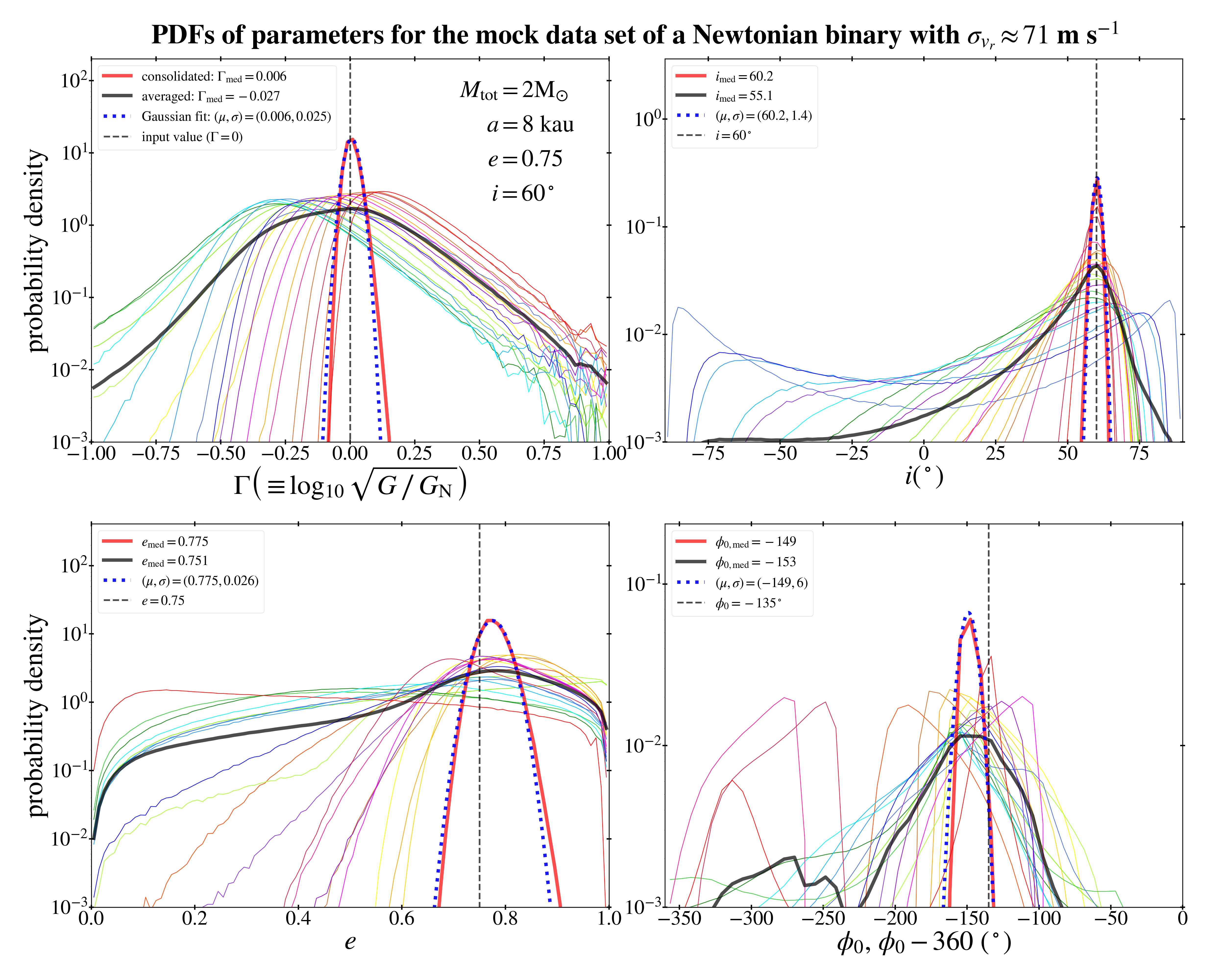}
      \caption{Same as panels in Figures~\ref{fig:probdist_Gam_mockNewton}, \ref{fig:probdist_inc}, \ref{fig:probdist_eccen}, and \ref{fig:probdist_phi0} but for Bayesian inferences with a smaller uncertainty of $\approx 71$~m\,s$^{-1}$ for $v_r$.}
      \label{fig:probdist_Rverr50}
\end{figure*}

Thus, unlike $i$ and $e$, $\phi_0$ appears to be not perfectly recovered by the procedure for the above mock observations. A couple of relevant remarks need to be made here. First, as Figure~\ref{fig:probdist_Gam_mockNewton} shows, the relatively small biases in $\phi_0$ do not affect the correct recovery of $\Gamma$ in both data sets. This can be understood by the fact that only Equation~(\ref{eq:vmod}) has the direct dependence on $\Gamma$, and therefore the small bias in $\phi_0$ has little impact. Second, the bias is larger when the relative precision of the RVs is poorer, i.e., for the data set with $a=8$~kau. It is interesting to check whether the bias is reduced if the assumed uncertainties of the RVs are smaller for the same data. Figure~\ref{fig:probdist_Rverr50} shows the PDFs of the four parameters for the mock binary with $a=8$~kau, but assuming 50\,m\,s$^{-1}$ for RV uncertainties so that $\sigma_{v_r}\approx \sqrt{2}\times 50\approx71$\,m\,s$^{-1}$ for the relative RV $v_r$. The bias in $\phi_0$ is now reduced to $\approx -14^\circ$, while $\Gamma$ and the other parameters remain recovered correctly with smaller uncertainties as expected. 

Therefore, numerical experiments indicate that the procedure of Bayesian inferences with their consolidation can be used to measure $\Gamma$. Moreover, for a data set of one binary as in close binaries, the procedure can recover the parameters $e$ and $i$ correctly, and $\phi_0$ as well if the relative precision of velocities is good enough, which can be satisfied for close binaries. In close binaries, gravity is Newtonian (i.e., $\Gamma=0$ is fixed), so the procedure can be used to measure the total mass of the binary.

The above numerical experiments are designed to illustrate individual PDFs of parameters and develop/test the method of consolidation. For this reason, the mock data used in the above experiments have accurate values of velocities predicted by the Newtonian model. The assumed uncertainties of the velocities are only used to derive the PDFs of the model parameters. However, measured velocities will have errors due to measurement uncertainties, and thus errors need to be added to the model-predicted velocities for realistic simulations. For wide binary samples derived from the Gaia DR3 database to be considered in the next subsection, the projected components $v_{x^\prime}$ and $v_{y^\prime}$ have negligibly small errors, but the RV $v_r$ has large errors. Realistic simulations with the errors of $v_r$ added to the model-predicted values will be considered when the results of Bayesian modeling for real data are presented in Section~\ref{sec:result}.

\subsection{Binary Sample} \label{sec:sample}

This study requires a sample of pure binaries that are free of hidden close components.  {Selecting and using pure binaries was originally suggested and considered by \cite{hernandez2022} and \cite{hernandez2023} in the context of statistical tests of gravity with $v_p$.} Here I apply a multistep screening process to ensure that no (or extremely few) kinematic contaminated systems survive. I start with the initial selection criteria that were used by \cite{chae2024a,chae2024b}. The initial selection is based on the requirement that binaries in a bin of sufficiently high acceleration must agree with the expectation of Newtonian dynamics in a statistical sense, specifically in terms of the median or average properties. Such a selection \citep{chae2024a,chae2024b} can be achieved by requiring sufficiently good precision of PMs, parallaxes, and RVs for both components and dynamical consistency between them.  {Note that availability of Gaia-measured RVs (with a reasonable precision) particularly helps remove systems with hidden close companions because it implies a good single-star solution for the spectroscopic observations.} However, as already acknowledged in \cite{chae2024a,chae2024b}, there can be (a small number of) individual exceptions that elude such a statistical selection.\footnote{Minor bumps in the high-velocity tails of normalized projected velocity distributions in Figures 15 - 17 of \cite{chae2024c} may indicate such exceptions.} 

Here I include some of $\delta_A<-28^\circ$ (hereafter $\delta$ refers to the decl. and the subscript $A$ refers to the brighter component of the pair) binaries as opposed to \cite{chae2023a,chae2024a} who excluded the $\delta_A<-28^\circ$ data altogether (if distance is greater than 80~pc) because the dust extinction model adopted by \cite{chae2023a} did not cover that portion of the southern sky. I include those binaries with $\delta_A<-28^\circ$ that satisfy either distance $d_A<100$~pc or Galactic latitude $|b|>60^\circ$ as they are unlikely to be significantly affected by dust extinction (see Figure~3 of \citealt{chae2023a}). Another minor change is that I now select stars with $3.8< M_G < 13.4$ (rather than $4< M_G < 14$ where $M_G$ is the Gaia $G$-band absolute magnitude) in agreement with Figure~1 of \cite{penoyre2022b}. 

\begin{table*}
  \caption{Samples of wide binaries}\label{tab:sample}
\begin{center}
  \begin{tabular}{lccc}
  \hline
 sample   & $N_{\rm{binary}}$ &  key selection criteria & reference/comments  \\
 \hline
 base & 4276  &  Relative errors: PM  $< 0.005$, dist  $< 0.005$, RV $<0.2$  & extended from \cite{chae2024a,chae2024b} \\
 statistical (for this work)  & 652 &  $\sigma_{v_r} < 500$~m~s$^{-1}$, fly-by \& triple+ removed  &   $\sigma_{v_r}$: estimated uncertainty of $v_r$   \\
 nominal/standard  & 312 & $\sigma_{v_r} < 380$~m~s$^{-1}$, {\tt ruwe} $<1.3$ &  \\
 alternative  & 246 & $\sigma_{v_r} < 330$~m~s$^{-1}$, {\tt ruwe} $<1.3$ &  \\
 \hline
\end{tabular}
\end{center}
Note. (1)  The base sample is defined as a sample with $f_{\rm{multi}}\rightarrow 0$ ($f_{\rm{multi}}$ is the fraction of hierarchical systems) in the Newtonian regime (it may include individual exceptions of hierarchical systems). This sample is revised and extended from \cite{chae2024a,chae2024b} to include some southern-sky binaries. (2) The base sample was selected from the \cite{elbadry2021} catalog. Thus, it is free from flyby (chance alignment) and resolved triple (or higher-order) cases by their criteria. Here I use stricter criteria to remove any potentially suspicious systems as described in the text. Out of 4276 systems, there are 68 potential flybys and 35 triple+ systems. (3) The nominal sample is used to obtain nominal results, and the alternative sample and other samples with varied {\tt ruwe} and $\sigma_{v_r}$ limits are considered to explore possible systematic uncertainties. 
\end{table*}

The above selection returns a sample of 4276 wide binaries, which is larger than the \cite{chae2024b} sample of 3557 wide binaries thanks to the inclusion of some of the southern-sky binaries. This sample will be referred to as the ``base sample.'' Table~\ref{tab:sample} gives a summary of the definition of the samples considered in this work, including the base sample. This sample is free of fly-bys (chance alignments) and resolved triples or higher-order multiples (to be referred to as triple+) by the criteria of \cite{elbadry2021}. Here I consider a stronger criterion to remove any potentially suspicious systems because individual modeling requires individually pure binaries. The neighboring space around each binary is defined by the following limits of the transverse and longitudinal separations:
\begin{enumerate}[(i)]
    \item the sky-projected displacement from the brighter component satisfies $\Delta R_{\rm{sky}}<\sqrt{20} s$ where $s$ is the projected separation between the pair, and
    \item the distance ($d$) of an object from the Sun satisfies $|d-d_A|<4\sqrt{20}s+2\sqrt{\sigma_{d_A}^2+\sigma_{d_B}^2}$, where $d_A$ ($d_B$) is the distance of the brighter (fainter) component from the Sun and $\sigma_{d_A}$ and $\sigma_{d_B}$ are their measurement uncertainties.
\end{enumerate}
In condition (ii), a factor of 4 is inserted considering the projection effect. Any star outside this neighboring space is expected to exert a Newtonian gravitational force less than 5\% of the mutual gravitational force between the pair. Thus, a requirement that no additional objects are found within this space ensures that external perturbations are negligible. 

I search the entire Gaia catalog of all objects within 200~pc downloaded from the Gaia DR3 data archive\footnote{https://gea.esac.esa.int/archive/}. For 35 systems out of the base sample, one or more stars are found beyond the known pair within the space. I identify these as potential resolved triple+ systems. For 68 systems, the fainter component of the pair as originally identified by \cite{elbadry2021} is not found within the space defined above. I identify these as potential flybys. In total, 103 (2.4\%) of the 4276 systems have potential kinematic contaminants. I flag these systems in the base sample and exclude them in the following sample selection.

\begin{figure*}
      \centering
      \includegraphics[width=1.\linewidth]{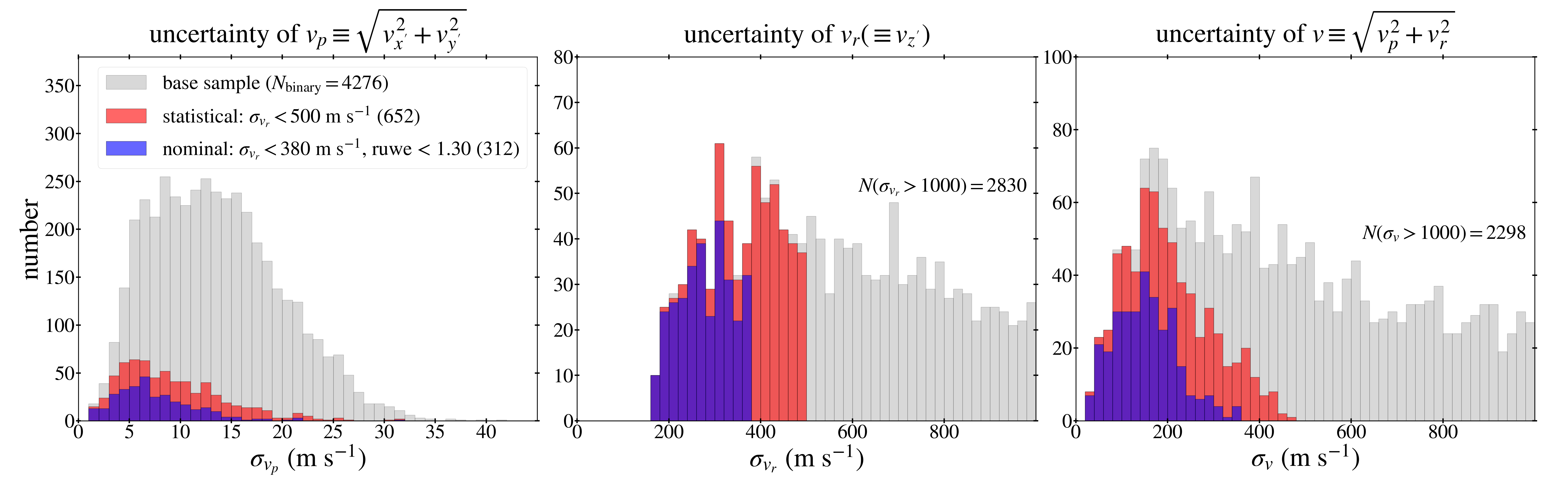}
      \caption{This figure shows the distributions of uncertainties of the relative velocities in the pairs for the samples of wide binaries listed in Table~\ref{tab:sample}. The velocity components are defined in Figure~\ref{fig:orbit_simulation2}. For the nominal sample to be used to obtain the main results on gravity, the median of $\sigma_{v_r}$ is $\approx 280$~m\,s$^{-1}$ similar to that used in the simulations of Figure~\ref{fig:orbit_simulation2} while $\sigma_{v_{x^\prime}}$ and $\sigma_{v_{y^\prime}}$ are negligibly small. For 76\% of binaries in the nominal sample uncertainties of $v(=|\mathbf{v}|)$ are smaller than 200~m\,s$^{-1}$, meaning that gravity can be constrained individually to some extent, though not so strongly.}
      \label{fig:verr}
\end{figure*}

This work requires accurate and precise values for all three components of the relative velocity between the pair. This requirement is met for the two components of the projected velocity $v_p$ from the Gaia DR3 database: most projected velocities have uncertainties $\sigma_{v_p}\la 35$~m\,s$^{-1}$ as shown in Figure~\ref{fig:verr}. Here $v_p$ and $\sigma_{v_p}$ are calculated using Equations~(5) and (9) of \cite{chae2024a}. RVs $v_r$ from the Gaia DR3 database have much larger uncertainties $\sigma_{v_r}>168$~m\,s$^{-1}$. The uncertainty of $v=|\mathbf{v}|=\sqrt{v_p^2+v_r^2}$ ($\sigma_v$) is estimated from Monte Carlo distributions of $v_p$ and $v_r$ with their estimated uncertainties ($\sigma_{v_p}$ and $\sigma_{v_r}$) assuming normal distributions.

As $\sigma_{v_r}$ increases, $\sigma_v$ also increases and individual Bayesian modeling becomes increasingly less feasible and model parameters become more poorly constrained (and thus the results become less meaningful). As shown in Figure~\ref{fig:orbit_simulation2}, wide binaries of low internal acceleration have relative velocities between the pair as low as a few hundreds meters per second or lower. Thus, for wide binaries of low internal acceleration, $\sigma_{v_r}$ significantly larger than a few hundreds meters per second are not likely to be useful. I put an initial hard cut $\sigma_{v_r}<500$~m\,s$^{-1}$, which leaves only 652 binaries (15.6\% of the base sample). This will be referred to as the ``statistical sample''  for Bayesian 3D modeling of this work.

Because reasonable precision is required for $v_r$ and a significant number of binaries is required for an unbiased inference of gravity (as will be shown below in Section~\ref{sec:result}), some compromise between the sample size and the tolerance of $\sigma_{v_r}$ is inevitable. Based on numerical experiments, I take $380$~m\,s$^{-1}$ as a nominal choice for the upper limit of $\sigma_{v_r}$ but also consider $330$~m\,s$^{-1}$ as an alternative choice. For $\sigma_{v_r}$ limits higher than $380$~m\,s$^{-1}$, the bias caused by RV scatter seems to outweigh the increase in number statistics. For $\sigma_{v_r}$ limits lower than $330$~m\,s$^{-1}$, the number statistics in the MOND regime is poor. 

For the nominal sample, the median $\sigma_{v_r}$ is around $280$~m\,s$^{-1}$ similar to that used in the simulations of Section~\ref{sec:demo}. Most systems in the nominal sample satisfy $\sigma_v \la 250$~m\,s$^{-1}$. Also, the nominal sample has the virtue $\sigma_{v_p}\la 15$~m\,s$^{-1}$ for most systems. 

\begin{figure}
      \centering
      \includegraphics[width=1.\linewidth]{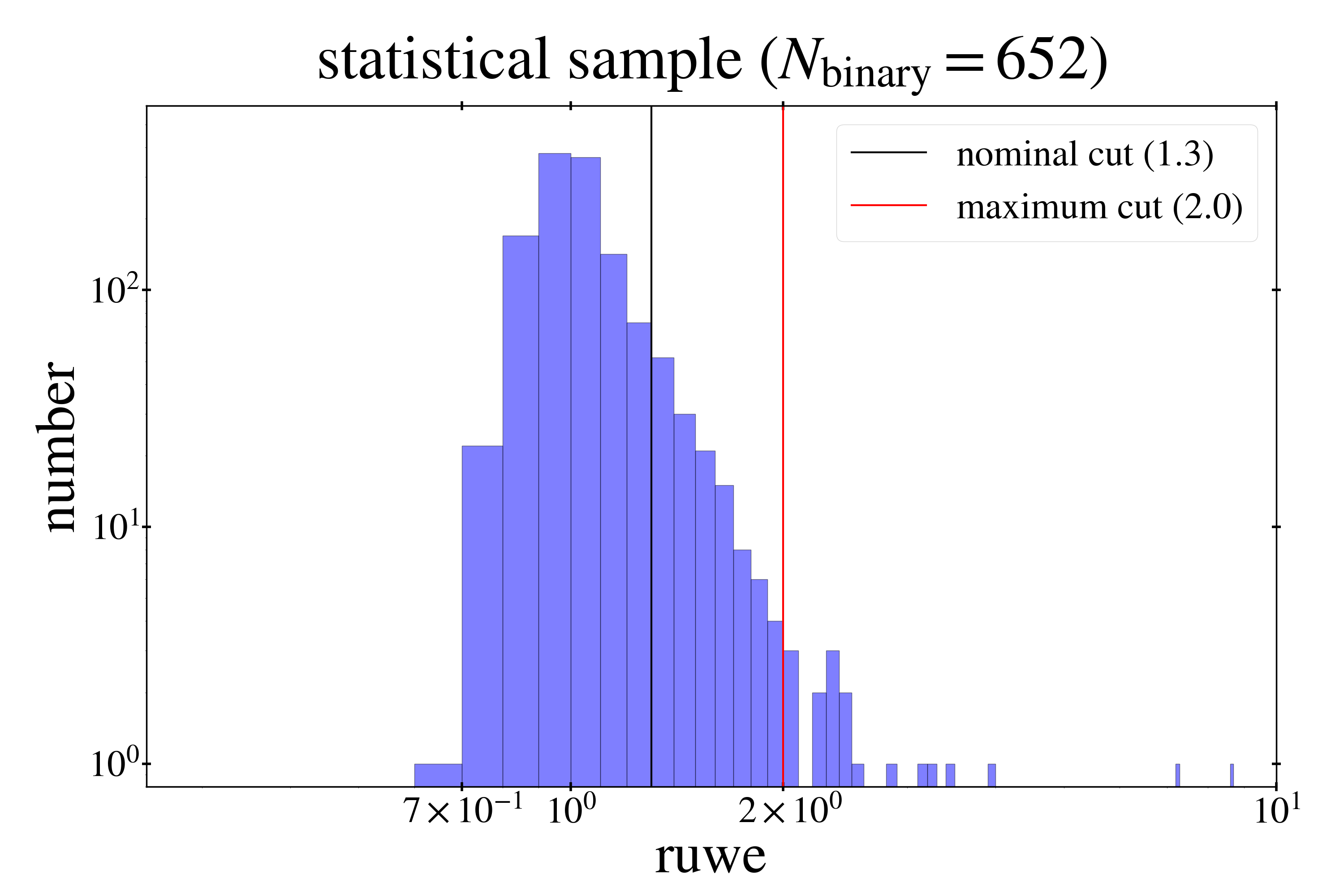}
      \caption{This figure shows the distribution of the Gaia {\tt ruwe} values for the stars in the statistical sample. The distribution is approximately symmetrical about $1$ within $1.3$, which is used as the cut (indicated by the black vertical line) to define the nominal sample. The maximum cut of $2.0$ indicated by the red vertical line is used as the absolute maximum in defining various alternative samples.}
      \label{fig:ruwe}
\end{figure}

The statistical sample with relatively precise $v_r$ values may still contain exceptional cases with kinematic contamination due to, e.g., unseen low-mass star(s) or massive Jovian planet(s) orbiting any component(s) of the binary system. A popular indicator of such close kinematic contamination is the Gaia renormalized unit weight error ({\tt ruwe}) parameter\footnote{https://gea.esac.esa.int/archive/documentation/}. For general stars in the Gaia DR3 database a value significantly greater than $1.0$ (e.g., {\tt ruwe} $>1.4$) could indicate that the source is nonsingle or otherwise problematic for the astrometric solution. Figure~\ref{fig:ruwe} shows the distribution of {\tt ruwe} values for 1304 stars in the statistical sample; 92.2\% (82.6\%) satisfy {\tt ruwe} $<1.4$ ($<1.2$). Nevertheless, it is possible that some of stars with {\tt ruwe} $<1.4$ (or some other value) are unresolved nonsingle stars.

Various simulations and observational studies suggest that {\tt ruwe} values ranging from $1.15$ - $1.6$ (e.g., \citealt{belokurov2020,bryson2020,fitton2022,penoyre2022a,castroginard2024}) can help remove some (unresolved) close binaries. Since such a {\tt ruwe} value does not provide a hard cut on close binaries (see, e.g., Figure~1 of \citealt{fitton2022}) and the statistical sample is purer than the base sample (which is already pure in a \emph{statistical} sense), I consider varying the upper limit of allowed {\tt ruwe} values to explore the possible effects of the {\tt ruwe} limit. For this purpose, I take $1.30$ as the nominal/standard choice for the upper limit of {\tt ruwe} (see Figure~\ref{fig:ruwe}). I consider other limits from {\tt ruwe} $=1.1$ - $2.0$ to explore possible systematic effects. Based on the nominal choices of both $\sigma_{v_r}$ and {\tt ruwe}, I define the nominal/standard sample of 312 binaries as given in Table~\ref{tab:sample}.

\begin{figure}
      \centering
      \includegraphics[width=1.03\linewidth]{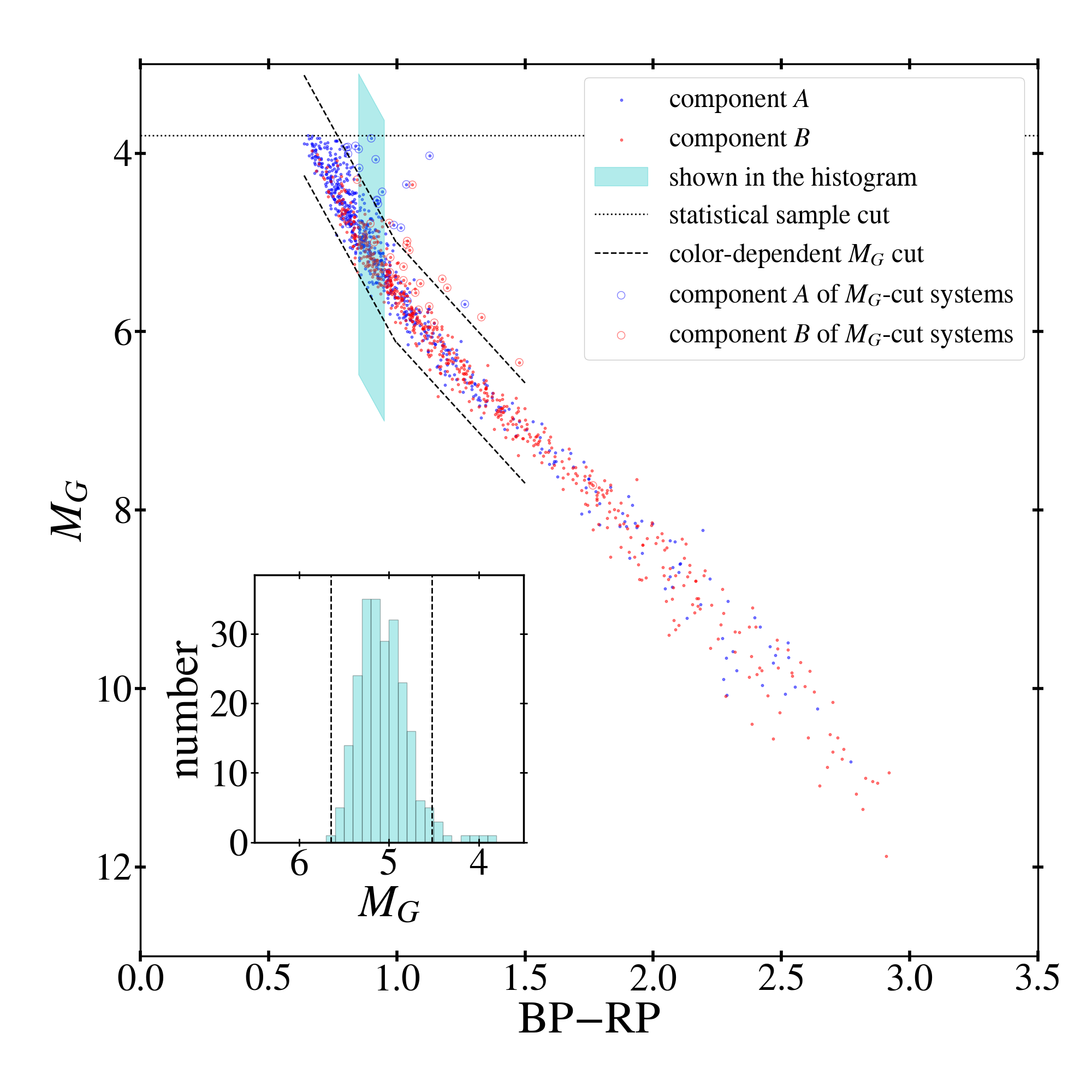}
      \caption{A color-magnitude diagram is displayed for the 1304 stars in the statistical sample of wide binaries. Here BP$-$RP is a Gaia color and $M_G$ is the Gaia $G$-band absolute magnitude without correction of any dust extinction. Most stars fall on a well-defined locus of main sequence stars. The thin vertical strip is used to investigate the distribution of $M_G$ at a fixed color. Based on the distribution shown in the inset, color-dependent $M_G$ cuts are identified for BP$-$RP $<1.5$. For 22 binaries, one or both stars are above the upper cut. These binaries will not be used to investigate gravity.} 
      \label{fig:CM}
\end{figure}

The color-magnitude (CM) diagram for stars is another useful diagnostic for suspicious unresolved close binaries because colors and luminosities of such systems deviate from the the main-sequence locus. Figure~\ref{fig:CM} shows the CM diagram for the stars in the statistical sample using the Gaia BP$-$RP color and the $G$-band absolute magnitude $M_G$. Most stars fall on a well-defined main sequence. For 22 binaries out of 652, one or both components lie $> 2\sigma$ above the scatters of $M_G$ at BP$-$RP. These are potentially suspicious systems as they may be abnormal beyond statistical fluctuations and will not be included in inferring gravity.

\begin{figure*}
      \centering
      \includegraphics[width=0.9\linewidth]{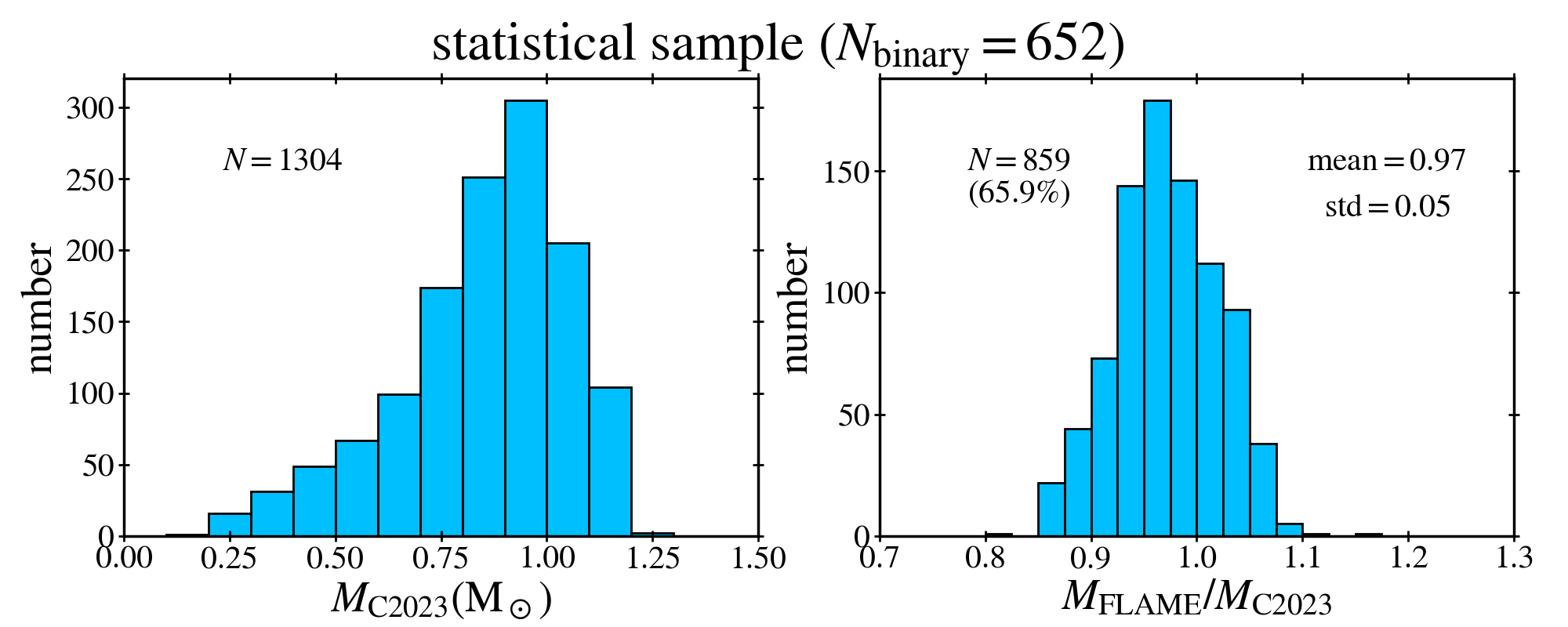}
      \caption{  {The left panel shows the distribution of the \cite{chae2023a} masses ($M_{\rm{C2023}}$) for all stars of the statistical sample. The right panel compares the Gaia FLAME masses ($M_{\rm{FLAME}}$) with $M_{\rm{C2023}}$ for a subset of stars that have $M_{\rm{FLAME}}$.} } 
      \label{fig:mass}
\end{figure*}

 {For the binary stars of the selected samples stellar masses are estimated using the standard $M_G$-mass relation obtained by \cite{chae2023a} (the first choice in Table~1 of that work) where $M_G$ is the $G$-band absolute magnitude corrected for dust extinction as described in \cite{chae2023a}. The left panel of Figure~\ref{fig:mass} shows the distribution of the \cite{chae2023a} masses of 1304 stars in the statistical sample. For those masses that are mainly $\ga 0.5 M_\odot$, the uncertainties of 5\% may well encompass alternative masses as shown in Figure~7 of \cite{chae2023a}. For alternative masses, it is relevant to consider those provided by the Gaia DR3 Final Luminosity Age Mass Estimator (FLAME). FLAME masses are independent and individual results for selective stars from stellar evolution modeling based on a host of Gaia photometric, spectroscopic, and astrometric data. FLAME masses are available for only 859 out of the 1304 stars. The right panel of Figure~\ref{fig:mass} shows the ratio of the FLAME mass to the \cite{chae2023a} mass for the subset of stars. The mean and standard deviation of the ratios are $0.97$ and $0.05$ indicating that the two masses are indeed consistent with each other up to about 5\% scatter. Because FLAME masses are not available for all stars, the masses in this work refer to the \cite{chae2023a} values unless stated otherwise. }

\begin{figure*}
    \centering
    \includegraphics[width=1.\linewidth]{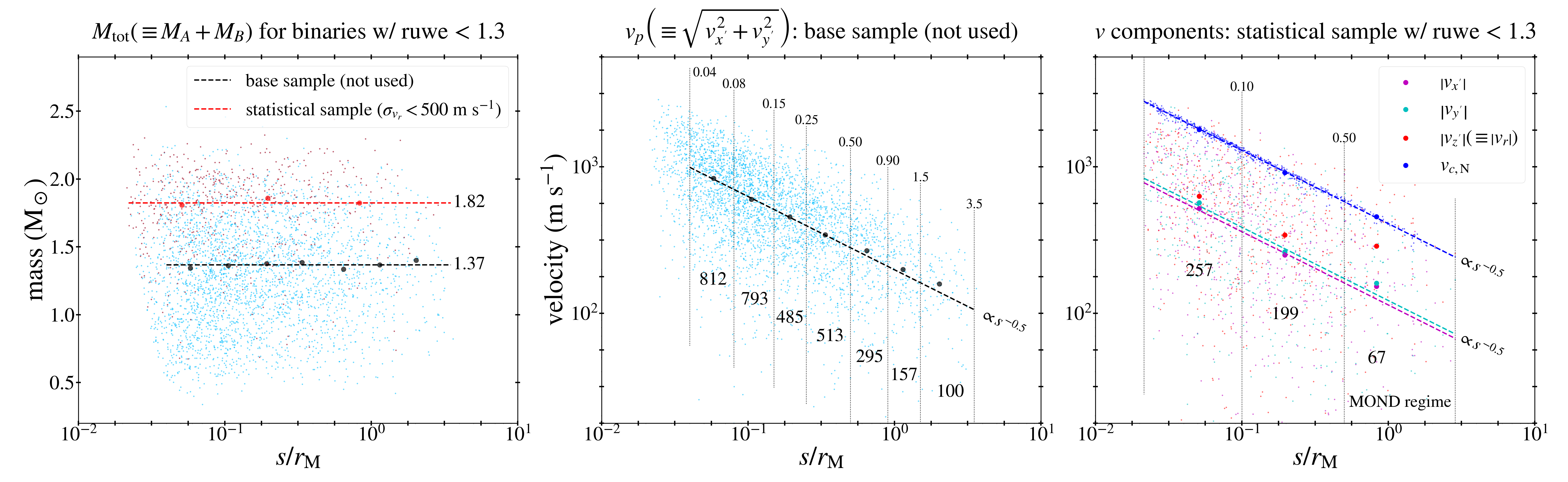}
    \caption{The left panel shows the profiles of $M_{\rm{tot}}$ as a function of $s/r_{\rm{M}}$, where $s$ is the projected separation and $r_{\rm{M}}$ is the MOND radius (Equation~(\ref{eq:MONDradius})), for the base sample and the statistical sample listed in Table~\ref{tab:sample}. While both profiles are flat, the median is much higher in the statistical sample. This indicates that stars with more precisely measured RVs are likely to be brighter. The middle panel shows only the profile of $v_p$ for the base sample as the measured $v_r$ values are too imprecise for general wide binaries (see Figure~\ref{fig:verr}). It can be seen that $v_p$ is boosted with respect to the Keplerian extrapolation in the last three bins consistent with the findings of \cite{chae2024a,chae2024c} and \cite{hernandez2024a}. The right panel shows the profiles of the three velocity components for the binaries in the statistical sample with the limit {\tt ruwe} $<1.3$. In the last bin of the MOND regime ($s/r_{\rm{M}}>0.5$), boosts of the three velocity components can be seen. But, notice that the medians of $|v_r|$ in the middle and last bins are markedly larger than $|v_{x^\prime}|$ and $|v_{y^\prime}|$, which are statistically indistinguishable each other. The additional boost can be understood as a consequence of the relatively large measurement errors of the RVs.}
    \label{fig:vprofile}
\end{figure*}

Figure~\ref{fig:vprofile} shows the distributions of $M_{\rm{tot}}$ and velocity components with normalized separation $s/r_{\rm{M}}$ for the base and statistical samples summarized in Table~\ref{tab:sample}. Here $r_{\rm{M}}$ is the MOND radius at which the Newtonian acceleration equals the MOND critical acceleration $a_0$, i.e., 
\begin{equation}
    r_{\rm{M}}\equiv\sqrt{\frac{G_{\rm{N}}M_{\rm{tot}}}{a_0}}
    \label{eq:MONDradius}
\end{equation}
where $a_0 = 1.2\times 10^{-10}$~m\,s$^{-2}$ is estimated by observations (e.g.\ \citealt{mcgaugh2016}). The left panel of Figure~\ref{fig:vprofile} shows that $M_{\rm{tot}}$ does not vary statistically with $s/r_{\rm{M}}$. This means that Newtonian gravity would predict a statistically Keplerian velocity profile with $s/r_{\rm{M}}$ as long as the eccentricity does not appreciably vary with $s/r_{\rm{M}}$ in a statistical sense. The middle panel shows the profile of $v_p$ for the base sample. Note that all $v_p$ values from the base sample have very small errors and thus all of them are used to obtain the profile. It agrees well with the Keplerian profile of $v_p \propto s^{-1/2}$ for the first four bins but indicates some boost with respect to the Keplerian prediction for the last three bins. This is in line with the findings of \cite{chae2024a,chae2024c}. 

The right panel of Figure~\ref{fig:vprofile} shows the profiles of the three velocity components $|v_{x^\prime}|$, $|v_{y^\prime}|$, and $|v_{z^\prime}|$ for the statistical sample with relatively precise RVs with $\sigma_{v_r}<500$~m\,s$^{-1}$ and {\tt ruwe} $< 1.3$. Here, only three bins are considered because the sample size is small. As in the middle panel, high-precision values of $|v_{x^\prime}|$ and $|v_{y^\prime}|$ follow well the Keplerian profile for the first two bins but are somewhat boosted in the `MOND regime' bin only. However, lower-precision values of $|v_{z^\prime}|$ exhibit large deviations from the Keplerian profile except for the first bin, which must be dominated by much larger uncertainties of the Gaia RVs compared to the sky-projected velocities.

\subsection{Mock binary samples} \label{sec:mock}

In Section~\ref{sec:demo} each set of mock observations (i.e., a mock sample) are generated for one mock binary with specific parameters. For comparison with the samples of snapshot-observed wide binaries that have different parameters, here I describe a procedure to generate samples of mock snapshot-observed wide binaries satisfying Newtonian or pseudo-Newtonian gravity with a generalized gravitational constant $G=\gamma_g G_{\rm{N}}$. 

As in \cite{chae2023a,chae2024a,chae2024c}, mock binaries are generated using the observed binaries by replacing the observed positions, PMs, and RVs with simulated values as described below. Because all three components of the relative velocity between the pair need to be generated in the observer's frame, I consider an arbitrary transformation from the orbital plane to the observer's frame with Euler angles as described in Appendix~\ref{sec:Euler}. Since what matters are the relative positions and motions between the pair, the positions and velocities of the component $A$ will be fixed at the observed values.
\begin{enumerate}
\item The sky-projected separation $s$ for each binary system is the same as the Gaia-measured value. The distance of the system from the Sun is taken to be the error-weighted mean ($d_M$) of the Gaia-measured values for the two components. The total mass $M_{\rm{tot}}$ is the same as the estimated value as described in Section~\ref{sec:sample}. The eccentricity ($e$) of the orbit is taken randomly from the \cite{hwang2022} maximum likelihood range for the binary, as described in Section~2.4 of \cite{chae2024c}.
\item In the orbital plane of Figure~\ref{fig:orbit} and Figure~\ref{fig:Euler}(a), $\phi_0$ (at the periastron) is taken randomly from $(-\pi,\pi)$. The normalized time $t/P$ (see Equation~(\ref{eq:integralphi}) and the remarks given before it) is taken randomly from $(0,1)$ and Equation~(\ref{eq:integralphi}) is used to obtain the phase angle $\phi$.
\item The normalized radius $\hat{r}\equiv r/a$ is given by $\hat{r}=(1-e^2)/(1+e\cos(\phi-\phi_0))$.  
\item In principle, the three Euler angles ($\alpha,\beta,\gamma$) defined in Appendix~\ref{sec:Euler} need to be assigned. The first and third angles ($\alpha$ and $\gamma$) represent rotations about the initial and final ``$z$-axes''. However, when $\phi_0$ is randomly taken as is the case here, $\alpha$ can be fixed at zero. Let $\gamma = \theta$ with $\theta$ taken randomly from $(-\pi,\pi)$. The second angle $\beta$ represents a rotation about the intermediate ``$x$-axis''. This is an inclination as it has been referred to. Let $\beta=i$ with $i =\cos^{-1}(u)$ where $u$ is taken randomly from $(0,1)$ so that the probability density of $i$ is $\sin i$. The two angles $i$ and $\theta$ (with a random assignment of $\phi_0$) specify a general transformation matrix $\mathbf{\Lambda}=\mathbf{\Lambda}(i,\theta)$ (Equation~(\ref{eq:EulerMat_2angles})) between the ``$xyz$'' frame (with the orbit lying on its $xy$ plane) and the observer's ``$x^\prime y^\prime z^\prime$'' frame.
\item The physical separation between the pair is given by $r= s/\sqrt{\cos^2\phi+\sin^2\phi\cos^2 i}$.
\item The semi-major axis of the orbit $a$ is given by $a=r/\hat{r}$.
\item The relative displacement ($\mathbf{r}$) and the relative velocity ($\mathbf{v}$) between the pair are given by Equation~(\ref{eq:relvec}) where $v$ is given by Equation~(\ref{eq:v3D}). In the observer's frame they are obtained by $\mathbf{r}^\prime = \mathbf{\Lambda}(i,\theta) \mathbf{r}$ and $\mathbf{v}^\prime = \mathbf{\Lambda}(i,\theta) \mathbf{v}$ as given in Equations~(\ref{eq:relpos_2angles}) and (\ref{eq:relvel_2angles}).
\item The mock relative R.A. and decl. of component $B$ with respect to $A$ are given by $\Delta\alpha = -x^\prime/d_M$ and $\Delta\delta = y^\prime/d_M$ in units of radians. These are converted to units of degrees to mimic the Gaia data. Although $z^\prime$ represents the difference in the distances of the two components, it has no use in the present simulations because the Gaia measurement errors are typically much larger than $|z^\prime|$.
\item The velocity components $v_{x^\prime}$ and $v_{y^\prime}$ are converted to the relative PMs of component $B$ with respect to $A$ at distance $d_M$ by reversing the left- and right-hand sides of Equation~(\ref{eq:relvel}).
\item The velocity component $v_{z^\prime}$ (i.e., $v_r$) gives the relative RV of component $B$ with respect to $A$ by reversing the left- and right-hand sides of Equation~(\ref{eq:vrobs}). These assigned values of RV$_A$ and RV$_B$ have no errors. However, unlike the Gaia-reported PMs giving accurate values of $v_{x^\prime}$ and $v_{y^\prime}$, the Gaia-reported errors of RV values are nonnegligible. In other words, their errors mean $\sigma_{v_r} \ga 170$~m\,s$^{-1}$ (Figure~\ref{fig:verr}) for $|v_r|=|\rm{RV}_B-\rm{RV}_A|$ which can be comparable to, or even larger than, $|v_{z^\prime}|$ for wide binaries with relatively large separation ($s \ga 5$~kau). Such large errors can bias $|v_r|$ statistically. Thus, the model-predicted RV$_A$ and RV$_B$ values are given Gaussian scatter with the Gaia measurement errors (or other errors as needed by the desired test).

\end{enumerate}

\begin{figure*}
    \centering
    \includegraphics[width=1.\linewidth]{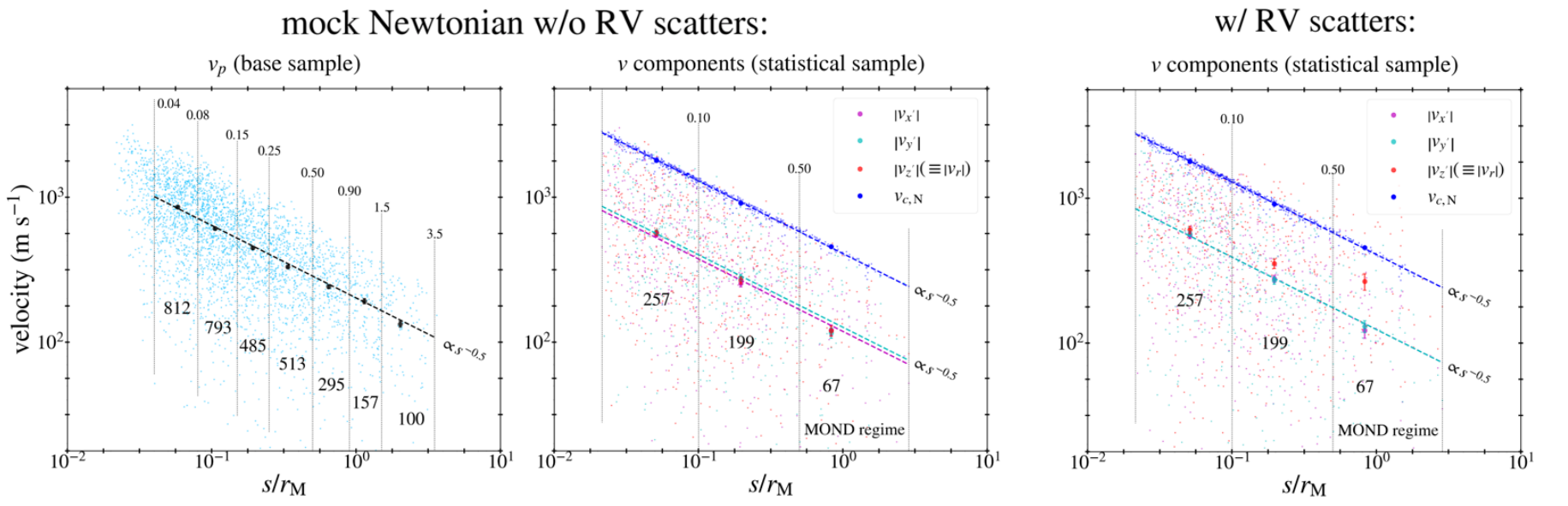}
    \caption{Velocity profiles in mock Newtonian samples are displayed. The left panel is similar to the middle panel of Figure~\ref{fig:vprofile} but for mock Newtonian wide binaries. The big black dots with error bars are the medians and their errors in the bins of $s/r_{\rm{M}}$ based on 100 mock samples of the base sample. The middle panel is similar to the right panel of Figure~\ref{fig:vprofile} but for 100 mock Newtonian samples of the statistical sample. Here $v_r$ has no measurement errors. Unlike the right panel of Figure~\ref{fig:vprofile}, all three velocity components are statistically indistinguishable one another. The right panel shows the profiles including measurement errors of $v_r$. It demonstrates the effects of RV scatters.}
    \label{fig:vprofile_mockNewton}
\end{figure*}

A sample of mock Newtonian binaries generated for the base sample shown in Figure~\ref{fig:vprofile} can be found in Figure~\ref{fig:vprofile_mockNewton}. The left panel shows the profile of $v_p(=\sqrt{v_{x^\prime}^2+v_{y^\prime}^2})$, which agrees well with the Keplerian profile of $v_p \propto s^{-1/2}$. The middle and right panels show the profiles of the three velocity components $|v_{x^\prime}|$, $|v_{y^\prime}|$, and $|v_{z^\prime}|$ in mock Newtonian samples generated without and with RV scatter for the statistical sample. The right panel clearly reveals the effects of RV scatter. While $|v_{x^\prime}|$ and $|v_{y^\prime}|$ follow a Keplerian profile, $|v_{z^\prime}|$ deviates increasingly at larger $s/r_{\rm{M}}$ due to the relatively larger effects of RV scatter for smaller $|v_r|$. Nevertheless, note that a comparison between the right panel of Figure~\ref{fig:vprofile} and that of Figure~\ref{fig:vprofile_mockNewton} indicates a boost in the ``MOND regime'' bin of the Gaia sample.  

The above procedure does not consider simulating hierarchical systems, i.e., cases where hidden close companions are present, as only pure binaries are meant to be used for Bayesian 3D modeling. However, for the purpose of testing the effects of hidden close companions, I consider producing mock samples including hierarchical systems. For that purpose, I use the procedure described in step \#{7} of Section 3.4 of \cite{chae2023a} (see Section~2.3.3 of \cite{chae2024c} for a brief description).

\section{Results} \label{sec:result}

\subsection{Overview and inspection of the individual Bayesian results} \label{sec:result_intro}

The 652 wide binaries in the statistical sample defined in Section~\ref{sec:sample} are individually modeled in the same manner as mock Newtonian binaries are modeled in Section~\ref{sec:demo}. Each binary is modeled with  {the thermal, the H2022, and the flat} priors on eccentricity.  {As will be shown below, the results with the thermal and H2022 priors are very similar while those with the flat prior are mildly different.} Here, I examine the properties of the Bayesian modeling outputs before inferring gravity based on them in the following subsections.  {In presenting results below, the thermal prior is implicit unless specified otherwise.}

\begin{figure*}
    \centering
    \includegraphics[width=1.\linewidth]{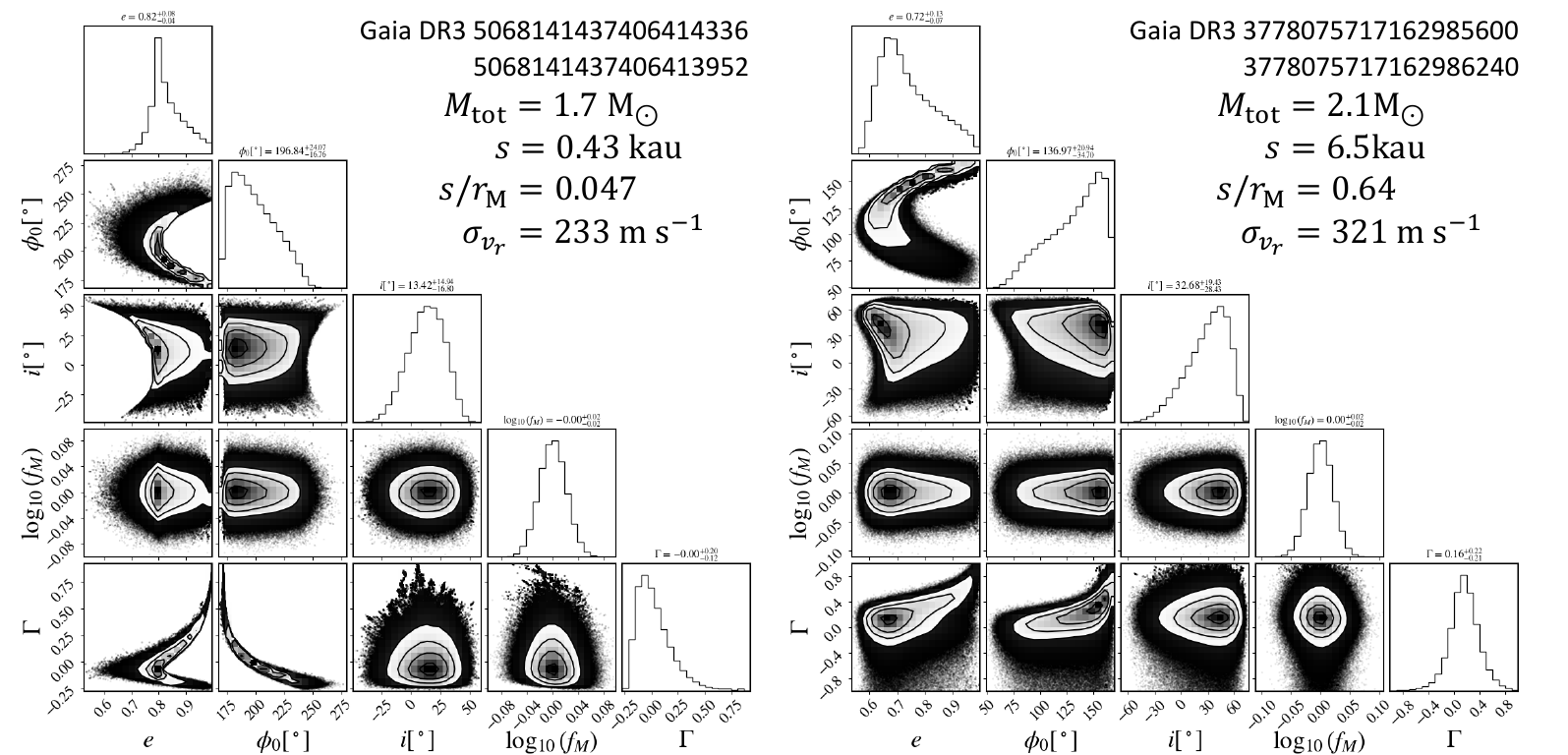}
    \caption{Each set of PDFs and contours is same as any set shown in Figure~\ref{fig:probdist_examples} but for a Gaia wide binary with the IDs and key observational properties indicated. These two are examples of relatively well-determined PDFs. }
    \label{fig:corner_gaia}
\end{figure*}

The properties of the posterior PDFs of the model parameters are diverse depending on the specifics of the data. The results for two binaries in different acceleration regimes can be found in Figure~\ref{fig:corner_gaia}. These are the cases where the model parameters are relatively well constrained. As noted in Section~\ref{sec:demo}, the PDFs of $\Gamma$ are quite broad and include the Newtonian value $\Gamma=0$ within reasonable confidence limits when the assumed gravity is Newtonian. This is what would be expected even when gravity is MONDian based on currently available models. For example, AQUAL/QUMOND predicts $\Gamma\approx 0.07$ in the low internal acceleration limit ($\la 10^{-10}$~m\,s$^{-2}$) due to the external field effect of the Milky Way based on numerical solutions \citep{chae2022a}. 

\begin{figure*}
    \centering
    \includegraphics[width=1.\linewidth]{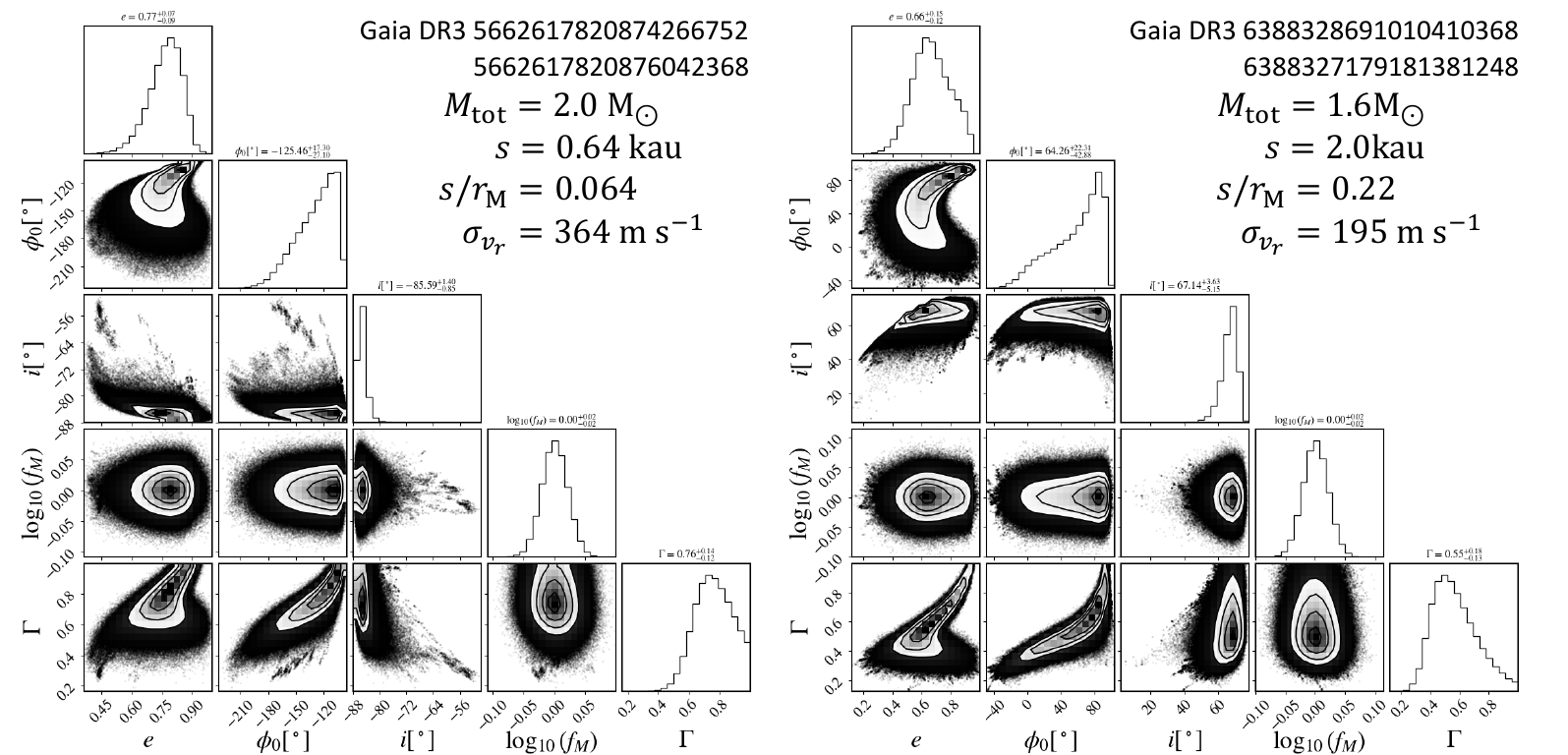}
    \caption{Same as Figure~\ref{fig:corner_gaia} but for wide binaries  {with abnormal PDFs of $\Gamma$}. The posterior PDFs of $\Gamma$ for these systems are unlike all the other in that one of them alone would strongly exclude Newtonian gravity. But, more reasonable possibility is that they suffer from kinematic contamination. All eight of such cases are listed in Table~\ref{tab:outlier}.}
    \label{fig:corner_outlier}
\end{figure*}

In all cases the individual PDFs of $\Gamma$ are much broader than the difference $\approx 0.07$ between Newton and MOND, and thus any specific result strongly inconsistent (e.g., $>3\sigma$) with the range $0\le\Gamma\le 0.07$ set by Newton and MOND may signal kinematic contaminants of hidden low-mass objects (e.g., faint stars or very massive Jovian planets) close to either or both of the stars,  {or some unknown issues with the RV data}\footnote{ {See \cite{katz2023} for a general description of the Gaia DR3 RV data.}}. Although the statistical sample was selected through a rigorous screening process, rare exceptions may be present. As a specific example, consider an unobserved companion star with $0.1M_\odot$ orbiting a host star with $1.0M_\odot$ at a separation of 10~au. For the sake of simplicity, assume a circular orbit. Then, the relative velocity is $v \approx 9900$~m\,s$^{-1}$ according to Equation~(\ref{eq:v3D}) and the velocity perturbation experienced by the host star is $\approx 9900\times 0.1/1.1 = 900$~m\,s$^{-1}$, which is significant. When the measured relative velocity between the two main stars is boosted by such a perturbation, the Bayesian modeling may return a PDF of $\Gamma$ that excludes the range $0\le\Gamma\le0.07$ with a high significance. This is confirmed with a mock sample including hierarchical systems.

\begin{table*}
  \caption{Wide binaries with  {abnormal PDFs of $\Gamma$} as inferred by Bayesian modeling}\label{tab:outlier}
\begin{center}
  \begin{tabular}{ccccccccc}
  \hline
 source id A  & source id B &  $M_{\rm{tot}}$ ($M_\odot$) &  $s$ (kau) & $s/r_{\rm{M}}$ & {\tt ruwe} A, B & $e$ & $\Gamma$ \\
 \hline
Gaia DR3 2533723670313663872 & 2533723464155501952 & 1.5 & 0.77 & 0.090 & 0.91, 0.88 & $0.96_{-0.02}^{+0.02}$ & $0.60_{-0.14}^{+0.18}$ \\
Gaia DR3 3262503032587954432 & 3262503032587954944 & 1.9 & 1.3  & 0.13  & 0.91, 0.89 & $0.49_{-0.24}^{+0.26}$ & $0.43_{-0.13}^{+0.19}$ \\
Gaia DR3 3999624183424427392 & 3999624183424190720 & 1.9 & 1.0  & 0.11  & 1.52, 1.33 & $0.95_{-0.02}^{+0.02}$ & $0.50_{-0.15}^{+0.19}$ \\
Gaia DR3 5626156262953609344 & 5626156267255201920 & 1.8 & 0.59 & 0.062 & 0.98, 0.97 & $0.95_{-0.05}^{+0.02}$ & $0.58_{-0.17}^{+0.19}$ \\
Gaia DR3 5662617820874266752 & 5662617820876042368 & 2.0 & 0.64 & 0.064 & 0.95, 1.03 & $0.77_{-0.09}^{+0.07}$ & $0.76_{-0.12}^{+0.14}$ \\
Gaia DR3 6388328691010410368 & 6388327179181381248 & 1.6 & 2.0  & 0.22  & 1.35, 1.06 & $0.66_{-0.12}^{+0.15}$ & $0.55_{-0.13}^{+0.18}$ \\
Gaia DR3 6830027182179257472 & 6830027143525634432 & 1.4 & 1.6  & 0.19  & 0.99, 1.09 & $0.71_{-0.17}^{+0.13}$ & $0.54_{-0.15}^{+0.19}$ \\
Gaia DR3 999837501401204992  & 999837501401205120  & 2.0 & 0.52 & 0.052 & 1.08, 1.03 & $0.90_{-0.07}^{+0.04}$ & $0.56_{-0.15}^{+0.18}$ \\
  \hline
\end{tabular}
\end{center}
Note. (1) The quoted values of $e$ and $\Gamma$ are from the posterior PDFs: examples are given in Figure~\ref{fig:corner_outlier}. If these were pure wide binaries without hidden kinematic contaminants  {and the RV values and uncertainties were reliable}, one of these systems alone would rule out Newtonian gravity. (2) Most stars have normal {\tt ruwe} values; ${\tt ruwe}<1.3$ is satisfied by all except for three stars, which have $1.3\la {\tt ruwe} \la 1.5$ that are still not so large. This indicates that kinematically abnormal systems  {based on Gaia data} not necessarily have large {\tt ruwe} values.
\end{table*}

Any kinematic perturbers may cause outliers only in the positive direction of $\Gamma$. This is indeed the case for the Bayesian results of the statistical sample. To be concrete, I define the lower ($\sigma_{\rm{low}}$) and upper ($\sigma_{\rm{up}}$) uncertainties of $\Gamma$ with the 16th and 84th percentiles in its PDF. Then, $\Gamma_{\rm{med}}+3\sigma_{\rm{up}}>0$ is satisfied in all cases, but $\Gamma_{\rm{med}}-3\sigma_{\rm{down}}<0.02$ (here $0.02$ is used to allow for a median effect of RV uncertainties in a transition regime between the Newtonian and the MOND regimes) is not satisfied by eight wide binaries. Two examples of these outliers can be found in Figure~\ref{fig:corner_outlier}. Clearly, any of these results alone would not only rule out the range $0\le\Gamma\le 0.07$, but would also be inconsistent with the PDFs of many other wide binaries. The properties of the eight wide binaries with suspected kinematic contaminants  {or abnormal RV data} are summarized in Table~\ref{tab:outlier}. These systems will not be included in the inference of consolidated gravity in the following subsections. In doing so, it is guaranteed that all individual binaries are consistent with Newtonian gravity within $\approx 3\sigma$ and the consolidated gravity is not dominated by a few radically different ones, but determined by the collective trend of the majority.  {See, however, Section~\ref{sec:meaning} for discussions on alternative interpretations.}

\begin{figure*}
    \centering
    \includegraphics[width=1.0\linewidth]{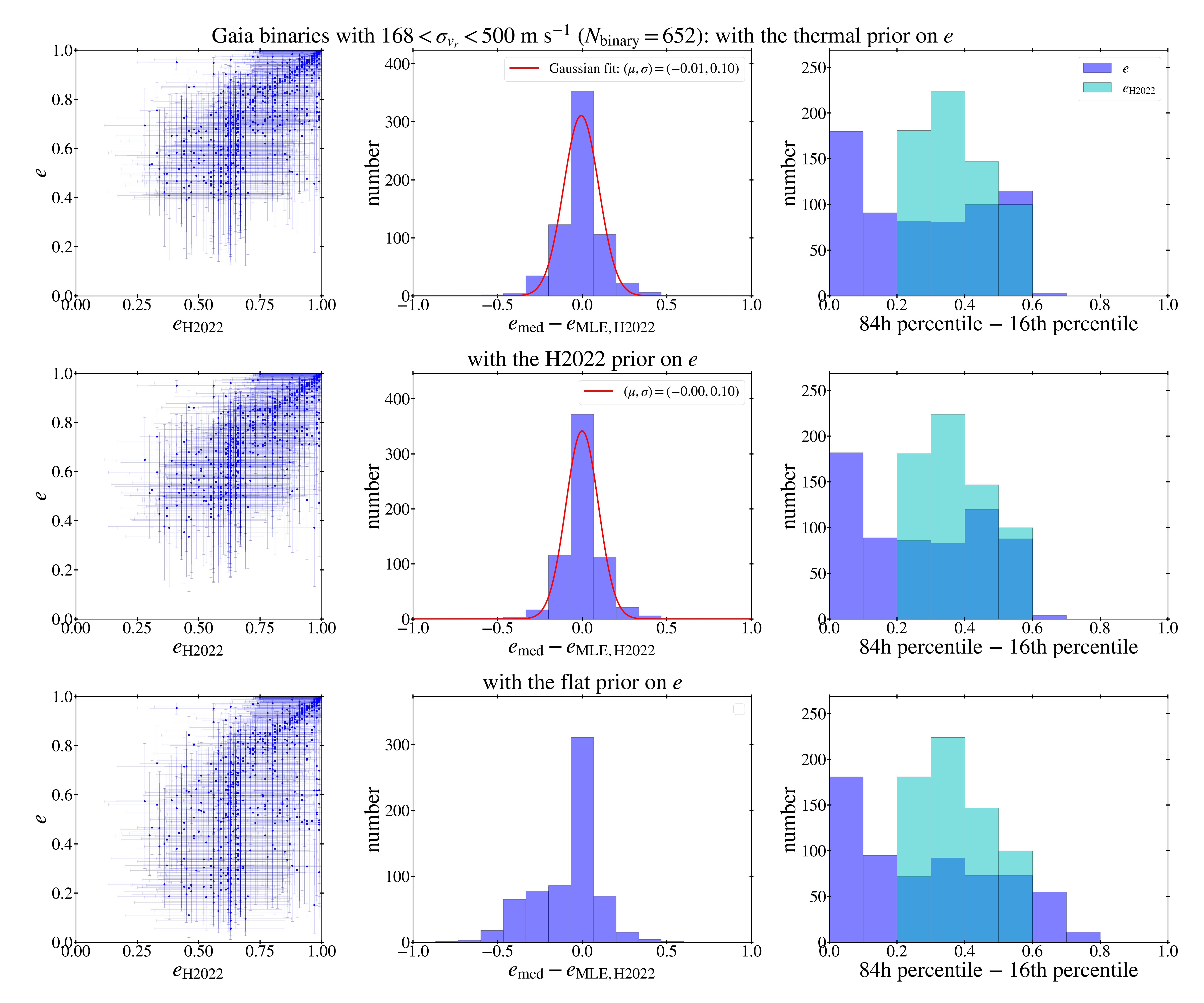}
    \caption{Bayesian inferred eccentricities for the statistical sample from this work ($e$) are compared with those ($e_{\rm{H2022}}$) from \cite{hwang2022}.  {The upper, middle, and lower rows show respectively the eccentricities inferred with the thermal, the H2022, and the flat priors.}  In the  {upper and middle rows}, there is an excellent match between this work and \cite{hwang2022} as the middle column compares the median values ($e_{\rm{med}}$) from this work with the maximum likelihood estimate (MLE) values ($e_{\rm{MLE,H2022}}$) from \cite{hwang2022}. The right column compares ``$1\sigma$'' full widths between this work and \cite{hwang2022}. Because this work uses the additional information of radial velocities, the widths from this work are overall narrower.}
    \label{fig:eccen_distribution}
\end{figure*}

Properties of the inferred orbital and geometric parameters are of interest not only to check whether they are reasonable but also in their own right. Eccentricities of wide binaries are of particular interest. For all wide binaries in the \cite{elbadry2021} catalog from which the statistical sample was drawn, \cite{hwang2022} have inferred eccentricities with an independent Bayesian method using only the sky-projected relative displacements and relative velocities (without RVs). Figure~\ref{fig:eccen_distribution} compares the eccentricities from this work with those from \cite{hwang2022}.  {The eccentricities with the thermal prior agree excellently with the \cite{hwang2022} values, as do those with the H2022 prior. However, those with the flat prior show some discrepancy.} This is not surprising given that the statistical properties of observationally inferred eccentricities for wide binaries (see Figure~24 of \cite{chae2023a}) are much more consistent with the  {thermal or H2022 distribution than} the flat distribution. Moreover, as will be shown in Section~\ref{sec:result_newt} below, a test with the Newtonian control sample indicates that the flat prior can cause some bias in the Bayesian inference of gravity when the RVs are not sufficiently precise. The right column of Figure~\ref{fig:eccen_distribution} compares the widths of the PDFs between this work and \cite{hwang2022}. Because this work uses an additional constraint from RVs, it is expected that the widths from this work will be narrower. Indeed, this is the case. In particular, the results with the thermal  {or H2022} prior exhibit narrower widths more clearly.

\begin{figure}
    \centering
    \includegraphics[width=1.0\linewidth]{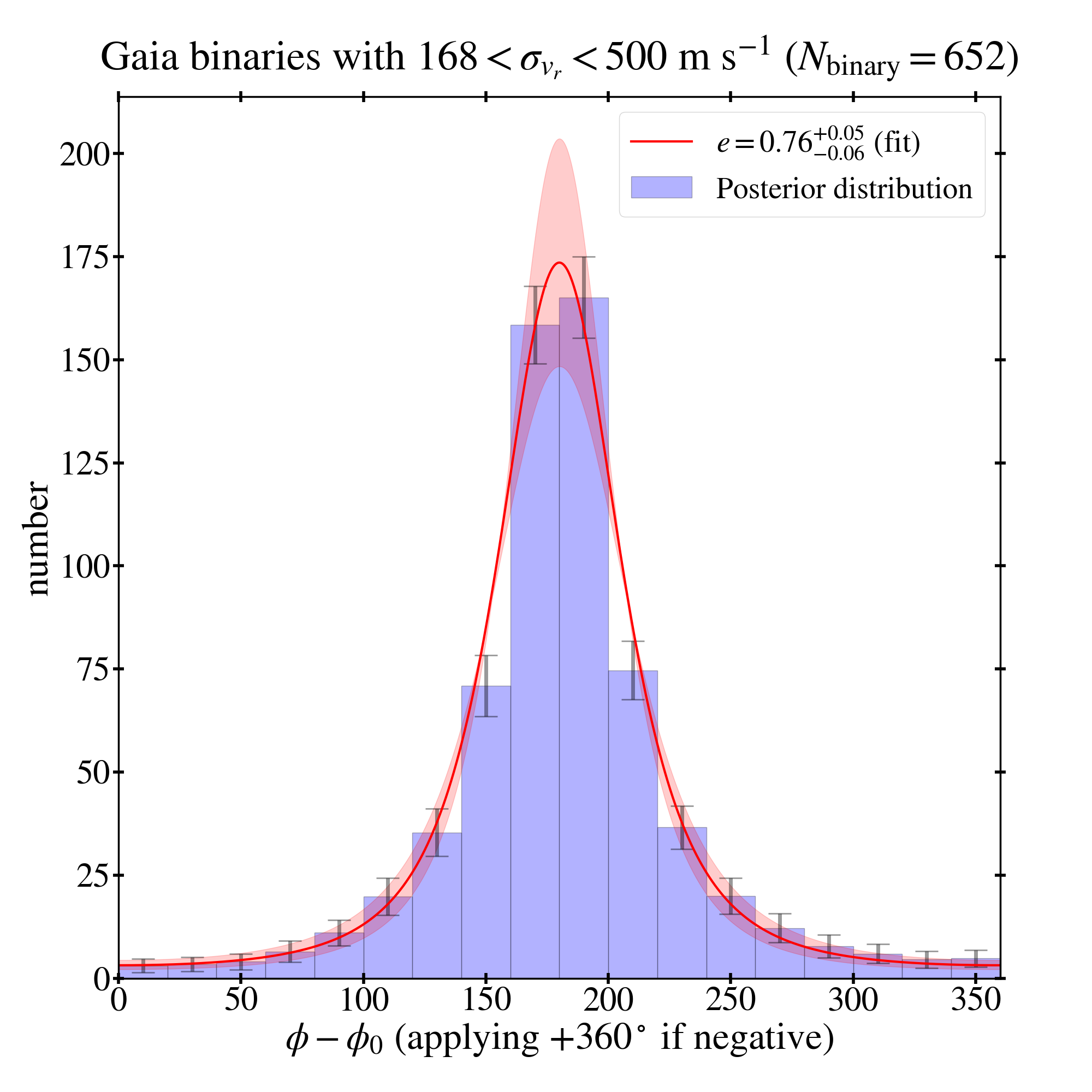}
    \caption{Bayesian inferred values of parameter $\phi_0$ (the longitude of the periastron, see Figure~\ref{fig:orbit}) are distributed with respect to the observed phase $\phi=\tan^{-1}((\Delta y^\prime/\Delta x^\prime)/\cos i_{\rm{obs}})$ with $i_{\rm{obs}}=\tan^{-1}(-v_r/v_{y^\prime})$. The light blue histogram shows the median of 400 distributions where each distribution is derived from one set of values for the 652 binaries of the statistical sample, each of which is taken randomly from the MCMC-produced $2\times 10^6$ values for each binary. Error bars refer to scatters in the 400 distributions. The red curve is a fit of Equation~(\ref{eq:prphi0}) with $e=0.76$ to the median distribution while the light shaded area represents the uncertainties of $e$. As expected, the posterior distribution of $\phi_0$ agrees well with the prior distribution of Equation~(\ref{eq:prphi0}) with a reasonable value of $e$.}
    \label{fig:delphi_distribution}
\end{figure}

For elliptical orbits, random snapshot observations will not have a uniform occurrence rate of $\phi-\phi_0$ where $\phi$ is the observed phase $\phi=\tan^{-1}((\Delta y^\prime/\Delta x^\prime)/\cos i_{\rm{obs}})$ with $i_{\rm{obs}}=\tan^{-1}(-v_r/v_{y^\prime})$ and $\phi_0$ is the longitude of the periastron. Thus, it is interesting to check the posterior distribution of $\phi-\phi_0$ for the statistical sample where $\phi_0$ is drawn from the PDF of each wide binary. Figure~\ref{fig:delphi_distribution} shows the distribution of $\phi-\phi_0$ for the 652 wide binaries of the statistical sample. The histogram represents the median of many distributions where each distribution represents values drawn once from all wide binaries. The error bars represent the scatter from one draw to another. This posterior distribution agrees remarkably well with the prior distribution given by Equation~(\ref{eq:prphi0}) with a reasonable representative value of $e=0.76_{-0.06}^{+0.05}$, indicating that snapshot observations are truly random in normalized time $t/P$ (where $t$ is the time from the passage of the periastron and $P$ is the orbital period).

\begin{figure*}
    \centering
    \includegraphics[width=0.7\linewidth]{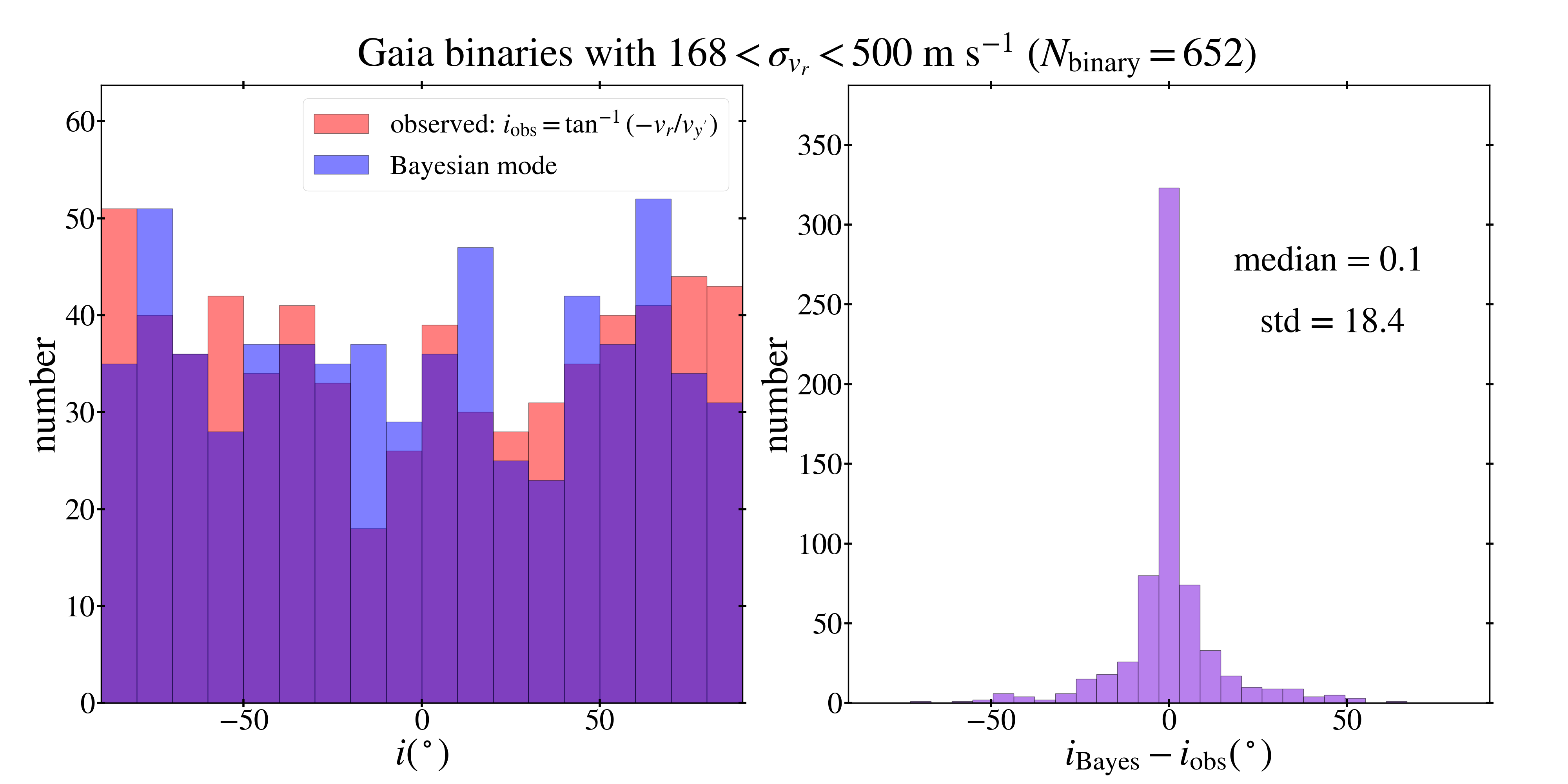}
    \caption{This left panel shows the distribution of the inclination parameter $i$ (see Figure~\ref{fig:orbit}) for the statistical sample. The red histogram shows the distribution of the observed values $i_{\rm{obs}}=\tan^{-1}(-v_r/v_{y^\prime})$ while the blue histogram shows the distribution of the modes in the posterior PDFs of $i$. Here the mode rather than the median is considered because $i$ can have a very broad range even including the opposite sign (see Figure~\ref{fig:probdist_inc}). The right panel shows the distribution of the differences between the two.}
    \label{fig:inc_distribution}
\end{figure*}

Figure~\ref{fig:inc_distribution} compares the distribution of $i_{\rm{obs}}=\tan^{-1}(-v_r/v_{y^\prime})$ with that of the modes in the posterior PDFs. Here, the modes rather than the medians are considered because the mode better agrees with the input inclination as can be seen in the simulations shown in Figure~\ref{fig:probdist_examples}. The two distributions agree statistically well with each other.

The posterior distribution of $\log_{10}f_M$ always agrees well with the prior input distribution because no observational constraints have any dependence on it. Allowing for a finite width of its PDF rather than fixing $\log_{10}f_M=0$ serves merely as a way to take into account the uncertainties of mass in the inference of $\Gamma$. 

To sum up, the Bayesian outputs of the parameters $e$, $\phi_0$, $i$ and $\log_{10}f_M$ are all reasonable, individually and statistically. As for the gravity parameter $\Gamma$, only eight wide binaries listed in Table~\ref{tab:outlier} return unacceptable PDFs given any theoretically conceivable range. These are interpreted as exceptional systems that hide kinematic contaminants  {and/or abnormal RV data}. These systems will not be used to derive consolidated values of $\Gamma$ in the main analyses.  {Alternative interpretations are discussed in Section~\ref{sec:meaning}.}

Since the goal of this work is to test gravity in low-acceleration regimes ($g_{\rm{N}}\la 10^{-9}$~m\,s$^{-2}$), binaries are split by their internal acceleration between the pair. Binaries in a relatively strong-acceleration regime ($g_{\rm{N}}\ga 10^{-8}$~m\,s$^{-2}$) are considered in Section~\ref{sec:result_newt} while binaries in low-acceleration regimes are considered in Section~\ref{sec:result_mond}. 

For the purpose of splitting binaries by internal acceleration, I use the normalized separation $s/r_{\rm{M}}$ where $s$ is the projected separation and $r_{\rm{M}}$ is the MOND radius. Since the median 3D separation is approximately given by $r \approx (4/\pi)s$ as can be verified numerically with orbits such as those shown in Figure~\ref{fig:orbit_simulation2}, the Newtonian acceleration at $s/r_{\rm{M}}$ can be estimated by
\begin{equation}
    g_{\rm{N}} \approx a_0 \left(\frac{\pi}{4}\right)^2 \left(\frac{s}{r_{M}} \right)^{-2}.
    \label{eq:gNstilde}
\end{equation}

The binaries in the statistical sample cover the range $0.02< s/r_{\rm{M}}< 2.86$. I consider the following acceleration regimes defined by $s/r_{\rm{M}}$: a Newtonian regime $0.02<s/r_{\rm{M}}<0.08$ ($10^{-7.9}\la g_{\rm{N}} \la 10^{-6.7}$~m\,s$^{-2}$), a transition regime $0.15<s/r_{\rm{M}}<0.50$ ($10^{-9.5}\la g_{\rm{N}} \la 10^{-8.5}$~m\,s$^{-2}$), a MOND regime $0.50<s/r_{\rm{M}}<2.86$ ($10^{-11.0}\la g_{\rm{N}} \la 10^{-9.5}$~m\,s$^{-2}$), and a transition+MOND regime $0.15<s/r_{\rm{M}}<2.86$ ($10^{-11.0}\la g_{\rm{N}} \la 10^{-8.5}$~m\,s$^{-2}$). The systems listed in Table~\ref{tab:outlier} or outside the magnitude cut line shown in Figure~\ref{fig:CM} will not be included in inferring the main results on gravity below. 

\subsection{Consolidated gravity in a Newtonian regime} \label{sec:result_newt} 

Binaries whose internal accelerations are stronger than $\approx 10^{-8}$~m\,s$^{-2}$ are expected to obey Newtonian dynamics whether in standard gravity or nonstandard gravity theories, in the context of orbital motions. Thus, a sample of binaries in such a Newtonian regime provides a control sample to check the validity of the Bayesian methodology with real data, although the methodology was already verified with mock data in Section~\ref{sec:demo}.

\begin{figure*}
    \centering
    \includegraphics[width=1.0\linewidth]{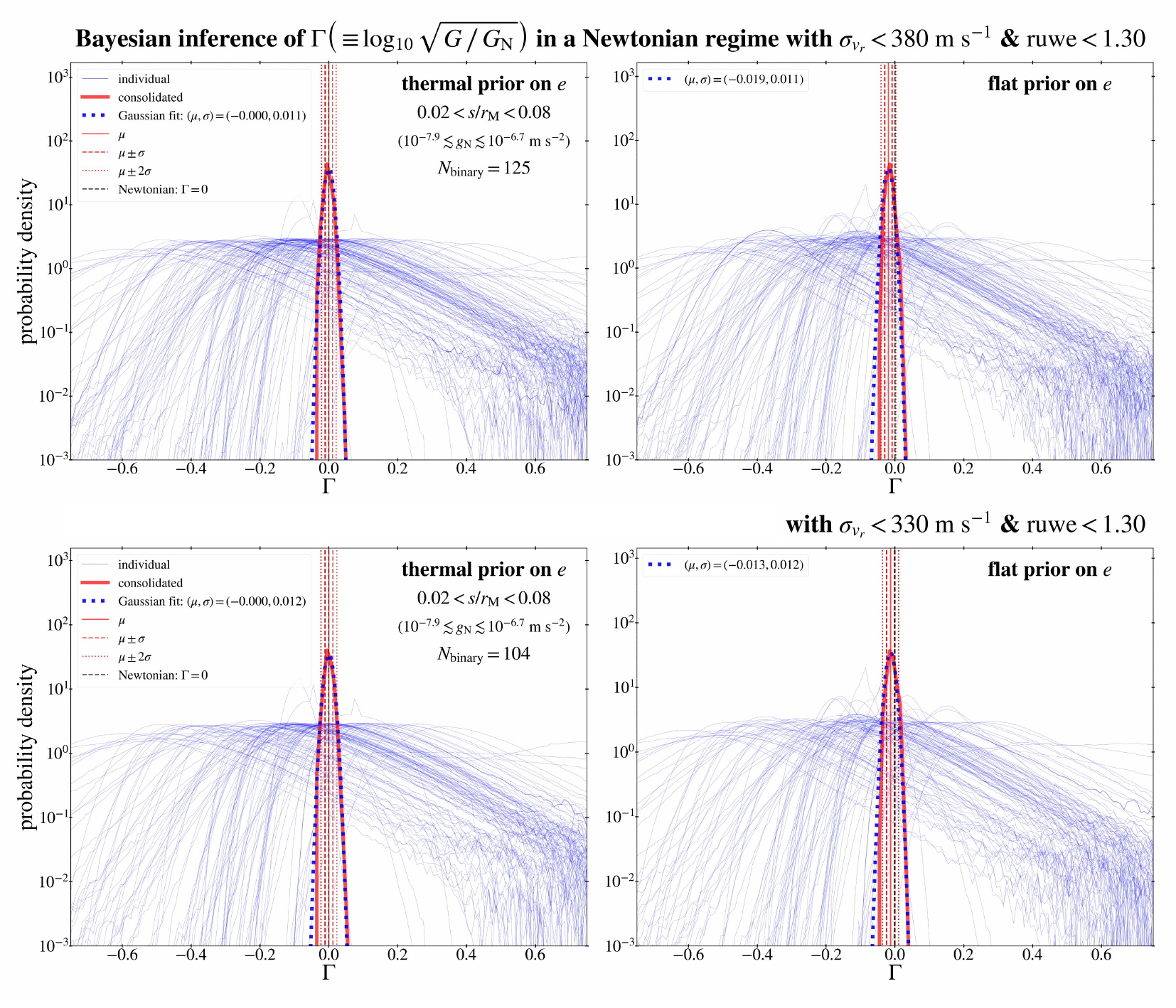}
    \caption{Each panel shows the individual posterior PDFs of the gravitational anomaly parameter $\Gamma$ (thin blue curves) and their consolidated distribution (thick red curve) for wide binaries in a Newtonian (i.e.\ relatively strong internal gravitational acceleration) regime with $0.02<s/r_{\rm{M}}<0.08$ where $s$ is the sky-projected separation and $r_{\rm{M}}$ is the MOND radius (Equation~(\ref{eq:MONDradius})). Thick blue dotted curve is a Gaussian function matching to the consolidated distribution. The upper/lower row is for the nominal/alternative sample while the left/right column is with the thermal/flat prior on eccentricity. The consolidated distributions with the thermal prior nearly perfectly match Newtonian gravity while those with the flat prior show mild discrepancies (still acceptable within $2\sigma$). }
    \label{fig:Gam_probdist_Newton}
\end{figure*}

Figure~\ref{fig:Gam_probdist_Newton} shows the results on the consolidated PDF of $\Gamma$ with the wide binaries in the Newtonian regime defined at the end of Section~\ref{sec:result_intro}. The results with the thermal and flat priors on eccentricity ($e$) for the standard and alternative samples are shown. The results with the thermal prior agree remarkably well with the Newtonian expectation $\Gamma=0$, while those with the flat prior show a moderate bias toward the negative side. The bias with the flat prior is not surprising given that the implicit median of $e$ is biased, while at least some current data are not precise enough to strongly constrain $e$ regardless of the prior. 

\begin{figure*}
    \centering
    \includegraphics[width=1.0\linewidth]{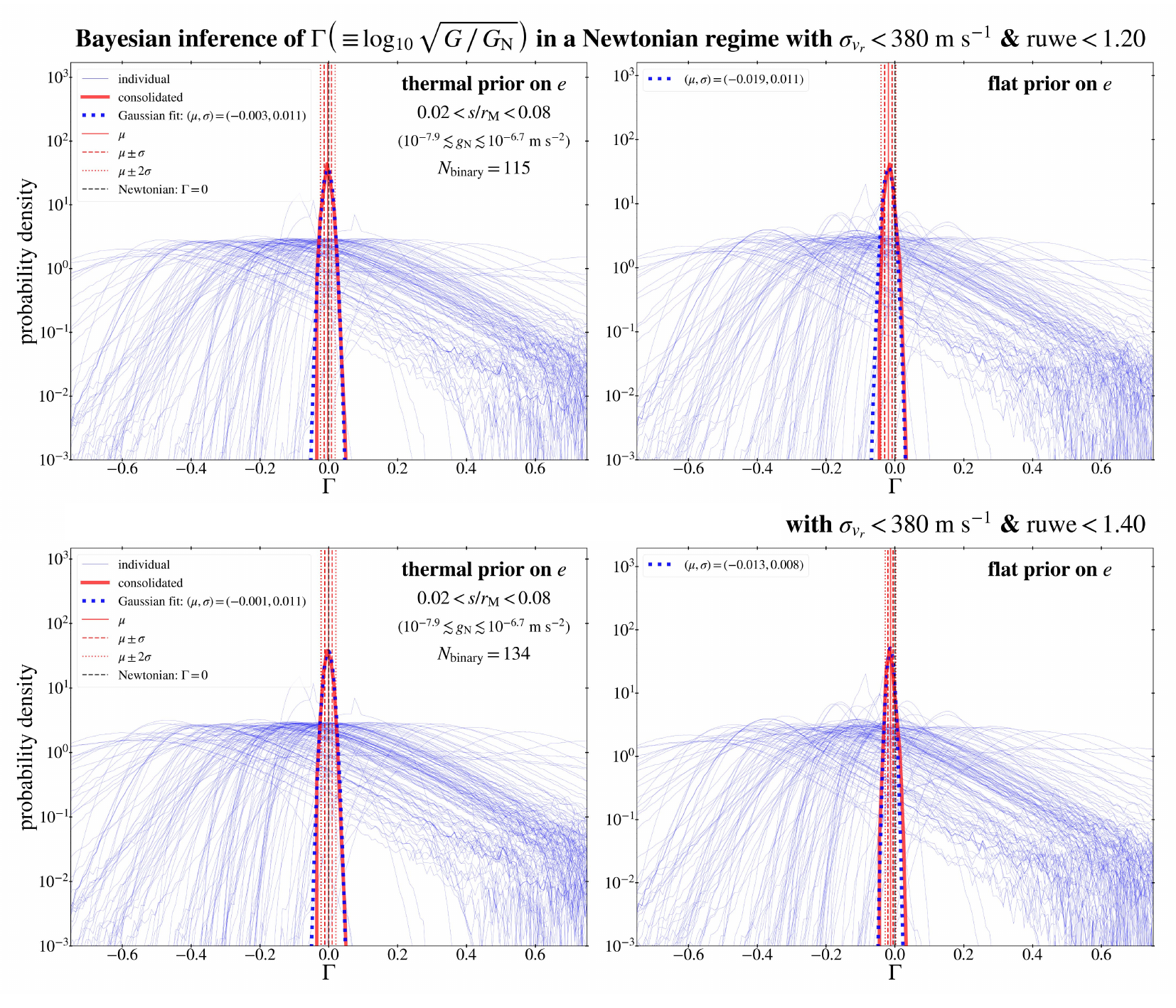}
    \caption{Each row is same as the upper row of Figure~\ref{fig:Gam_probdist_Newton} but with different limits of {\tt ruwe} in selecting the binaries.}
    \label{fig:Gam_probdist_Newton_ruwe}
\end{figure*}

Possible systematics can be explored by varying the sample selection criteria or the prior inputs.  {First of all, the results with the H2022 prior on eccentricities are essentially the same as those with the thermal prior because $\alpha$ in $p(e)=(1+\alpha)e^\alpha$ is close to $1$ for the Newtonian regime under consideration.} Varying the limit of $\sigma_{v_r}$ has no impact on the results as the numbers of wide binaries in the nominal and alternative samples are already sufficiently large and all $\sigma_{v_r}$ values in the statistical sample are small enough compared with relatively large internal relative velocities in the pairs. Varying the limit of {\tt ruwe} from the nominal limit $1.3$ has no impact either. Figure~\ref{fig:Gam_probdist_Newton_ruwe} shows the results with {\tt ruwe} $<1.2$ or $<1.4$. Thus, it is concluded that the measured value is $\Gamma = 0.000\pm 0.011$ or $\gamma_g=1.00\pm 0.05$ in the Newtonian regime based on the nominal sample and any systematics are negligible.

\subsection{Consolidated gravity in transition and MOND regimes} \label{sec:result_mond}

Compared with the Newtonian regime, wide binaries in the transition ($0.15<s/r_{\rm{M}}<0.50$) and MOND ($0.50<s/r_{\rm{M}}<2.86$) regimes have smaller internal relative velocities, and the Gaia measurement uncertainties of the RVs can be comparable to (or even smaller than) them. Also, the numbers of wide binaries are smaller than in the Newtonian regime. Therefore, larger statistical uncertainties of $\Gamma$ are expected. Moreover, as evident in Figure~\ref{fig:vprofile}, the Gaia-reported $|v_r|$ values are systematically biased upward (compared with the other precise velocity components) due to the relatively large uncertainties of RVs. The bias in $|v_r|$ means a (though smaller) bias in $v(=|\mathbf{v}|)$, which will induce a systematic bias in the consolidated PDF of $\Gamma$. This systematic bias will be estimated below using mock Newtonian samples.

\begin{figure*}
    \centering
    \includegraphics[width=1.0\linewidth]{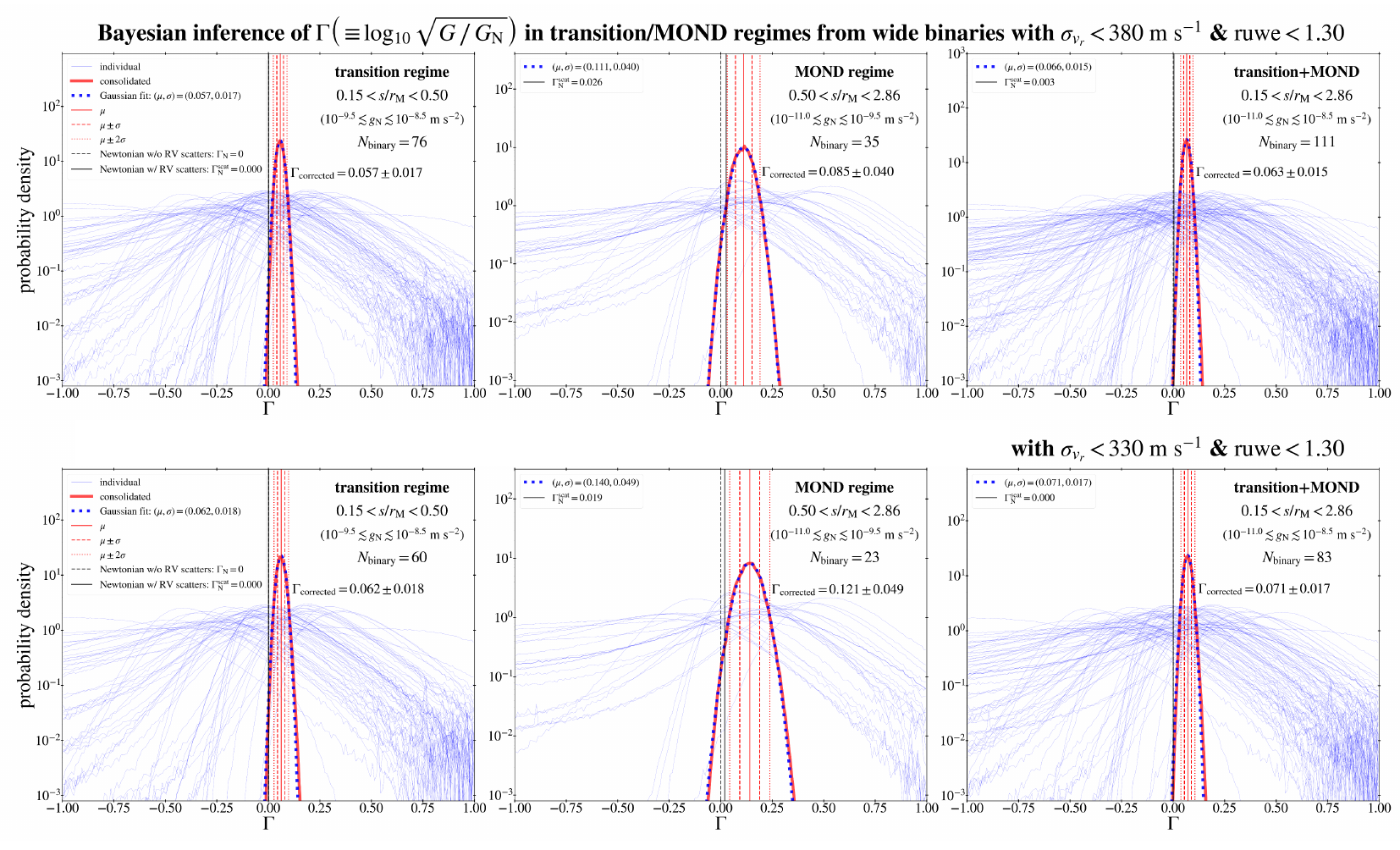}
    \caption{Each column is similar to the left column of Figure~\ref{fig:Gam_probdist_Newton} but for the transition, the MOND, and the combined regimes from left to right. Unlike Figure~\ref{fig:Gam_probdist_Newton}, a Newtonian prediction with RV scatters for each regime is estimated (see the text) and indicated by the black solid vertical line and subtracted from the consolidated distribution of $\Gamma$ to obtain $\Gamma_{\rm{corrected}}$ corrected for the bias due to RV scatters. The bias correction is appreciable only for the MOND regime because $v_r$ values are not sufficiently large compared with the measurement errors only for widely separated binaries with $s/r_{\rm{M}}\ga 0.5$. In contrast with the Newtonian regime shown in Figure~\ref{fig:Gam_probdist_Newton}, the consolidated distributions in these low-acceleration regimes show clear discrepancies with Newtonian gravity. }
    \label{fig:Gam_probdist_transmond}
\end{figure*}

Figure~\ref{fig:Gam_probdist_transmond} shows the consolidated distributions of $\Gamma$ in the transition, the MOND, and the combined regimes of internal acceleration for the nominal and alternative samples. The relevant systematic bias (or shift) of the PDF that will be described below is also indicated and subtracted from each distribution to obtain a corrected value of $\Gamma$. Note that the bias is appreciable only in the MOND regime. 

Figure~\ref{fig:Gam_probdist_transmond} reveals two remarkable features. First, in striking contrast with the results in the Newtonian regime, the consolidated PDFs of $\Gamma$ clearly deviate from $\Gamma=0$ in all cases. The discrepancy with Newton is $\approx 3.4\sigma$ in the transition regime, while it is $>2\sigma$ in the MOND regime, based on either the nominal ($\sigma_{v_r}<380$~m\,s$^{-1}$ \& {\tt ruwe} $<1.30$) or alternative ($\sigma_{v_r}<330$~m\,s$^{-1}$ \& {\tt ruwe} $<1.30$) sample. When all PDFs from both regimes are combined, the discrepancy with Newton is $\approx 4.2\sigma$ based on either sample. 

Second, the deviation is larger in the MOND regime than the transition regime, indicating a gradual deviation from Newton in the high-acceleration regime ($\ga 100 a_0$) to modified gravity in the low-acceleration regime ($\la a_0$). This trend and the estimated value $\Gamma_{\rm{corrected}}=0.085\pm 0.040$ or $\gamma_g=1.48_{-0.23}^{+0.30}$ (after correcting for the bias described below)\footnote{Only in this subsection, the symbol ﻿$\Gamma_{\rm{corrected}}$ is used to indicate the correction for the bias. Elsewhere in the paper, the symbol $\Gamma$ refers to $\Gamma_{\rm{corrected}}$ if the bias correction is present.} in the MOND regime based on the nominal sample agree well with the generic prediction $\Gamma \approx 0.07$ (or $\gamma_g \approx 1.4$) of MOND Lagrangian gravity models (e.g., \citealt{bekenstein1984,milgrom2010}). The value $\Gamma_{\rm{corrected}}=0.121\pm 0.049$ or $\gamma_g=1.75_{-0.32}^{+0.50}$ based on the alternative sample is also consistent with $\Gamma \approx 0.07$. These results through the new Bayesian methodology match well with several recent results \citep{chae2023a,chae2024a,chae2024c,hernandez2023,hernandez2024a} based on various statistical analyses utilizing $v_p$ only.

\begin{figure*}
    \centering
    \includegraphics[width=0.9\linewidth]{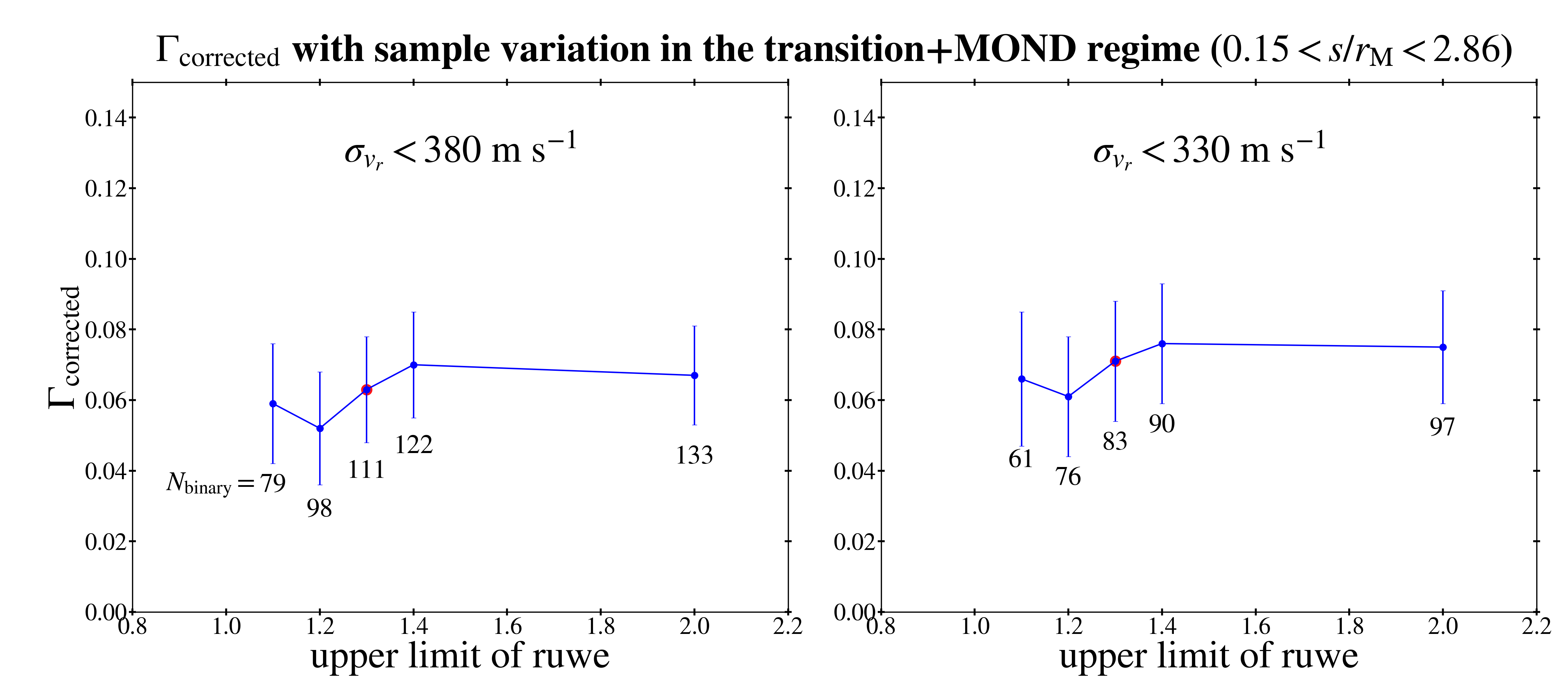}
    \caption{The Bayesian inference of the gravitational anomaly parameter $\Gamma$ for the transition+MOND regime shown in the right column of Figure~\ref{fig:Gam_probdist_transmond} is further investigated with samples varied with the upper limit of {\tt ruwe}. What is shown is $\Gamma_{\rm{corrected}}$ corrected for the small bias due to RV scatters. }
    \label{fig:scaling_Gamma}
\end{figure*}

The above results are based on the two samples with {\tt ruwe} $<1.30$. More results with alternative cuts of {\tt ruwe} can be found in Figure~\ref{fig:scaling_Gamma}. Clearly, for the binaries that have already passed a number of criteria, the {\tt ruwe} cut does not seem to matter up to $2.0$.

Now I describe the systematic bias in $\Gamma$ induced by the RV uncertainties. To estimate it, I use mock Newtonian binaries produced with the procedure described in Section~\ref{sec:mock}.  {The reason for this is twofold. First, to estimate the bias in $\Gamma$, the unbiased value must be accurately known. While the unbiased value is exact (that is, zero) in Newtonian gravity, it is not so in any modified gravity. Second, based on all previous studies of wide binary gravity tests, any effective gravity experienced by wide binaries is not expected to be drastically different from Newton but rather pseudo-Newtonian or something similar. Thus, at least as a first-order approximation, Newtonian gravity can be used for the purpose of estimating the bias due to large uncertainties of RVs. However, I will also consider mock binaries produced with the observed Gaia binaries themselves by giving scatter to the observed RVs. It will be shown that the biases estimated with mock Newtonian and mock Gaia samples are similar. The nominal estimates will be based on mock Newtonian samples. }

For each Gaia binary of the nominal sample ($N_{\rm{binary}}=111$) in the range $0.15<s/r_{\rm{M}}<2.86$, I produce 100 random binaries with the same values of $s$ and $M_{\rm{tot}}$ (and thus the same $s/r_{\rm{M}}$ as well), and with the same endowed uncertainties of RVs but random otherwise. PDFs of $\Gamma$ are derived for these mock binaries. Thus, a total of $100 \times 111$ different mock PDFs of $\Gamma$ are in hand. Now, $N$ sets of mock PDFs of $\Gamma$ corresponding to the set of 111 Gaia binaries are obtained using the mock PDFs in a bootstrap manner. Here, each set is composed of 111 PDFs, each of which is drawn from the set of the ready-made 100 mock PDFs for each Gaia binary. I consider $N=10^5$, meaning that the 100 mock PDFs for each Gaia binary are used multiple times. Although any two sets can share some PDFs in doing so, all sets are different and can be taken as random Newtonian realizations. 

\begin{figure*}
    \centering
    \includegraphics[width=1.0\linewidth]{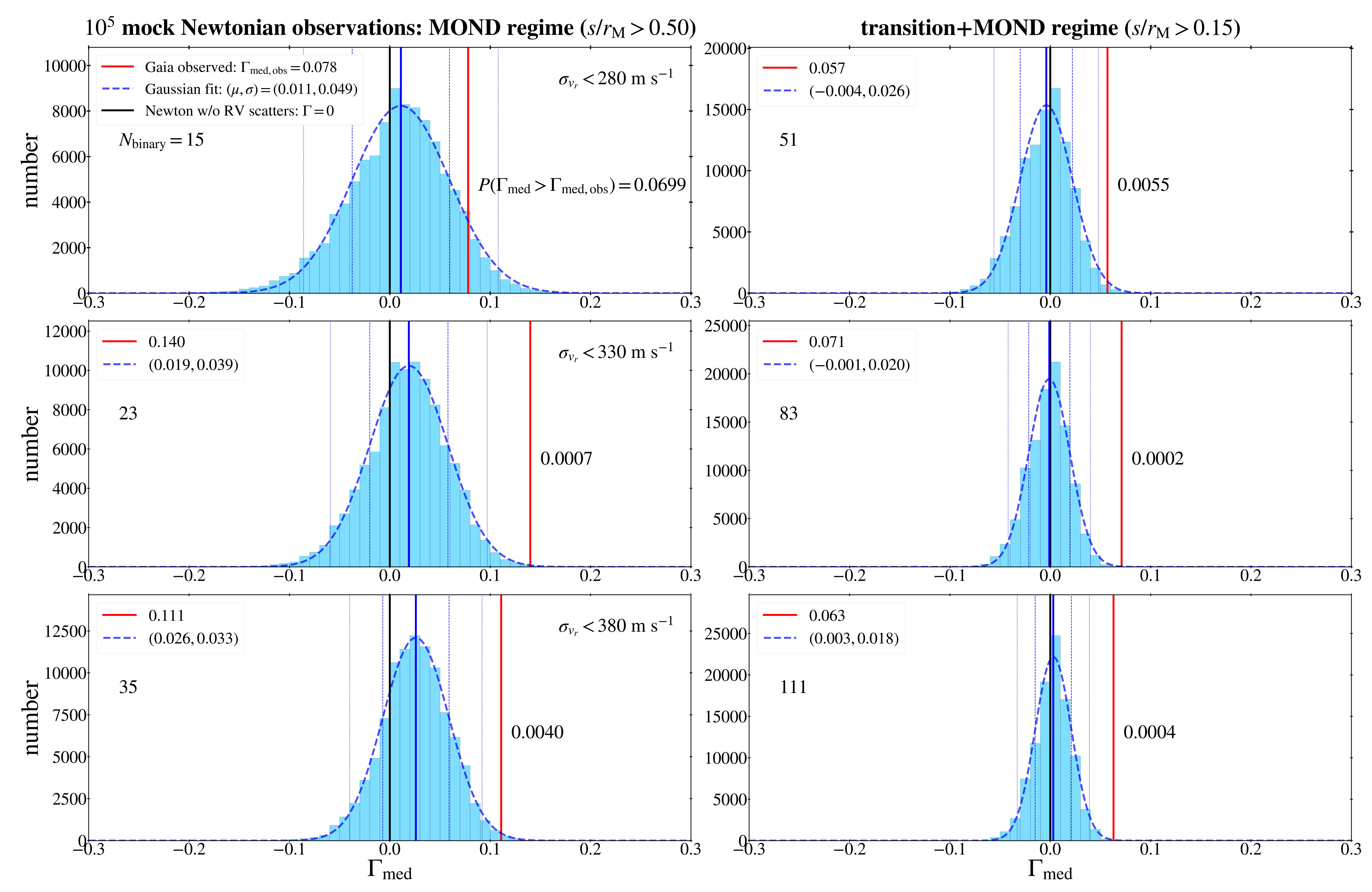}
    \caption{$N=10^5$ mock Newtonian samples corresponding to each Gaia binary sample are used to obtain $N$ consolidated distributions of $\Gamma$. In practice, a bootstrap strategy is employed to save the computing time; only a pool of 100 mock Newtonian binaries are produced for each Gaia binary and these $N_{\rm{binary}}$ pools are used to provide different $N$ samples in a bootstrap way (see the text for the details). Each histogram shows the distribution of the medians ($\Gamma_{\rm{med}}$) of the $N$ consolidated distributions of $\Gamma$. The left column considers three samples in the MOND regime with different upper limits of $\sigma_{v_r}$. The samples with $\sigma_{v_r}<380$~m\,s$^{-1}$ and $<330$~m\,s$^{-1}$ are the nominal and alternative samples shown in Figure~\ref{fig:Gam_probdist_transmond}. The smaller sample with $\sigma_{v_r}<280$~m\,s$^{-1}$ is considered only here for the purpose of illustration. As the upper limit of $\sigma_{v_r}$ increases from top to bottom, the histogram exhibits two systematic variations: (1) the median represented by the vertical blue line is gradually shifted to the right from top to bottom panel due to the gradually increased scatters of $v_r$, (2) the width of the histogram decreases gradually as the sample size increases. The value $\Gamma_{\rm{med,obs}}$ for each Gaia sample (derived in Figure~\ref{fig:Gam_probdist_transmond}) is indicated by the vertical red line. $P(\Gamma_{\rm{med}}>\Gamma_{\rm{med,obs}})$ is the Newton-predicted probability that the observed value or higher would occur by chance. The right column shows the results for the Gaia samples in the transition+MOND regime. The very small values of $P(\Gamma_{\rm{med}}>\Gamma_{\rm{med,obs}})$ for the nominal and alternative samples show the difficulty of Newtonian gravity. Note, however, that the widths of the histograms in these Newtonian simulations are somewhat different from the widths of the corresponding consolidated distributions for the Gaia samples shown in Figure~\ref{fig:Gam_probdist_transmond}  {(See the text for further details.)}. }
    \label{fig:hist_Gam_mockNewton}
\end{figure*}

It turns out that the bias can be appreciable only for binaries in the MOND regime. Thus, I quantify biases in the MOND and MOND+transition regimes. Figure~\ref{fig:hist_Gam_mockNewton} shows the distributions of $\Gamma_{\rm{med}}$ (median of $\Gamma$) in the $N$ consolidated distributions of $\Gamma$ for the MOND  and MOND+transition regimes (one consolidated distribution giving one $\Gamma_{\rm{med}}$). For each regime, I consider three samples by varying the upper limit of $\sigma_{v_r}$. The histograms are well approximated by Gaussian distributions. As expected, the width of the distribution gets narrower as the sample size increases.

For the MOND regime, the Newton-predicted distributions are shifted to the right, and the systematic shift increases (although the width of the distribution gets narrower) as the upper limit of $\sigma_{v_r}$ increases. For the MOND+transition regime, the distributions seem to be dominated by more numerous and narrower PDFs in the transition regime, so that the shifts are very small. Each distribution represents a predicted probability distribution of mock observations in a Newtonian world. That is, if gravity were Newtonian, an observed value of $\Gamma_{\rm{med}}$ would fall within a likely range. However, what happens is that the observed values fall outside the predicted likely ranges and thus seem very unlikely in Newtonian gravity as the calculated survival probabilities are indicated in the figure. Here, I note that the widths of the simulated distributions  {agree up to 20\% with} those of the corresponding observed consolidated distributions shown in Figure~\ref{fig:Gam_probdist_transmond}.  {The widths of the simulated distributions are somewhat narrower in the MOND regime but somewhat wider in the transition+MOND regime.} 

\begin{figure*}
    \centering
    \includegraphics[width=1.0\linewidth]{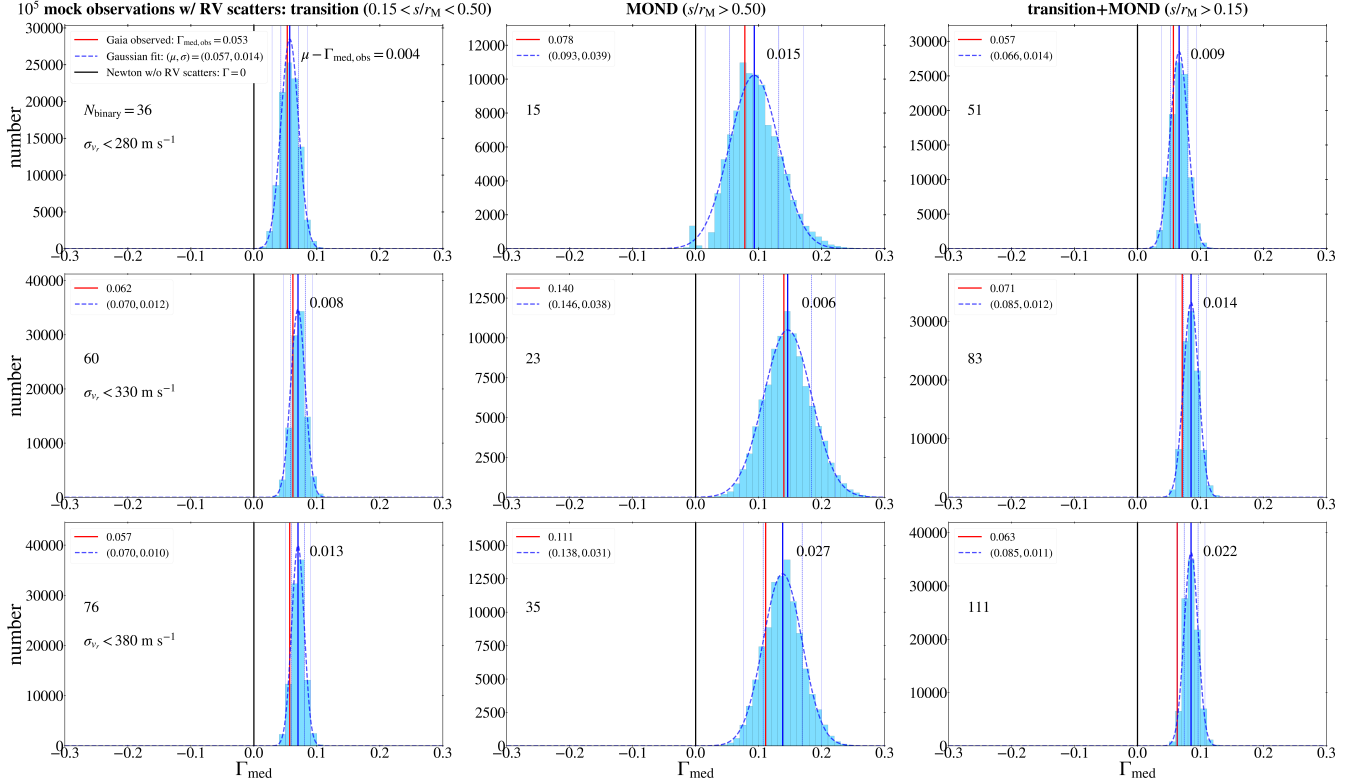}
    \caption{  {Similar to Figure~\ref{fig:hist_Gam_mockNewton} but for $N=10^5$ mock ``Gaia'' samples that are produced by adding to the Gaia RV values random scatters using their reported uncertainties. Thus, in this case the benchmark gravity is not Newton but the Gaia-implied one (uncorrected for the bias being investigated) that is of course uncertain. Unlike Figure~\ref{fig:hist_Gam_mockNewton}, the transition regime is exhibited in the left column since there occur small shifts. Because the benchmark gravity models indicated by red vertical lines are subject to uncertainties, the shifts are also unstable as opposed to the Newtonian case in Figure~\ref{fig:hist_Gam_mockNewton} where the shifts are gradual from top to bottom. The widths of the distributions are similar to, or somewhat different from, the corresponding ones in Figure~\ref{fig:Gam_probdist_transmond} and Figure~\ref{fig:hist_Gam_mockNewton}. See the text for further details. } }
    \label{fig:hist_Gam_mockGaia}
\end{figure*}

 {Simulations using Gaia RVs themselves are also carried out in a very similar way. Here the Gaia-reported RVs are used to produce mock RVs by adding scatter using the reported nominal uncertainties. Thus, in this case the uncorrected gravity based on the Gaia data provides the input gravity as opposed to Newtonian gravity in the above case. Because the input gravity is not known precisely in this case, there is a minor difference in the details of the procedure. That is to say, because the Gaia-reported RVs are already scattered quantities due to measurement errors, adding additional scatter can, in few cases, produce too wildly scattered RVs that lead to extremely unlikely PDFs of $\Gamma$, e.g., more than $4\sigma$ away from the ``input'' gravity (note that there was no such concern in the case of the Newtonian simulations because RVs exactly follow the underlying gravity, or $\Gamma=0$, before adding scatter). Those extreme individuals are excluded when $\Gamma_{\rm{med}}$ is derived by consolidation in a mock sample.}

 {Figure~\ref{fig:hist_Gam_mockGaia} shows the distributions of $\Gamma_{\rm{med}}$ from mock Gaia observations. The biases or shifts of $\Gamma$ are overall similar to the case ofthe  mock Newtonian observations shown in Figure~\ref{fig:hist_Gam_mockNewton}, but show some differences in detail. Unlike the Newtonian case, the transition regime is displayed in the left column because some (though small) shifts occur. In the MOND regime, the shifts and the widths of the distributions are overall similar to those in the Newtonian case. The exception is the shift in the middle row. This appears to be due to the relatively large input value of $\Gamma$ while the median of the distribution is similar to the bottom row. In the transition+MOND regime, the widths are significantly narrower than in the Newtonian case, but the widths from the mock Newtonian and mock Gaia cases bracket the widths of the consolidated distributions shown in the right column of Figure~\ref{fig:Gam_probdist_transmond}. In line with Figure~\ref{fig:Gam_probdist_transmond} and Figure~\ref{fig:hist_Gam_mockNewton}, Figure~\ref{fig:hist_Gam_mockGaia} indicates that Newtonian gravity is inconsistent with the Gaia data, if any, even more strongly than in the other figures.  }

\begin{figure*}
    \centering
    \includegraphics[width=1.0\linewidth]{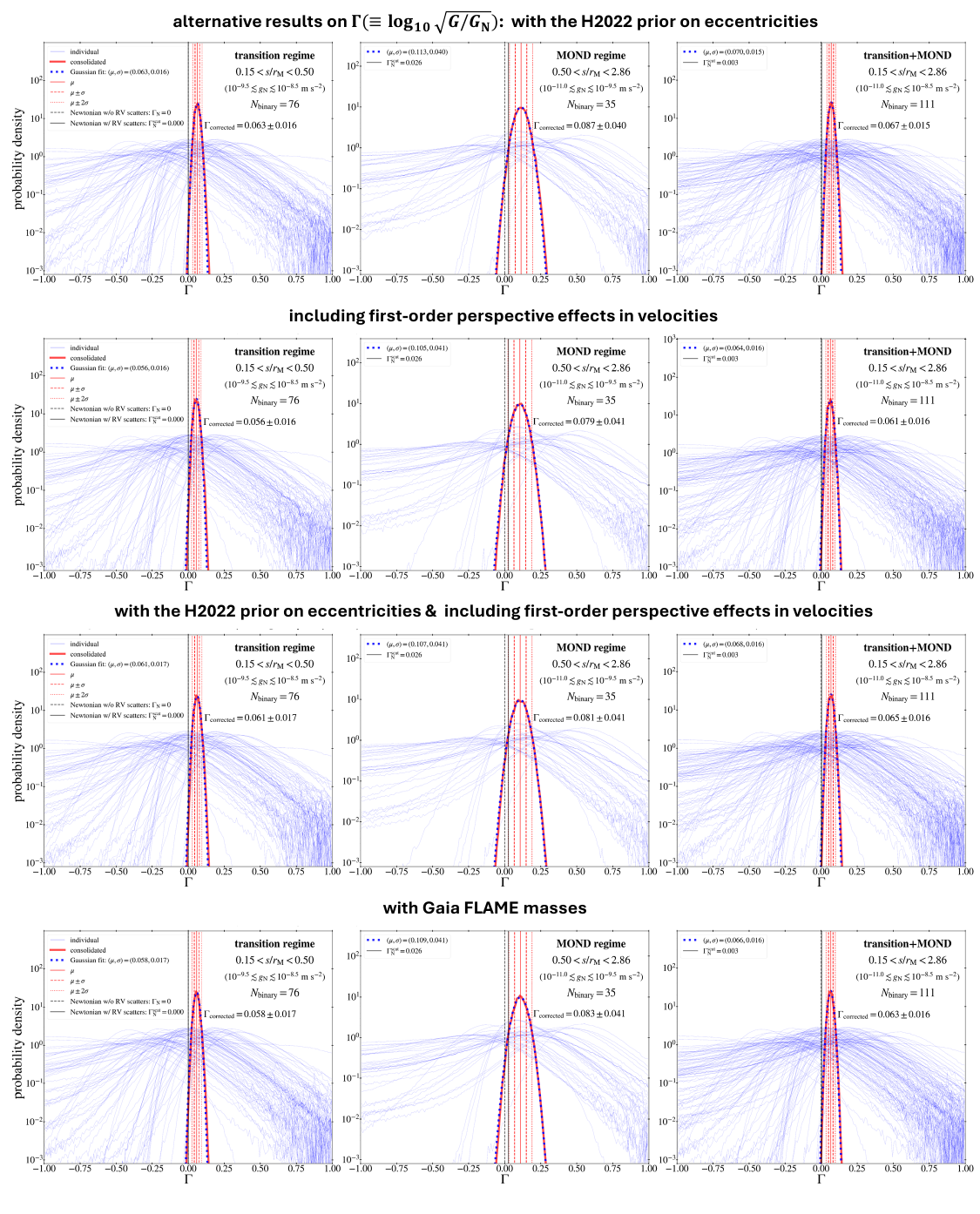}
    \caption{  {Each row is same as the upper row of Figure~\ref{fig:Gam_probdist_transmond} but with a correction/variation (or corrections/variations) to the nominal inputs as indicated. See the text for the details. } }
    \label{fig:Gam_probdist_alternative}
\end{figure*}

 {So far all results are based on the nominal/standard inputs. Here, reasonable corrections or variations to the inputs are considered to estimate known (but ignored) systematics. They include the H2022 prior on eccentricities, perspective effects in observed velocity components, and Gaia FLAME masses. Note that perspective effects are calculated (see Appendix~\ref{sec:pers}) assuming that two stars in a pair are at the same distance, and thus I refer to them as first-order estimates. When the tiny distance difference is not measurable, it appears most reliable to assume that it is zero (see Appendix~\ref{sec:pers} for further details). Note also that FLAME masses are available for about two thirds of the stars, and thus nominal masses are replaced only for those stars.} 

 {Figure~\ref{fig:Gam_probdist_alternative} shows results on consolidated gravity in the MOND and transition regimes with the alternative inputs. Notice that I also consider the case with both the H2022 prior and the perspective effects. The H2022 prior slightly increases $\Gamma$ compared with the thermal prior while correcting velocities for the perspective effects slightly decreases it. Thus, if both are considered, $\Gamma$ is essentially unchanged from the nominal results. FLAME masses do not lead to any tangible change either, as expected from Figure~\ref{fig:mass}. In summary, any ignored known systematics lead to negligibly small variations of $\Gamma$ compared with its statistical uncertainties based on the present data.}

\section{Discussion} \label{sec:disc}

\subsection{Meaning of the results and an exploration of possible systematic errors} \label{sec:meaning}

In this work, isolated (both in projection and line-of-sight) wide binaries are carefully selected (Section~\ref{sec:sample}) and used for Bayesian 3D modeling of orbits and gravity. This work is based on two key assumptions that (1) the instantaneous relative motion between the stars in any observed isolated binary is at a phase of an unknown elliptical orbit and (2) the instantaneous motion can be modeled with pseudo-Newtonian gravity with a generalized effective gravitational constant $G_{\rm{eff}}$. 

When the center of mass of a binary system orbits in the Galaxy, the relative motion between the two stars will suffer from perturbations due to long-range chance encounters with other stars and the Galactic tidal field (see \citealt{jiang2010}). Such perturbations can accumulate over a long period of time since the birth of the binary system (e.g., several giga years), leading to an evolved statistical property of binary populations as studied by \cite{jiang2010} in Newtonian gravity. The maximum projected separation for the wide binaries in the nominal sample (Table~\ref{tab:sample}) is $\approx 25$~kau. This corresponds to $\approx 0.074 r_J$ where $r_J \approx 1.65$~pc is the Jacobi radius for the median total mass of $1.82 M_\odot$ based on Equation~(43) of \cite{jiang2010}. According to the Newtonian simulation results shown in Figure~7 of \cite{jiang2010}, presently observed binaries with $s/r_J \la 0.074$ would be expected to follow Keplerian behavior well. This indicates that the assumption of elliptical orbits is valid as long as Newtonian gravity is tested with the wide binaries of this work. Thus, the disagreement (Figure~\ref{fig:Gam_probdist_transmond}) with Newton in low-acceleration regimes (along with the agreement in a strong-acceleration regime) is an unacceptable result in Newtonian gravity.  {Note also that the local dark matter density ($\la 0.01 M_\odot\text{ pc}^{-3}$: see \cite{read2014}) inferred assuming standard gravity cannot make any contribution to the gravitational anomaly.}

However, in MOND or other modified gravity it is unclear how good the two assumptions will be and how to interpret the Bayesian-inferred values of $\Gamma$ ($> 0$) (i.e.,  $G_{\rm{eff}} > G_{\rm{N}}$). In MOND gravity (e.g.\ \citealt{bekenstein1984,milgrom2010}), the rather strong external field of $\approx 1.9 a_0$ of the Galaxy can warp the orbit and the orbit may not be closed when the external field is inclined with respect to the rotation axis \citep{brada2000}. In such gravity models, numerical simulations are needed to produce mock wide binaries and to infer $\Gamma$ in the instantaneous motions of the binaries. The values of $\Gamma$ can then be compared with those for the Gaia binaries.

Therefore, a specific modified gravity model cannot be accurately tested with the measured value of $\Gamma$. The agreement of the Bayesian-inferred values of $\Gamma = 0.085\pm0.040$ (MOND regime) and $\Gamma = 0.063\pm0.015$ (transition+MOND regime) with the generic prediction of MOND gravity ($\approx 0.07$) needs to be taken with a grain of salt. In this respect, it would be interesting to reconsider the wide binaries with wildly large values of $\Gamma$ (Table~\ref{tab:outlier}). These systems were rejected as systems with suspicious kinematic contaminants  {or abnormal RV data} because they appeared to be incompatible not only with Newtonian gravity but also with MOND gravity. However, it is unclear whether MOND gravity could accommodate these systems until realistic simulations are used to probe all possible orbital motions. 

 {Whatever the case with MOND gravity, if at least some of these systems are pure binaries without kinematic contaminants or abnormal RV data, standard gravity will be ruled out more severely. Perhaps, it is unlikely that all of them have kinematic contaminants or abnormal RV data for several reasons: (1) all of them already satisfy the presently highest data qualities and 13 of the 16 stars have {\tt ruwe} $< 1.2$; (2) Gaia DR3 astrometric solutions are based on data collected over a time span of 34 months, and thus any hidden close binaries with periods of the same order or shorter have already been flagged and precluded; (3) the availability of the Gaia-measured RVs with \emph{relatively} good precision indicates that the stars are likely to satisfy relatively good single-star solutions for spectroscopic observations; and (4) the posterior PDFs of all parameters are well-behaved and the fitted values of $e$ are particularly reasonable. Therefore, it may well be that at least some of these binaries \emph{individually} rule out standard gravity. }

\subsection{Wide binary results in the context of recent small-scale gravity tests} \label{sec:smallscale}

The inference of $\Gamma$ from the observed wide binaries with $s\la 0.1$~pc is a direct measurement without reference to any modified gravity. In inferring $\Gamma$, only the Newtonian calculation is performed to provide the reference point with respect to which $\Gamma$ is defined. As discussed in Section~\ref{sec:meaning}, numerical simulations are required to accurately test a specific nonstandard theory of gravity using the measured value of $\Gamma$. In this respect, a measurement should not be confused with a theory.

Isolated wide binary stars with a separation more than several kilo-astronomical units in the solar neighborhood are unique laboratories to probe low-acceleration gravity for two reasons. First, the dynamics is simple because only two point-like objects are involved (if suspicious impure systems are properly rejected in advance) and their motions can be measured accurately. Second, wide binaries can test the basic nature of gravity generic to a broad class of nonstandard gravity models, that is, whether gravity is boosted as a function of internal acceleration and how much. If the Sun had a companion star (or an observable bright component) several kilo-astronomical units away, the fictitious solar binary system would provide an ideal definite laboratory to directly probe gravity at low acceleration. 

Without a directly observable object at a distance of more than several kilo-astronomical units from the Sun, the observed comets in a relatively inner part of the solar system have been considered to test gravity models by calculating (or extrapolating) their dynamical evolution up to large distances (e.g., \citealt{krolikowska2020}). For example, \cite{vokrouhlicky2024} have inferred the distributions of $1/a$ (or, binding energy where $a$ is the semi-major axis of the orbit) for hundreds of long-period comets in Newtonian and AQUAL models. This is an example of testing specific theories through modeling based on modeling assumptions including the MOND transition/interpolating function. In contrast, as stated above, the inferred value of $\Gamma$ from wide binaries is a pure measurement, based on movements observed directly at large separations and without reference to any nonstandard theories of gravity. In other words, motions in the MOND regime are directly caught in action rather than extrapolated.

For certain theories of nonstandard gravitational dynamics, even planetary motions can be used to test them. Because planets are subject to high gravitational accelerations that are orders of magnitude higher than $a_0$, there is no gravitational boost in the sense of boosted acceleration or orbital velocity.\footnote{As shown in this work and previous work, wide binaries also confirm that there is no gravitational boost for internal gravitational acceleration higher than $\approx 10^2 a_0$.} However, specific modified gravity theories such as AQUAL and QUMOND predict that the gravitational potential of the Sun becomes nonspherical even in the inner solar system due to the the external field effect of the Galaxy. This nonsphericity can be quantified by a quadrupole moment \citep{milgrom2009} of the potential. 

The quadrupole moment can be viewed as a second-tier gravitational anomaly that depends on the details of a specific model, whereas an acceleration-dependent gravitational anomaly such as the radial acceleration relation in galaxies \citep{mcgaugh2016} and the $s/r_{\rm{M}}$-dependent value of $\Gamma$ in wide binaries from this work is a first-tier anomaly that is generic to the MOND (modified gravity) paradigm. \cite{hees2014} use 9 yr of Cassini range and Doppler measurements to infer a value of the quadruple moment that is consistent with zero. The \cite{hees2014} value of the quadrupole moment is used by \cite{desmond2024} to test some MOND models and MOND transition/interpolating functions. As \cite{desmond2024} demonstrate, the quadrupole moment can strongly constrain the MOND external field effect and/or the MOND interpolating function if the measured value of the quadrupole moment is confirmed and the MOND interpolating function can be well constrained in a wide acceleration range around $a_0$ by more accurate galactic rotation curves in the future.

 {Unlike the above two studies \citep{desmond2024,vokrouhlicky2024} on the dynamics of solar system bodies, \cite{brown2023} argue for MOND gravity. Kuiper Belt objects (KBOs) with semi-major axes greater than 250~au exhibit orbital anomalies that require an unidentified planet ($5-10$ times the mass of Earth) at an average distance of about 500~au in standard gravity, but may be alternatively explained by the quadrupole and octupole moments of MOND modified gravity \citep{brown2023} (but see \cite{vokrouhlicky2024} for possible issues with their results). Future studies of KBOs may be interesting. Also, it is interesting to note another recent small-scale test with the dynamics of open clusters (although it is not a study of the solar system). \cite{kroupa2024} find that asymmetrical tidal tails of open clusters strongly imply gravitational anomalies that agree with MOND gravity but strongly rule out standard gravity.}

At the heart of the debate between standard gravity (with dark matter) and MOND or modified gravity (without dark matter) in the Universe is whether a direct gravitational boost occurs at low accelerations $\la a_0$. The low-acceleration gravitational boost is the fundamental nature that is shared by all models belonging to the nonstandard paradigm, while solar system tests discussed above are more about the specifics of models. This means that even if solar system tests truly rule out certain modified gravity models, they cannot rule out the MOND or modified gravity paradigm as long as the low-acceleration gravitational anomaly is experimentally true. This fundamental nature cannot be so easily tested with galactic dynamics (although feasible in principle and will be done eventually) because the prediction of modified gravity needs to be distinguished from that of the cosmologically inferred dark matter density \citep{chae2022c}. Unlike galactic dynamics, wide binary internal dynamics does not care whether the dark matter density is zero or what general relativity infers cosmologically, and thus most directly tests the low-acceleration gravitational anomaly. What the results from this work and several recent studies \citep{chae2023a,chae2024a,chae2024c,hernandez2023,hernandez2024a,hernandez2025} show clearly is that the fundamental low-acceleration gravitational boost is present in the currently available data. 

\subsection{Future direction of wide binary gravity research}

Prior to this work, wide binary gravity research has been based only on the observed sky-projected velocity $v_p$, and all the previous methods are purely statistical, meaning that an individual binary (except for the specific study of $\alpha$ Centauri; see \citealt{banik2018,banik2019}) does not provide any (even weak) constraint on gravity. Each method considers/proposes a gravity-sensitive quantity $Y$, and derives and compares the observational estimates and theoretical predictions of the medians or distributions of $Y$ in bins of an independent variable $X$, which is usually a proxy of Newtonian acceleration $g_{\rm{N}}(\equiv G_{\rm{N}} M_{\rm{tot}}/r^2)$ rather than itself. The statistical methods from the literature are summarized in Table~\ref{tab:statgravity}.

\begin{table*}
  \caption{Statistical methods of testing gravity based only on sky-projected velocities of wide binaries }\label{tab:statgravity}
\begin{center}
  \begin{tabular}{ccccc}
  \hline
 method  & $Y$ (gravity sensitive) &  $X$ (independent) &  comments  & references    \\
 \hline
 stacked velocity profile &  $v_p$ &  $s$  &  pure binaries desired &  H12, H23, C24a, H25  \\
 normalized velocity profile &  $\tilde{v}(\equiv v_p/v_c(s))$ &  $s/r_{\rm{M}}$  & pure binaries desired  &  C24c  \\
 $\tilde{v}$-distribution &  histogram of $\tilde{v}$  &  $s$ (or $s/r_{\rm{M}}$) & $s/r_{\rm{M}}$ desired, but $s$ used by many  &  B18, P18, H24, C24c  \\
 acceleration plane test &  $g(\equiv v^2/r)$ &  $g_{\rm{N}}(\equiv G M_{\rm{tot}}/r^2)$  & Monte Carlo deprojection required  &  C23a, C23b  \\ 
 \hline
\end{tabular}
\end{center}
Note. (1) Variables: $s=$ sky-projected separation, $r_{\rm{M}}=$ MOND radius (Equation~(\ref{eq:MONDradius})), $v_p=$ magnitude of the sky-projected relative velocity, $v_c(s)(\equiv\sqrt{G_{\rm{N}} M_{\rm{tot}}/{s}}) =$ Newtonian circular velocity at $s$, $v(\equiv|\mathbf{v}|)=$ magnitude of the 3D relative velocity $\mathbf{v}$, $r=$ physical separation in 3D space. (2) References: H12 - \cite{hernandez2012}, H23 - \cite{hernandez2023}, C24a - \cite{chae2023a}, H25 - \cite{hernandez2025}, C24c - \cite{chae2024c}, B18 - \cite{banik2018}, P18 - \cite{pittordis2018}, H24 - \cite{hernandez2024a}, C23a - \cite{chae2023a}, C23b - \cite{chae2023b} (Python codes) 
\end{table*}

The statistical methods listed in Table~\ref{tab:statgravity} have certain limitations. Among others, the actual physical velocity $\mathbf{v}$ is not used and the individual binary-specific eccentricity is usually ignored. For the latter matter, only a series of studies \citep{chae2023a,chae2024a,chae2024c} by the same author used individual constraints on eccentricities from a Bayesian analysis \citep{hwang2022} of the angle between the projected relative displacement and the projected relative velocity. However, these studies are still not fully satisfactory because eccentricities are not self-determined for the gravity tested. These limitations are overcome by the Bayesian 3D modeling introduced in this study. Moreover, the Bayesian 3D modeling naturally takes into account all observational uncertainties including the individually allowed mass uncertainties in inferring gravity.

Therefore, I believe that the Bayesian 3D modeling approach initiated in this study will be the ultimate direction of gravity research with snapshot motions of wide binaries. This approach requires accurate measurements of all three components of $\mathbf{v}$. The pilot study carried out here was limited by the relatively large uncertainties of $v_r$. Fortunately, there exist a number of telescopes that can measure RVs accurately and observational campaigns are already taking place. However, I also think that at least some of the statistical methods listed in Table~\ref{tab:statgravity} will continue to be useful as approaches complementary to the Bayesian 3D approach, as they can be applied to much larger samples without measured RVs. 

Wide binary gravity tests up to the present work have paid most attention to the detection and confirmation of the low-acceleration gravitational anomaly. Now that multiple independent approaches give convergent results, future studies need to pay more attention to the accurate characterization of gravity as a function of acceleration. For that purpose, I envisage that RVs need to be accurately measured for several hundreds (or thousands) of wide binaries in the transition and MOND regime with separation $2\la s \la 50$~kau.  

As noted in Section~\ref{sec:meaning}, numerical evolution of wide binaries in a specific modified gravity model is needed to rigorously test the model with the Bayesian measured value of $\Gamma$. Thus, Bayesian 3D modeling with a large number of RV measurements needs to be accompanied by a large suite of numerical simulations.

\section{Conclusion and follow-up work} \label{sec:conc}

This work has developed a new Bayesian 3D modeling method for wide binary internal dynamics in the context of testing gravity in low acceleration $\la 10^{-9}$~m\,s$^{-2}$. It is shown with mock snapshot observations of wide binary internal dynamics that the method can individually infer PDFs of the model parameters that correctly encompass the input values of the gravity anomaly parameter $\Gamma$ (Equation~(\ref{eq:anomaly})) and the orbital parameters. Each PDF of $\Gamma$ is broad given the snapshot nature of the data, but PDFs of $\Gamma$ from a number of independent wide binaries in a similar acceleration (i.e., internal gravity) regime can be combined to result in a much narrower consolidated distribution. The consolidated distribution can be well approximated by the Gaussian function and accurately reproduces the input value of $\Gamma$. 

The Bayesian 3D modeling method is applied to a highest-quality sample of 312 pure wide binaries carefully selected from the Gaia DR3 database. This sample covers a broad dynamic range of $0.02 < s/r_{\rm{M}} < 2.86$, or $10^{-11.0} \la g_{\rm{N}} \la 10^{-6.7}$~m\,s$^{-2}$. Notably, this sample has uncertainties of $v_r$ (the relative radial velocity between the pair) in the range $168\la \sigma_{v_r} \la 380$~m\,s$^{-1}$ with a median of $\approx 280$~m\,s$^{-1}$. The following results are derived.
\begin{enumerate}
    \item With 125 wide binaries in a high-acceleration range (``Newtonian regime'') $0.02 < s/r_{\rm{M}} < 0.08$ ($10^{-7.9} \la g_{\rm{N}} \la 10^{-6.7}$~m\,s$^{-2}$), the Bayesian inference returns $\Gamma = 0.000\pm 0.011$ or $\gamma_g=1.00\pm 0.05$. Thus, internal dynamics of binaries is consistent with Newtonian gravity in acceleration regimes $\ga 100 a_0$ as far as the first-tier orbital motions are concerned. No first-tier gravitational anomaly is observed in this regime.
    \item With 76 wide binaries in an intermediate-acceleration range (``transition regime'') $0.15 < s/r_{\rm{M}} < 0.50$ ($10^{-9.5} \la g_{\rm{N}} \la 10^{-8.5}$~m\,s$^{-2}$), the Bayesian inference returns $\Gamma = 0.057\pm 0.017$ or $\gamma_g=1.30_{-0.09}^{+0.11}$. Newtonian gravity is in tension with these wide binaries at a significance of $3.4\sigma$.
    \item With 35 wide binaries in a low-acceleration range (``MOND regime'') $0.50 < s/r_{\rm{M}} < 2.86$ ($10^{-11.0} \la g_{\rm{N}} \la 10^{-9.5}$~m\,s$^{-2}$), the Bayesian inference returns $\Gamma = 0.085\pm 0.040$ or $\gamma_g=1.48_{-0.23}^{+0.30}$. Newtonian gravity is in tension with these wide binaries at a significance of $2.1\sigma$. While the value of $\Gamma$ appears to be higher than in the transition regime, the statistical uncertainty is much larger.
    \item With 111 wide binaries in a combined acceleration range (``transition+MOND regime'') $0.15 < s/r_{\rm{M}} < 2.86$ ($10^{-11.0} \la g_{\rm{N}} \la 10^{-8.5}$~m\,s$^{-2}$), the Bayesian inference returns $\Gamma = 0.063\pm 0.015$ or $\gamma_g=1.34_{-0.08}^{+0.10}$. Newtonian gravity is in tension with these wide binaries at a significance of $4.2\sigma$. 
    \item The inferred values of $\Gamma$ in the MOND and transition+MOND regimes are consistent with a generic value $\Gamma \approx 0.07$ or $\gamma_g \approx 1.4$ in the low-acceleration ($\la a_0$) regime predicted by classical theories of MOND gravity \citep{bekenstein1984,milgrom2010}. However, to test a specific theory accurately, numerical simulations of wide binary orbits in the theory are required.
    \item The magnitude of the gravitational anomaly quantified by $\Gamma$ agrees well with the results from three statistical methods \citep{chae2023a,chae2024a,chae2024c,hernandez2023,hernandez2024a,hernandez2025} based on much larger samples.
    \item  {Eight wide binaries with abnormal PDFs of $\Gamma$ are found (Table~\ref{tab:outlier}). If any of these does not have kinematic contaminants or abnormal RV data, it will \emph{individually} rule out standard gravity. These systems require further studies.}
    \item Eccentricities of orbits inferred by the Bayesian 3D approach agree well with those inferred by \cite{hwang2022} through an independent Bayesian approach that usees only projected quantities and thus does not use RVs. In addition, the uncertainties of the new measurements are smaller than in \cite{hwang2022} as a consequence of the additional information from RVs.
\end{enumerate}

The present results are limited by the relatively large uncertainties ($\sim 280$~m\,s$^{-1}$) of $v_r$ that are $\approx 10$ times larger than the uncertainties of the sky-projected velocity components. At the time of this writing, for a significant number of wide binaries in a MOND regime, RVs with﻿ uncertainties $\la 50$~m\,s$^{-1}$ are being gathered from various sources/observations.﻿ With these newly measured RVs, the precision of $\Gamma$ in the MOND regime will improve dramatically. In this study, stars brighter than $M_G=3.8$ were excluded to obtain a uniform sample of main-sequence stars by precluding possible contamination by bright subgiant or giant stars.  {However, as long as involved issues (such as higher probabilities of hiding close companions and availability of accurate masses) can be resolved,} it is interesting to consider stars brighter than $M_G=3.8$ in the future because brighter stars allow easier measurements of RVs.

\section*{Acknowledgments} 
 {I thank the anonymous referee for useful and constructive suggestions that led to improved and richer analyses and presentations.} I have discussed this work with a number of excellent scientists. In particular, during the summer of 2024 I have visited three institutions and had interesting discussions of this work based on preliminary results. Those scientists include Stacy McGaugh and Tobias Mistele at Case Western Reserve University, Arthur Kosowsky, David Turnshek, and Andrew Zentner at the University of Pittsburgh, and Federico Lelli, Laura Magrini, and Luca Pasquini (a visitor there) at Arcetri Astrophysical Observatory. Luca Pasquini is further thanked for helpful conversations on the RVs of stars. I thank Bill Janesh at CWRU for help with computing during the summer visit. I thank mathematicians Daniela Calvetti and Erkki Somersalo at CWRU for illuminating discussions on statistical consolidation. I thank Xavier Hernandez and Pavel Kroupa for discussions on wide binary gravity tests. I thank Byeong-Cheol Lee, Dongwook Lim, and David Turnshek for collaboration on measuring the RVs of stars. I thank Yong Tian and Youngsub Yoon at Sejong for assistance and undertaking new statistical analyses of wide binaries.  {Y. Yoon is further thanked for carrying out independent calculations of the perspective effects based on the equations of \cite{shaya2011} with which the calculations of Appendix~\ref{sec:pers} agree well.} Finally, I thank Moti Milgrom for encouragement and inspirations of gravity research. Initial results of this work were presented by an invited talk given at the Korean Society of High Energy Physics 2024 Fall Meeting (https://indico.kshep.kr/event/839/contributions/4051). This work was supported by the National Research Foundation of Korea (grant No. NRF-2022R1A2C1092306). 

\bibliographystyle{aasjournal}

\newpage

\appendix

\section{Relating the orbital plane to the observer's frame with the Euler angles} \label{sec:Euler}

\begin{figure*}[h]
    \centering
    \includegraphics[width=0.7\linewidth]{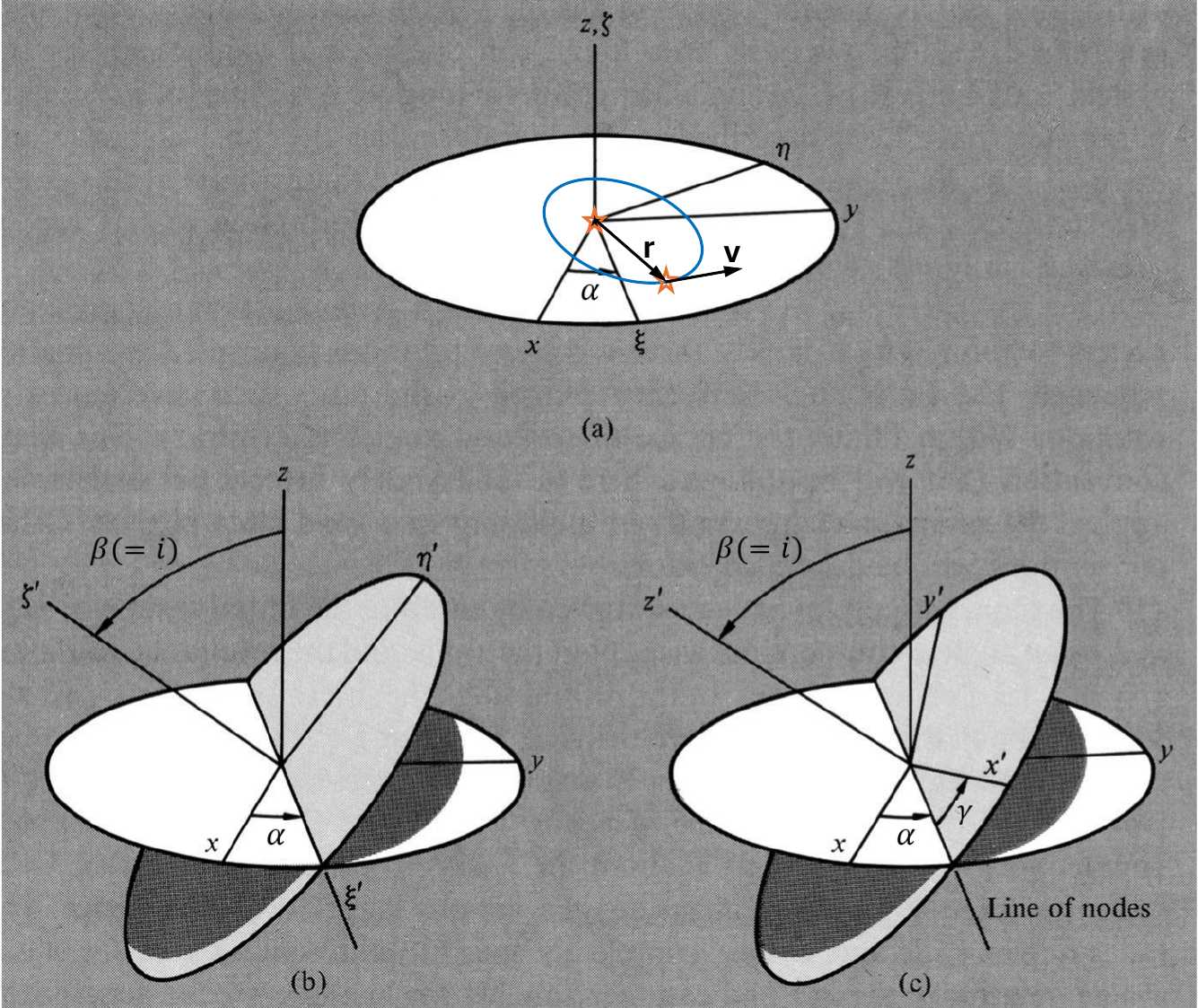}
    \caption{(Adapted from Figure~4.7 of \cite{goldstein2002}) The $xyz$ frame including the orbit lying on its $xy$ plane is related by three Euler angles ($\alpha$, $\beta$, $\gamma$) to the observer's $x^\prime y^\prime z^\prime$ frame where the $x^\prime y^\prime$ plane represents the observed portion of the sky. When $\alpha=\gamma=0$ and $\beta=i$, this figure is equivalent to Figure~\ref{fig:orbit}.  }
    \label{fig:Euler}
\end{figure*}

Figure~\ref{fig:Euler} shows an $xyz$ frame including an elliptical orbit lying on its $xy$ plane, arbitrarily oriented with respect to the observer's $x^\prime y^\prime z^\prime$ frame. The two frames can be related by three successive counterclockwise rotations with Euler angles $\alpha$, $\beta$, and $\gamma$, as shown in (a), (b), and (c) of Figure~\ref{fig:Euler}. The angles $\alpha$ and $\gamma$ are rotations about the $z$ and $z^\prime$ axes, while the angle $\beta$ is a rotation about an intermediate ``$x$-axis'' (the coinciding $\xi$ and $\xi^\prime$ axes in the figure). The three transformations from the initial $xyz$ frame, through the intermediate frames $\xi\eta\zeta$ and $\xi^\prime\eta^\prime\zeta^\prime$, to the final $x^\prime y^\prime z^\prime$ frame can be described by the multiplication of three rotation matrices. The product matrix is given (see Equation (4.46) of \cite{goldstein2002}) by
\begin{equation}
 \mathbf{\Lambda} = \mathbf{\Lambda}(\alpha,\beta,\gamma) = \left[\begin{array}{ccc}
          \cos\gamma\cos\alpha-\cos\beta\sin\alpha\sin\gamma  & \cos\gamma\sin\alpha+\cos\beta\cos\alpha\sin\gamma  &  \sin\gamma\sin\beta \\
          -\sin\gamma\cos\alpha-\cos\beta\sin\alpha\cos\gamma  & -\sin\gamma\sin\alpha+\cos\beta\cos\alpha\cos\gamma  &  \cos\gamma\sin\beta \\
           \sin\beta\sin\alpha  & -\sin\beta\cos\alpha &  \cos\beta  \\
    \end{array} \right].  
    \label{eq:EulerMat}  
\end{equation}
This matrix transforms the three components of an arbitrary vector from the $xyz$ frame to the $x^\prime y^\prime z^\prime$ frame.

In simulating mock snapshot observations of elliptical orbits in the $x^\prime y^\prime z^\prime$ frame, it is sufficient to fix $\alpha=0$ if the periastron is allowed to be arbitrary (i.e., $\phi_0$ is random in Figure~\ref{fig:orbit}) as is the case in this work. In this case, let $\beta=i$ and $\gamma=\theta$, then the transformation matrix takes a simpler form as
\begin{equation}
 \mathbf{\Lambda}(i,\theta) = \left[\begin{array}{ccc}
          \cos\theta  & \cos i \sin\theta  &  \sin\theta\sin i \\
          -\sin\theta  & \cos i\cos\theta  &  \cos\theta\sin i \\
            0  & -\sin i &  \cos i  \\
    \end{array} \right].  
    \label{eq:EulerMat_2angles}  
\end{equation}
Finally, in the special case that $\alpha=\gamma=0$ and $\beta=i$, which are shown in Figure~\ref{fig:orbit}, the above matrix becomes
\begin{equation}
 \mathbf{\Lambda}(i) = \left[\begin{array}{ccc}
          1  & 0  &  0 \\
          0  & \cos i  &  \sin i \\
          0  & -\sin i &  \cos i  \\
    \end{array} \right].  
    \label{eq:EulerMat_special}  
\end{equation}

Now consider the orbit shown in Figure~\ref{fig:Euler} representing the relative motion of one star from the other. In the $xyz$ frame, the relative displacement vector and the relative velocity vector are given by
\begin{equation}
 \mathbf{r} = r\left[\begin{array}{c}
        \cos\phi  \\
        \sin\phi  \\
         0 \\
    \end{array} \right]\text{ and }   
    \mathbf{v} = v \left[\begin{array}{c}
        |\cos\psi|{\rm{sgn}}(dx/d\phi)  \\
        \tan\psi |\cos\psi|{\rm{sgn}}(dx/d\phi)  \\
         0 \\
    \end{array} \right]
    \label{eq:relvec},  
\end{equation}
where $v=v(r)$, $\phi$, $\psi$, and $dx/d\phi$ can be found in Section~\ref{sec:demo}. In the $x^\prime y^\prime z^\prime$ frame the relative displacement vector and the relative velocity vector are then given by $\mathbf{r}^\prime=\mathbf{\Lambda}\mathbf{r}$ and $\mathbf{v}^\prime=\mathbf{\Lambda}\mathbf{v}$. For $\mathbf{\Lambda}=\mathbf{\Lambda}(i,\theta)$, these become
\begin{equation}
 \mathbf{r}^\prime = r\left[\begin{array}{c}
        \cos\phi\cos\theta+\sin\phi\cos i \sin\theta  \\
        -\cos\phi\sin\theta+\sin\phi\cos i \cos\theta  \\
         -\sin\phi\sin i \\
    \end{array} \right]
    \label{eq:relpos_2angles},  
\end{equation}
and    
\begin{equation}
    \mathbf{v}^\prime = v \left[\begin{array}{c}
        |\cos\psi|{\rm{sgn}}(dx/d\phi)\cos\theta + \tan\psi |\cos\psi|{\rm{sgn}}(dx/d\phi) \cos i \sin\theta  \\
       -|\cos\psi|{\rm{sgn}}(dx/d\phi)\sin\theta + \tan\psi |\cos\psi|{\rm{sgn}}(dx/d\phi)\cos i \cos\theta  \\
       -\tan\psi |\cos\psi|{\rm{sgn}}(dx/d\phi)\sin i   \\
    \end{array} \right].
    \label{eq:relvel_2angles}  
\end{equation}
For $\mathbf{\Lambda}=\mathbf{\Lambda}(i)$, Equations~(\ref{eq:positioncomponents}) and (\ref{eq:vcomponents}) are obtained. Note that $z^\prime$, $v_{z^\prime}$, $r_p (\equiv \sqrt{(x^\prime)^2 + (y^\prime)^2} = r\sqrt{\cos^2\phi +\sin^2\phi \cos^2 i} )$, and $v_p (\equiv \sqrt{(v_{x^\prime})^2 + (v_{y^\prime})^2} = v\sqrt{\cos^2\psi +\sin^2\psi \cos^2 i})$ in Equations~(\ref{eq:relpos_2angles}) and (\ref{eq:relvel_2angles}) do not depend on $\theta$ and are the same as those in Equations~(\ref{eq:positioncomponents}) and (\ref{eq:vcomponents}). This means that only if $r_p$ and $v_p$ (but not $x^\prime$ or $y^\prime$-components) are needed, $\mathbf{\Lambda}(i)$ suffices as in previous work \citep{chae2023a,chae2024a,chae2024c}.

\section{Perspective effects in proper motions and radial velocities} \label{sec:pers}

For a binary system, perspective effects \citep{shaya2011} refer to the apparent relative motions between two stars, observed as biased PMs and RVs, due to the motion of the barycenter with respect to the Sun in the physical 3D space. Perspective effects are expected to be very small for wide binaries in the solar neighborhood with projected separation $s<20$~kau \citep{elbadry2019}. Nevertheless, here I describe a direct numerical algorithm and use it to investigate the perspective effects. 

A binary is described using the notation introduced in Section~\ref{sec:bayes}. With the R.A. and decl. of two stars $A$ and $B$, the barycenter's azimuthal and polar angles $\theta$ and $\phi$ (which should not be confused with those used in Appendix~\ref{sec:Euler}) in the equatorial frame are given by
\begin{equation}
 \phi = (\alpha_A + \alpha_B)/2 \text{\hspace{1em}and\hspace{1em}} \theta = \pi/2 - (\delta_A + \delta_B)/2  ,
    \label{eq:spherical}
\end{equation}
where all angles are expressed in radians and I assume that the barycenter is the middle between the two stars for simplicity. The system velocity components in the R.A., decl., and radial directions are given by
\begin{equation}
\left. \begin{array}{cl}
  V_{\alpha} = & 4.7404\times 10^{-3}\times d_M (\mu^\star_{\alpha,A}+ \mu^\star_{\alpha,B})/2\text{  km s}^{-1} \\ 
  V_{\delta} = & 4.7404\times 10^{-3}\times d_M (\mu_{\delta,A}+\mu_{\delta,B})/2\text{  km s}^{-1} \\
  V_r = & ({\rm{RV}}_A+{\rm{RV}}_B)/2\text{  km s}^{-1}
\end{array} \right\}.
\label{eq:Vsys_spher}
\end{equation}
In the geocentric Cartesian $XYZ$ frame the velocity components are
\begin{equation}
\left. \begin{array}{cl}
  V_{X} = & V_r\sin\theta\cos\phi-V_\delta\cos\theta\cos\phi-V_\alpha\sin\phi \\ 
  V_{Y} = & V_r\sin\theta\sin\phi-V_\delta\cos\theta\sin\phi+V_\alpha\cos\phi \\
  V_Z = & V_r\cos\theta+V_\delta\sin\theta
\end{array} \right\}.
\label{eq:Vsys_Cart}
\end{equation}

Now I consider mock physical motions of two stars in the 3D space with the system velocity given by Equation~(\ref{eq:Vsys_Cart}) for time $\Delta t$ into the future (or past). The current positions of the stars $A$ and $B$ are identified by subscript 0 while the future positions by subscript 1. At the current time $t_0$, the azimuthal and polar coordinates are obtained by the relations $\phi=\alpha$ and $\theta = \pi/2-\delta$ (with the required conversions of units). The positions of the two stars at $t_0$ and future time $t_1 = t_0 +\Delta t $ are given in the Cartesian coordinates by
\begin{equation}
\left. \begin{array}{ccc}
  X_{i,0} &= & R_i \sin\theta_i \cos\phi_i \\ 
  Y_{i,0} &= & R_i \sin\theta_i \sin\phi_i \\
  Z_{i,0} &= & R_i \cos\theta_i 
\end{array} \right\} \text{\hspace{1em}and\hspace{1em}}
\left. \begin{array}{ccc}
  X_{i,1} &= & X_{i,0} + V_X \Delta t \\ 
  Y_{i,1} &= & Y_{i,0} + V_Y \Delta t \\
  Z_{i,1} &= & Z_{i,0} + V_Z \Delta t
\end{array} \right\},
\label{eq:pos_Cart}
\end{equation}
where $i=A,B$. For the two radii, I take $R_A = d_M$ (the error-weighted mean of the two observed distances) and consider the case that they are the same or $R_B = R_A +\mathcal{N}(0,2s)$ where $s$ is the observed projected separation between the two stars.

At $t_1$, the spherical coordinates of the two stars are given by
\begin{equation}
\left. \begin{array}{ccc}
  R_{i,1} &= & \left(X_{i,1}^2 + Y_{i,1}^2 + Z_{i,1}^2 \right)^{1/2} \\ 
  \theta_{i,1} &= & \cos^{-1}\left(Z_{i,1}/R_{i,1} \right) \\
  \phi_{i,1} &= & \tan^{-1}\left(Y_{i,1}/X_{i,1} \right)
\end{array} \right\}.
\label{eq:pos_spher}
\end{equation}
The differences in the angular and radial motions of the two stars from $t_0$ to $t_1$ are then given by
\begin{equation}
\left. \begin{array}{ccc}
  v_{\rm{pers},r} &= & \frac{R_{B,1}-R_{B,0}}{\Delta t}  - \frac{R_{A,1}-R_{A,0}}{\Delta t} \\ 
  \mu_{\rm{pers},\delta} &= & \frac{-(\theta_{B,1}-\theta_{B,0})}{\Delta t}  - \frac{-(\theta_{A,1}-\theta_{A,0})}{\Delta t} \\
   \mu^\star_{\rm{pers},\alpha} &= & \sin\theta \left( \frac{\phi_{B,1}-\phi_{B,0}}{\Delta t}  - \frac{\phi_{A,1}-\phi_{A,0}}{\Delta t} \right)
\end{array} \right\},
\label{eq:pos_spher}
\end{equation}
where $\theta$ refers to the mean polar angle given in Equation~(\ref{eq:spherical}). These are the perspective effects of star $B$ with respect to star $A$. To correct for these perspective effects, they are added to the observed RV and PM components of star $A$. 

\begin{figure*}
    \centering
    \includegraphics[width=1.0\linewidth]{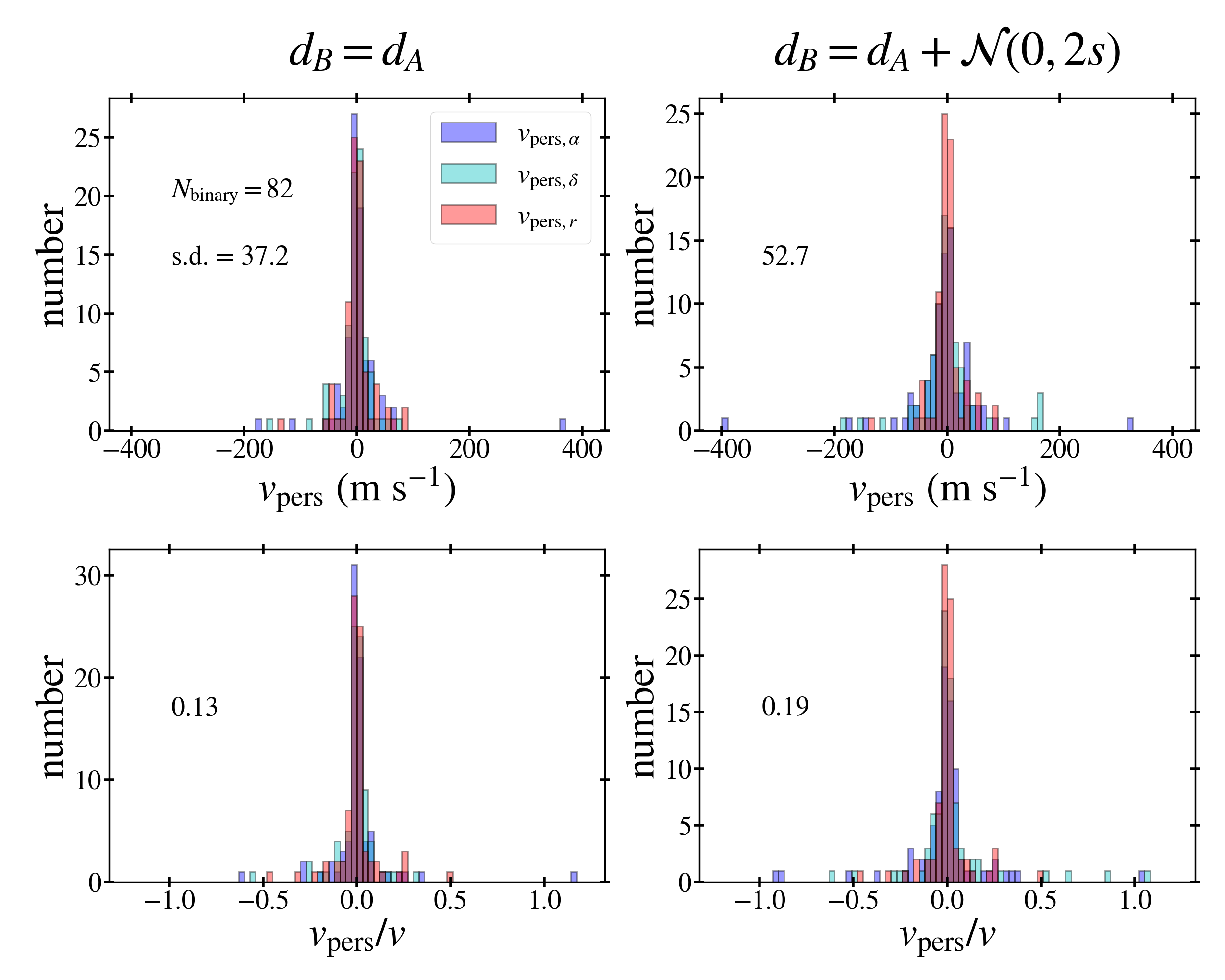}
    \caption{Perspective effects are shown for three relative velocity components due to PMs in R.A. and decl.\ directions and RVs of stars in wide binaries in the MOND regime with $s/r_{\rm{M}}>0.5$, where $s$ is the sky-projected separation between the two stars and $r_{\rm{M}}$ is the MOND radius. The results shown in the left column are obtained assuming that the two stars are at the same distance from the Sun while the right column shows the results assuming that two distances differ by normal deviates with a standard deviation of $2s$. The upper row shows perspective effects in physical units of m~s$^{-1}$ while the lower row shows the values normalized by physical 3D velocity magnitudes ($v$). Each panel indicates the standard deviation of $N_{\rm{binary}}$ perspective effects.}
    \label{fig:pers}
\end{figure*}

Perspective effects can be a concern only for wide binaries of relatively large separations. Figure~\ref{fig:pers} shows distributions of perspective effects for 82 wide binaries in the MOND regime ($s/r_{\rm{M}}>0.5$) from the statistical sample (see Table~\ref{tab:sample}). Here perspective effects of PMs are reexpressed as velocities. Because the system velocities (Equation~(\ref{eq:Vsys_spher})) are in random directions, the median perspective effects are zero, and only the dispersions are relevant here. For the case $d_B=d_A$, the standard deviation of all perspective velocities is $37.2$~m~s$^{-1}$, which is comparable to or much smaller than the Gaia DR3 measurement uncertainties of the velocities in the R.A., decl., and radial directions. The perspective velocities normalized by $v$ (which is the magnitude of the observed physical velocity) have a standard deviation of $0.13$. This does not necessarily mean that $v$ is systematically boosted by an appreciable amount because perspective effects can occur in the positive or negative direction for each velocity component. A proper estimate of any systematic bias in the inferred gravity that perspective effects may cause requires individual Bayesian modeling with the PMs and RVs corrected for the perspective effects. Comparison of the second row in Figure~\ref{fig:Gam_probdist_alternative} with the first row in Figure~\ref{fig:Gam_probdist_transmond} shows that the effect is very minor (with $\Delta\Gamma = -0.006$ in the MOND regime). 

For the case $d_B = d_A +\mathcal{N}(0,2s)$, the perspective effects have a standard deviation of $\approx 50$~m~s$^{-1}$, which is still comparable to or much smaller than the Gaia DR3 measurement uncertainties of the velocities in the R.A., decl., and radial directions. Moreover, because perspective effects can increase or decrease depending on the distance deviates in individual binaries, the large scatter does not necessarily lead to a greater reduction of the gravitational anomaly than in the case $d_B = d_A$. To see the effects of the distance deviate $\mathcal{N}(0,2s)$ on $\Gamma$, I have carried out Bayesian modeling of a statistical sample generated by correcting for the perspective effects in PMs and RVs with $d_B = d_A +\mathcal{N}(0,2s)$. Perhaps ironically, the magnitude of $\Delta\Gamma$ in the MOND regime can be less than in the case $d_B = d_A$. When the distance difference between the two stars is not measurable, it is perhaps most reliable to assume that it is zero as done in the main body of this paper.
 
\end{document}